\documentclass[epj]{svjour}
\usepackage{graphicx}
\usepackage{amsmath}
\usepackage{hyperref}
\usepackage{color}
\usepackage{cite}

\def\om{\omega}
\def\be{\begin{eqnarray}}
\def\ee{\end{eqnarray}}

\def\pt{\partial}
\newcommand{\lsim}{\stackrel{\scriptstyle <}{\phantom{}_{\sim}}}
\newcommand{\gsim}{\stackrel{\scriptstyle >}{\phantom{}_{\sim}}}
\def\s2{{*2}}

\def\const{{\rm const}}

\def\bup{{\rm b.up}}
\def\Elab{{\cal E}_{\rm lab}}

\def\om{\omega}
\def\pt{\partial}

\def\mev{{\rm MeV}}

\begin{document}

\title{RMF models with $\sigma$-scaled hadron masses and couplings
for description of
heavy-ion collisions below 2A~GeV}

\author{Konstantin A.~Maslov\inst{1,2} \and Dmitry N.~Voskresensky\inst{1, 2}}
\authorrunning{K.A.~Maslov, D.N.~Voskresensky}
\titlerunning{RMF models for description of HICs below 2$A$~GeV}

\institute{$^1$National Research Nuclear  University (MEPhI), Kashirskoe shosse 31, 115409 Moscow, Russia\\
$^2$Joint Institute for Nuclear Research,
		Joliot-Curie street 6,
		141980 Dubna, Russia}

\abstract{
Within the relativistic mean-field framework with hadron masses and coupling constants dependent on the mean scalar field we study properties of nuclear matter at finite temperatures, baryon densities and isospin asymmetries relevant for heavy-ion collisions at laboratory energies below 2$A$~GeV. Previously constructed (KVORcut-based and MKVOR-based) models for the description of the cold hadron matter, which differ mainly by the density dependence of the nucleon effective mass and symmetry energy, are extended for finite temperatures. The baryon equation of state, which includes nucleons and $\Delta$ resonances is supplemented by the contribution of the pion gas described either by the vacuum dispersion relation or with taking into account the $s$-wave pion-baryon interaction. Distribution of the charge between components is found. Thermodynamical characteristics on $T-n$ plane are considered. The energy-density and entropy-density isotherms are constructed and a dynamical trajectory of the hadron system formed in heavy-ion collisions is described. The effects of taking into account the $\Delta$ isobars and the $s$-wave pion-nucleon interaction on pion differential cross sections, pion to proton and $\pi^-/\pi^+$ ratios are  studied. The liquid-gas first-order phase transition is  studied within the same models in isospin-symmetric and asymmetric systems. We demonstrate that our models yield thermodynamic characteristics of the phase transition compatible with available experimental results. In addition, we discuss the scaled variance of baryon and electric charge in the phase transition region. Effect of the non-zero surface tension on spatial redistribution of the electric charge is considered for a possible application to  heavy-ion collisions at low energies.}	

\maketitle

	\section{Introduction}
Knowledge of the equation of state (EoS) of cold dense hadronic matter
is required for the description of  atomic nuclei and neutron stars after minutes-hours since their formation,
and EoS of warm and hot hadron matter is required for the description of supernovae
and heavy-ion collisions. Nowadays there exists a vast number of EoSs and a large
set of experimental and observational constraints, which an appropriate EoS should
fulfill \cite{Klahn:2006ir}. No one of existing EoSs satisfies all the  known constraints.  Flexible phenomenological approaches to  EoSs are introduced within relativistic mean-field (RMF) models with density dependent couplings, see \cite{Typel,Typel2005,Voskresenskaya,Oertel:2016bki} and refs. therein, and with hadron
masses and coupling constants dependent on the mean scalar field \cite{Kolomeitsev:2004ff}.
 The latter model has been generalized to the description of the isospin-symmetric hot
 hadronic matter including various baryon resonances and bosonic excitations   
  \cite{Khvorostukhin:2006ih,Khvorostukhin:2008xn,Khvorostukhin:2010aj} and was  applied
  to the description of heavy-ion collisions in a broad range of collision energies. Isospin-asymmetric matter (IAM) was not considered in mentioned works. Bosonic excitations were considered in the  ideal gas model.

 Recent measurements of  masses of the most massive binary pulsars demonstrated that
the maximum compact star mass, predicted by an EoS, should exceed  $2 M_{\odot}$. It was found that PSR  J1614-2230
has the mass $M = 1.928 \pm 0.017 \, M_\odot$ \cite{Demorest:2010bx,Fonseca:2016tux}
and  PSR~J0348+0432 has the mass $M = 2.01 \pm 0.04 \, M_\odot$~\cite{Antoniadis:2013pzd}.
These measurements rule out many soft EoSs of the purely nucleon matter. However  additional degrees of freedom may appear in dense neutron-star interiors, such as hyperons,  $\Delta$ isobars and meson condensates. This leads to an additional  softening of the EoS of the
 beta-equilibrium matter (BEM) resulting in a decrease of the maximum  neutron-star mass. On the other hand,  description of the particle flow in heavy-ion collisions requires a rather soft EoS
 of the isospin-symmetric matter (ISM) \cite{Danielewicz:2002pu}.
 Thereby it is challenging
 to construct an EoS, which would simultaneously satisfy the maximum neutron-star mass constraint together with the flow constraint.

 The working model with hadron masses and coupling constants dependent on the mean scalar field was constructed in \cite{Kolomeitsev:2004ff} and labeled  in \cite{Klahn:2006ir} as KVOR model. It satisfies the flow constraint for ISM and yields the maximum neutron-star mass
 $M\simeq 2.01 M_{\odot}$ for BEM, however only if no baryons other than nucleons are included into consideration. In our subsequent works
 \cite{Maslov:2015msa, Maslov:2015wba, Kolomeitsev:2016ptu,Kolomeitsev:2017gli} we constructed
 RMF models of the cold hadronic matter of arbitrary isospin composition  with effective hadron masses and coupling constants
 dependent on the scalar field with hyperons and $\Delta$ resonances taken into account
 \cite{Kolomeitsev:2016ptu}, as well as with the charged $\rho$ condensate
 \cite{Kolomeitsev:2017gli},  which successfully pass the maximum neutron-star mass constraint
  and the flow constraint
simultaneously  with many other constraints.

Various characteristics  of  heavy-ion collisions at
energies below a few $A$~GeV  have been extensively studied within the expanding fireball  framework
\cite{Siemens:1978pb,Friedman:1981qm,Mishustin:1983nv} and then within the ideal hydrodymamics and in various transport models, \textit{cf.} \cite{Arsene:2006vf,Buss:2011mx} and refs. therein. Densities reached at such collision energies are typically $\lsim 3 n_0$, where $n_0$
is the nuclear saturation density,
and temperatures are below the pion mass $m_\pi \simeq 140$ MeV. Only EoS for ISM has been studied within the expanding fireball  framework
\cite{Siemens:1978pb,Friedman:1981qm,Mishustin:1983nv}.
Reference \cite{Friedman:1981qm} used a
variational theory of nuclear matter for a description of nucleons in ISM and \cite{Voskresensky:1997vq}
 used the original Walecka RMF model  \cite{Walecka:1974qa}, whereas pions were considered with taking into account $p$-wave pion-baryon medium polarization effects. Then \cite{Voskresensky:1989sn,Migdal:1990vm,Voskresensky:1993ud} exploited  a modified Walecka
RMF model of \cite{Cubero:1987pr} for the nucleon ISM. A comparison with the data
on pion and nucleon spectra available at that time demonstrated advantages
of such description. At the freeze-out stage
an influence of the effects of the nuclear polarization on the pion spectra was considered within the prompt breakup
model  \cite{Senatorov:1989cg}.
 However, being  extended to describe BEM,
such EoSs do not satisfy  modern data on the high masses
of cold compact stars.

In this work we generalize the models with $\sigma$-scaled hadron masses
and couplings developed
for the description of the cold BEM in
\cite{Maslov:2015msa, Maslov:2015wba, Kolomeitsev:2016ptu,Kolomeitsev:2017gli}, now
for the case of the ISM and IAM formed in heavy-ion collision reactions for collision energies $\lsim 2A$~GeV, so the reached baryon  densities are $\lsim 3n_0$ and
temperatures are below $m_\pi$.
 In heavy-ion collisions the strangeness is approximately conserved. Thereby, the hyperon contribution to the thermodynamical values,  $\propto e^{-2m_H^*/T}$, where $m_H^*$ is the hyperon effective mass, can be neglected. Effect of boson $\sigma$, $\omega$, $\rho$ excitations  can be also disregarded for $n\lsim 3 n_0$ and $T<m_\pi$. The temperature dependence can be then  included only in  the nucleon and $\Delta$ kinetic energy terms and in pion quantities. We use a simplified  expanding fireball framework. In a subsequent work we plan to check the validity of our EoSs in actual hydrodynamical calculations. Up to now simulations of heavy-ion collisions have been done within ideal hydrodynamics with various EoSs of isospin symmetric matter with pions treated as particles obeying the vacuum dispersion law, {\em cf.} \cite{Ivanov:2016xev}.  As the first step, in the present work pions will be treated either as   ideal gas of the particles obeying the vacuum dispersion law or as the  quasiparticle gas with the $s$-wave pion-baryon interactions included. The latter contribute only for IAM. More involved effects of the $p$-wave pion-baryon interaction will be disregarded.

In the heavy-ion collisions at very low collision  energies
(${\cal{E}}_{\rm lab}\lsim (200-300)A$~MeV)
 in the expansion stage of the matter
at nucleon  densities  $n(t)< n_0$ and at low temperatures, $T(t)\lsim (15-20)$ MeV,
there may occur a spinodal instability during
the first-order liquid-gas (LG) phase transition  \cite{Ropke:1982vzx}.
Possible effects of the supercooled gas and superheated liquid phases, as well as
the effects of the spinodal instabilities, have been discussed in
\cite{Ropke:1982vzx,Schulz:1983pz,BS,Panagiotou:1984rb}. The nuclear LG transition
phenomenon remained an arena of intense research both on
theoretical and experimental sides during subsequent years, \textit{cf.}
\cite{Muller:1995ji,Li:1997px,Ducoin:2005aa, Alam:2017krb}.
Occurrence of a
negative specific heat at constant pressure was reported, as the first
experimental evidence of the LG phase transition in heavy-ion collision
reactions \cite{DAgostino:1999dod,Schmidt:2000zs}.  For a review of this interesting
topic see \cite{Chomaz:2003dz} and more recent works
\cite{Skokov:2008zp,Skokov:2009yu,Skokov:2010dd,Voskresensky:2010gf,Borderie:2018fsi}. Isospin dependence was studied
in \cite{Colonna:2002ti,Ducoin:2005aa}. Effects of a finite surface tension were disregarded. Below we  apply our  models with effective hadron masses and coupling constants dependent
on the scalar field also to describe
the LG first-order phase transition occurring at a low temperatures and densities. First we assume the surface tension  to be zero and then include effects of the non-zero surface tension, which may result in preparation of the pasta-like structures in heavy-ion collisions.
	
The manuscript is organized as follows. In  section \ref{Preliminaries} we present the RMF model
with $\sigma$-scaled hadron masses and couplings generalized for description of the hadron matter at arbitrary isospin composition  for $T\neq 0$, including pion gas.
For specificity we further use the KVORcut03-based and MKVOR*-based models of EoS
\cite{Kolomeitsev:2016ptu}   in the region of the baryon densities $\lsim 3n_0$ and temperatures $T< m_\pi$.
In Sect. \ref{fireball} we present results obtained in a simplified expanding fireball model. This simplified description allows us to demonstrate many qualitative and quantitative effects. In Sect.  \ref{liquidgas} we focus on the description of the region of
the LG phase
transition first considering ISM and then IAM. Consideration is first performed within the RMF framework and then effects of fluctuations are discussed. Importance of the surface tension effects will be then emphasised. Then in Sect. \ref{Conclus} we formulate  our conclusions. For completeness in Appendix \ref{Appen} we indicate  effects of the polarization of the medium due to the $p$-wave pion-baryon interaction, which were disregarded in the present study.


\section{RMF models with $\sigma$-scaled hadron masses and couplings. EoS of hadron matter
	in the region
	$T<m_\pi$, $n<3n_0$, $0.4\leq Z/A\leq 0.5$}\label{Preliminaries}

We use the framework proposed in \cite{Kolomeitsev:2004ff} and then developed
further in \cite{Maslov:2015msa,Maslov:2015wba,Kolomeitsev:2016ptu} for $T=0$ and an arbitrary isospin composition and generalized in
\cite{Khvorostukhin:2006ih,Khvorostukhin:2008xn,Khvorostukhin:2010aj}
for the case of $T\neq 0$, but only for ISM.
In the present work we  focus on the description of matter produced in heavy-ion reactions at collision energies less than few $A$~GeV.
Thereby we study the ISM, for which  $Y_Z =Z_{\rm tot}/A_{\rm tot} \simeq 0.5$, and   the IAM matter, when $0.4\lsim Y_Z <0.5$, where $Z_{\rm tot}$ is the total charge of the colliding nuclei and $A_{\rm tot}$ is the corresponding baryon number.

Our model is a generalization  on the case $T\neq 0$ of the non-linear Walecka model
with effective coupling
constants and hadron masses
\begin{align}
g_{Mb}^* = g_{Mb} \chi_{Mb}(\sigma)\,,\quad m_i^* = m_i \Phi_i(\sigma)
\end{align}
dependent on the scalar field $\sigma$. Here $M = (\sigma, \omega, \rho)$
denotes mesons, for which we use the mean-field solutions of the equations of motion, $b = (N, \Delta)$ lists the included baryon species, nucleons
$N=(p, n)$ and $\Delta$ isobars, $i=(M, b)$. We  neglect a  contribution of hyperons and anti-baryons, which are tiny for collision energies under consideration, $\propto e^{-2m^*_H/T}$ for hyperons due to the strangeness conservation \cite{Khvorostukhin:2006ih, Khvorostukhin:2008xn} and $\propto e^{-2m^*_b/T}$  for anti-baryons. In absence of the hyperon occupations there is no contribution of the $\phi$ meson mean field.  Besides, we include  pions $\pi =\{\pi^-, \pi^0, \pi^+\}$, as lightest among pseudo-Goldstone particles. Other pseudo-Goldstone particles $G=(\pi, K, \eta)$ and their heavier partners $K^*$ and $\eta'$ are not included, since their contributions, $\propto e^{-m_G /T}$, remain  tiny for $T<m_\pi$.
The quantities $\chi_{Mb}(\sigma)$ and $\Phi_i(\sigma)$ are the dimensionless scaling functions have been fitted in \cite{Maslov:2015msa,Maslov:2015wba,Kolomeitsev:2016ptu} for the best  description of the cold baryon matter.

Using mean-field solutions for meson fields we present the energy density of the hadronic system as \cite{Maslov:2015msa,Maslov:2015wba,Kolomeitsev:2016ptu}
\begin{gather}
E[{\{n_b\}},  f,T] =  \frac{m_N^4 f^2}{2 C_\sigma^2 }
\eta_\sigma(f)+\frac{1}{2 m_N^2} \Big[\frac{C_\om^2 n_V^2}{ \eta_\om(f)}
+\frac{C_\rho^2 n_I^2}{\eta_\rho(f)}\Big] \nonumber\\ +\sum_b E_{\rm kin}^b+E_{\rm pion}\,,
\label{edensity}  \\
n_V=\sum_b x_{\om b} n_b,\,\,
n_I=\sum_b x_{\rho b} t_{3 b} n_b, \nonumber\\
E_{\rm kin}^b = \gamma_b \int\limits_0^{\infty}
\frac{p^2 dp}{2 \pi^2} f_b (p) \sqrt{p^2 + m_b^{*2}}\,, \,\, \gamma_b = (2 s_b + 1),
\nonumber \\
f_b(p) =\frac{1}{e^{(\sqrt{p^2 + m_b^{*2}} -\mu_b^{*})/T)}+ 1}\,,
\,\, n_b= \gamma_b \int\limits_0^{\infty}
\frac{p^2 dp}{2 \pi^2} f_b (p), \nonumber\\
\mu_b^{*}=\mu_b -\frac{C_\om^2 n_V x_{\om b}}{ m_N^2 \eta_\om(f)}
-\frac{C_\rho^2 n_I x_{\rho b}t_{3b}}{ m_N^2 \eta_\rho(f)},\label{eq::mu_eff}
\end{gather}
where $s_b$ is the baryon spin,
$\mu_b = \mu_B - Q_b \mu_Q$ is the chemical potential for the given baryon species $b$, $\mu_B$ is the baryon-charge chemical potential, $\mu_Q$ is the chemical potential of a negative electric charge,  $Q_j$ is the electric charge of a particle $j$, $t_{3b}=Q_b-1/2$
is the isospin projection of baryon $b$.

For given $\mu_B$ and $\mu_Q$ eqs. (\ref{eq::mu_eff}) can be solved to find the particle densities and their effective chemical potentials $\mu_b^*$. Then the definitions of the total baryon density $n = \sum_b n_b$ and charge density $n_Q = \sum_b Q_b n_b + \sum_{\pi} Q_\pi n_\pi$ are treated as equations for finding the chemical potentials $\mu_B, \mu_Q$ for given $n, n_Q$.
Here $n_\pi$ are the pion number densities
\begin{gather}
n_\pi = \int\limits_{0}^{\infty} \frac{p^2 dp}{2 \pi^2} f_\pi(p), \,\, f_\pi (p) = \frac{1}{e^{(\omega_\pi (p) + \mu_\pi ) / T} - 1},
\end{gather}
where $\omega_\pi (p)$ is the dispersion relation of a pion and $\mu_\pi  = Q_\pi \mu_Q$ is its chemical potential. For ISM $\mu_\pi  = 0$. The term
\begin{gather}
E_{\rm pion} = \sum_{\pi} \int\limits_{0}^{\infty} \frac{p^2dp}{2\pi^2} \omega_\pi (p) f_\pi (p)
\end{gather}
is the contribution of $\pi$ species to the energy density.

For any set of baryon concentrations and the temperature equation $\pt E[{\{n_b\}},  f,T] / \pt f = 0$ is solved to find the equilibrium value of the scalar field.

The effective hadron masses and scaling functions of mesons enter the volume part of the thermodynamic quantities only in combinations
$
C_M = {g_{MN} m_N}/{m_M} $,
\begin{gather}
\eta_\om (f) = {\Phi^2_\om (f)}/{\chi^2_{\om N}(f)}\,,\quad \eta_\rho (f)
= {\Phi^2_\rho (f)}/{\chi^2_{\rho N}(f)}\,,\\
\eta_{\sigma}(f)=\frac{\Phi_{\sigma}^2[\sigma(f)]}{\chi_{\sigma N}^2[\sigma(f)]}
+ \frac{ 2 \, C_{\sigma}^2}{m_N^4 f^2}  {U}[\sigma(f)]\,,\nonumber
\end{gather}
where
the
self-interaction potential ${U}(\sigma)$ employed in standard RMF models is included into the definition of the scaling function $\eta_{\sigma}(f)$.
The scaling functions of the  mass are
$$\Phi_M =1-f\,,\quad
\Phi_b(f) = 1 - x_{\sigma b} \xi_{\sigma b}{m_N}f/{m_b}\,,$$
where
$$\xi_{\sigma b}=\chi_{\sigma b}/\chi_{\sigma N} \,,\quad
f =g_{\sigma N}\chi_{\sigma N}\sigma/m_N \,,$$
and thereby $\Phi_N =\Phi_M$. We
suppose that $\chi_{\omega b}(f) = \chi_{\omega N}(f)$\,, $\chi_{\rho b}(f)
= \chi_{\rho N}(f)$. Explicit expressions for the scaling functions $\eta_{M}(f)$
are presented in \cite{Kolomeitsev:2016ptu}.

The coupling constant ratios are introduced as $x_{Mb}=g_{Mb}/g_{MN}$. 
The vector-meson coupling constants to $\Delta$s are chosen following the quark SU(6) symmetry:
\begin{gather}
x_{\om \Delta}=x_{\rho\Delta}=1, \quad x_{\phi\Delta}=0\,.\nonumber
\label{gHm}
\end{gather}
The $\Delta$ coupling constants with the scalar field are
deduced from the values of the  optical potentials  $U_{b}$ in ISM at
the saturation density $n=n_0$ given by
\begin{gather}
U_{b}(n_0) = \frac{C_{\omega}^2 x_{\omega b} n_0}{\eta_\om (f(n_0)) m_N^{2}}
- (m_N-m_N^{*} (n_0))x_{\sigma b}\,.
\end{gather}
The value of the $\Delta$
potential $U_\Delta(n_0)$ is poorly constrained by the data. As in \cite{Kolomeitsev:2016ptu}, we  use
$U_\Delta(n_0) = -50 \, \mev$ as the most realistic estimate. Models including $\Delta$s
will be denoted by  "$\Delta$" suffix.

Also we  assume that the size of the
system under consideration is such that the volume part of the thermodynamic quantities
of our interest  is much larger than the surface part.
Moreover,  we first disregard finite-size Coulomb effects compared to the
strong-interaction effects. The former effects will be discussed in Sect. \ref{liquidgas}.  Focusing on the description of heavy-ion collisions we  do not
include lepton terms.

In ref.~\cite{Maslov:cut} we demonstrated that within an RMF model the EoS becomes
stiffer for $n >n^* >n_0$, if a growth of the scalar field as a function of the
density is quenched and the nucleon effective  mass becomes weakly dependent on the
density for $n >n^*$. In~\cite{Maslov:cut}  such a quenching was achieved by adding
to the scalar potential a rapidly rising function of $f$  at $f>f^*$, where $f^*$ is
$f$ corresponding to $n=n^*$. We called it the cut-mechanism.  In Ref.~\cite{Maslov:cut}
the cut-mechanism is realized in the $\sigma$ sector.
We focus now on two models based on KVORcut03 and MKVOR* models proposed
in~\cite{Maslov:2015msa,Maslov:2015wba,Kolomeitsev:2016ptu}. These models proved to satisfy many constraints on the hadronic EoS. In neutron-star matter for large densities the hyperons and $\Delta$ baryons appear in our models.
These two models utilize the cut-mechanisms  in the $\om$ and $\rho$ sectors, respectively.
The cut mechanism in $\rho$ sector is implemented in MKVOR-based models in order
simultaneously to keep the EoS not too stiff in ISM (to satisfy the flow constraint from
heavy-ion collisions~\cite{Danielewicz:2002pu}) and to do the EoS as stiff as possible in
the BEM to safely fulfill the constraint on the maximum mass of
neutron stars. The $\rho$ mean  field is coupled to the isospin density that makes the
$f$-saturation mechanism very sensitive to the composition of the BEM. Incorporating  $\Delta$
baryons  we use the MKVOR* extension of the MKVOR model \cite{Kolomeitsev:2016ptu}, which prevents the effective nucleon mass from vanishing at high density.

Free parameters of the model are fitted to reproduce  properties of the cold nuclear matter
near the saturation point. These parameters are defined as the coefficients of
the Taylor expansion of the energy per nucleon for $T=0$,


\begin{table*}[t]
	\caption{Coefficients of the energy expansion (\ref{Eexpans})  for KVORcut03 and MKVOR models.}
	\begin{center}
		\begin{tabular}{ccccccccc}
			\hline
			\raisebox{-.23cm}[0cm][0cm]{EoS}& $\mathcal{E}_0$ & $n_0$ & $K$ & $m_N^*(n_0)$ &$J$ & $L$ &$K'$ & $K_{\rm sym}$
			\\ \cline{2-9}
			&           [MeV] & [fm$^{-3}$] & [MeV] & $[m_N]$ & [MeV] & [MeV] &[MeV] & [MeV]
			\\ \hline
			KVORcut03 & $- 16$ & 0.16 & 275 &  0.805 &  32 & 71 &  422& -86\\
			MKVOR & $- 16$ & 0.16 & 240 &  0.730 &  30 & 41 &  557 & -158\\
			\hline
		\end{tabular}
	\end{center}
	\label{tab:sat-param}
\end{table*}
\begin{align}
& \mathcal{E} = \mathcal{E}_0 + \frac{1}{2}K\epsilon^2
-\frac{1}{6}K'\epsilon^3 +\beta^2\widetilde{\mathcal{E}}_{\rm sym}(n) +
...\,,
\nonumber\\
& \widetilde{\mathcal{E}}_{\rm sym}(n)={J} + L\epsilon +\frac{K_{\rm sym}}{2}\epsilon^2+\dots\,,
\label{Eexpans}
\end{align}
in terms of small $\epsilon=(n-n_0)/3n_0$ and $\beta=(n_n-n_p)/n$ parameters. The parameters for the MKVOR* and MKVOR models are identical.
The properties of the KVORcut03 and MKVOR*
models, which we exploit in this work, at the nuclear saturation density $n_0$ are illustrated
in Table~\ref{tab:sat-param}, where we collect coefficients of the expansion of the nucleon binding energy per nucleon near $n_0$.

For the difference of the neutron and proton chemical potentials we get
\begin{equation}
\mu_n -\mu_p =\frac{\partial E[n_p,n_n]}{\partial n_n}-\frac{\partial E[n_p,n_n]}{\partial n_p}\, \equiv \mu_Q\,.
\label{munp}
\end{equation}

The pion quasiparticle spectrum is  determined as a solution of the dispersion equation \cite{Migdal:1990vm,Voskresensky:1993ud}
\begin{equation} \omega^2 =m_\pi^2 +k^2 + {\mbox{Re}}\Pi (\omega, k, n_b, T)\,,\label{piPol}
\end{equation}
where  $\Pi (\omega, k, n_b,T)$ is the pion polarization operator in the baryon medium.
In this work we will  consider pions as an ideal gas of quasiparticles  including for IAM only their $s$-wave interaction with baryons, being
determined by  the so called Weinberg-Tomozawa term. For ISM the $s$-wave $\pi N$ interaction is suppressed \cite{Baym:1975tm,Migdal:1978az,Friedman:2019zhc, Migdal:1990vm,Voskresensky:1993ud}.   Including only $s$-wave pion-nucleon interactions the retarded pion
polarization operator is given by
$$\Pi^{\pi^-}_s =(n_n -n_p)\omega/(2f_{\pi}^2),\quad \Pi^{\pi^0}_s =0\,,$$
where $f_\pi \simeq 92.4$MeV is the
pion weak decay constant,  \textit{cf.}
\cite{Migdal:1990vm,Kolomeitsev:2002gc}, and the $\pi^{\mp}$ spectrum  is thereby as follows
\begin{gather}
\omega_{{\pi}^{\mp}} (k) =\pm \frac{n_n-n_p}{4f_{\pi}^2}+\sqrt{m_{\pi}^2 +k^2 +
	\left(\frac{n_n-n_p}{4f_{\pi}^2}\right)^2}, \label{WT}\\
\omega_{{\pi}^{0}} (k) \equiv \om_k =\sqrt{m_{\pi}^2 +k^2}.\nonumber
\end{gather}
Models with pion quasiparticles treated following eq. (\ref{WT})  will be labeled by ``$\pi_{\rm WT}$'' suffix and models with pions described by the vacuum dispersion law we label by ``$\pi_{\rm free}$'' suffix, respectively.  Important role of the $p$-wave pion-baryon   interactions has been
intensively studied in
\cite{Baym:1975tm,Migdal:1978az,Migdal:1990vm}.
This issue will be briefly reviewed in the Appendix.

\begin{figure*}
	\centering
	\includegraphics[width=0.48\linewidth]{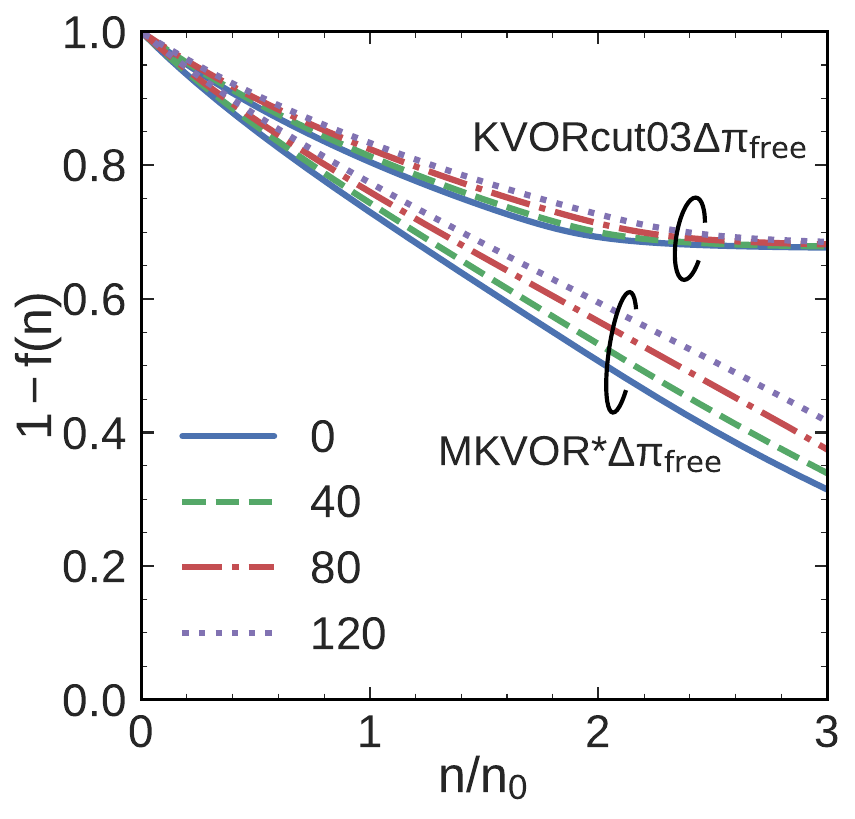}
	\includegraphics[width=0.48\linewidth]{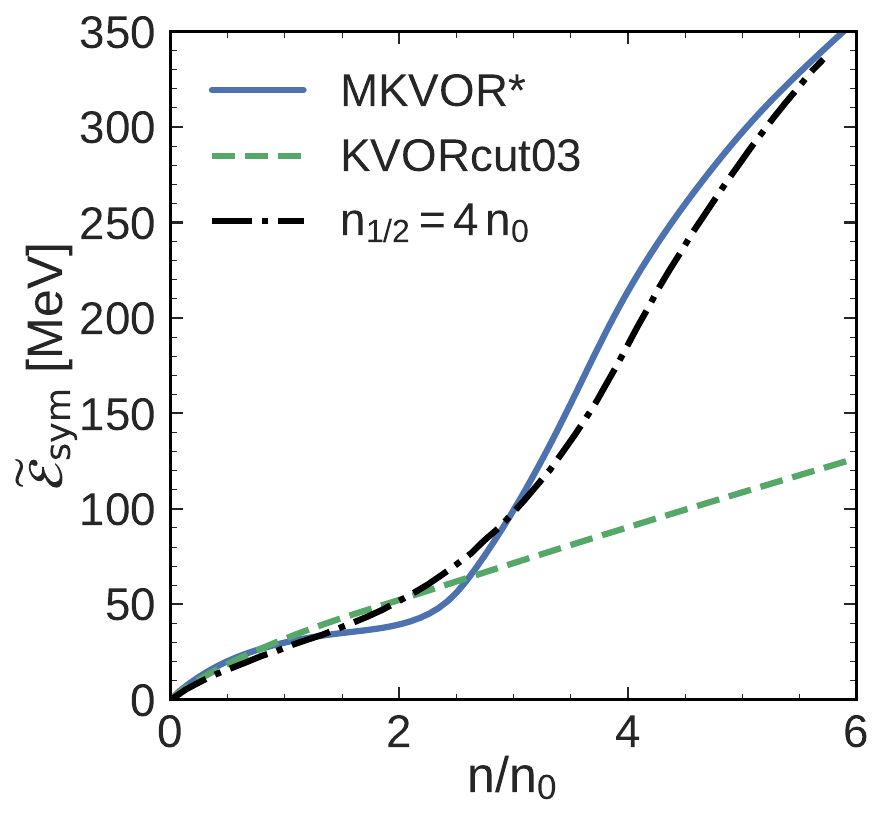}
	\caption{Left panel: The effective mass scaling function $\Phi_N=\Phi_M = 1-f$ for the models KVORcut03$\Delta\pi_{\rm free}$ and
		MKVOR*$\Delta\pi_{\rm free}$ for ISM
		as a function of the baryon density $n$
		for various temperatures indicated in the legend in MeV. Right panel: The symmetry energy coefficient in models KVORcut03 (dashed line) and
		MKVOR* (solid line) for ISM. For comparison by dash-dotted line is shown the symmetry energy coefficient obtained in model \cite{Ma:2018xjw}  with a topology change mimicking the baryon-quark continuity taking place at $n=n_{1/2}=4n_0$. }
	\label{NuclEfMass}
\end{figure*}

One of the main differences between the KVORcut-based and MKVOR*-based models we consider here is the behavior of the nucleon effective mass with  the density. Therefore, on left panel in fig. \ref{NuclEfMass}  we show the  baryon density dependence of the scaling function
$\Phi_N(f) = m_N^* /m_N = m^*_M /m_M = 1-f$ in ISM  calculated for various temperatures in the models KVORcut03$\Delta\pi_{\rm free}$ and
MKVOR*$\Delta\pi_{\rm free}$.
We see that the density dependence of this quantity is significant for
$n\lsim 3 n_0$,
whereas the temperature dependence is  moderate for $T<m_\pi$.
In the KVORcut03$\Delta\pi_{\rm free}$ model  $m^*_\sigma \gsim 400$~MeV
in the density and temperature interval
under consideration, whereas in the MKVOR*$\Delta\pi_{\rm free}$ model we have $m^*_\sigma \gsim 200$~MeV.\footnote{We should note that  $m_\sigma^*$ here is the mass-coefficient
	of the $\sigma$ mean field rather than
	the effective mass of $\sigma$ excitations, $m_\sigma^{\rm{part}*}$, \textit{cf.}
	\cite{Khvorostukhin:2006ih,Khvorostukhin:2008xn}.}  Thus the effective masses of the
$\sigma,\omega,\rho$ mesons  remain significantly larger  than $m_\pi$, and the thermal contribution of
the $\sigma$, $\omega$, $\rho$, $\phi$ excitations, which is $\propto e^{-m_M^*(n,T)/T}$, can be neglected for temperatures and densities we consider here.
 The curves computed for $\pi_{\rm free}$ and $\pi_{\rm WT}$ models prove to be visually almost not distinguishable for $0.4< Y_Z <0.5$. Therefore below we mainly focus   consideration on $\pi_{\rm free}$ models. Besides, we note that the $Y_Z$ dependence of $1-f$  proves to be very weak in the interval $0.4\leq Y_Z \leq 0.5$ of our interest. The curves computed for $Y_Z = 0.4$  and for $Y_Z =0.5$ are visually almost not distinguishable.

The temperature dependence of the nucleon energy is $\propto [1  + (T/\epsilon_{{\rm F},N})^2]$ and thereby
it is essential already for $T <  \epsilon_{{\rm F},N}$,
where $\epsilon_{{\rm F},N}$ is the nucleon Fermi energy
($\epsilon_{{\rm F},N}\sim 40(n/n_0)^{2/3}$~MeV
for $n\sim n_0$, $Y_Z=0.5$ and for the Landau nucleon effective mass $\simeq m_N$). The energy of the pion ideal gas is $\propto e^{-m_\pi /T}$ and
becomes significant for $T\gsim 0.5 \, m_{\pi}$. The contribution of the $\Delta$s to the energy
is suppressed compared to the nucleon one as $4 e^{-(m^*_\Delta -m_N^* + Q_\Delta \mu_Q)/T}$. All these contributions are included in our work. The first-order phase transition to the $\Delta$ resonance matter considered in
\cite{Kolomeitsev:2016ptu}
for $T=0$
does not occur for $T<m_\pi$, $n\lsim 3n_0$, for the $\Delta$ optical potential $U_\Delta (n_0)=-50$ MeV that we use in this work.
In  \cite{Khvorostukhin:2008xn} within the model, where pions are treated with the vacuum dispersion law, the effect of  the nonzero $\Delta$ width was estimated as not significant. Therefore in what follows  within our RMF-based model the $\Delta$ resonances will be treated as  quasiparticles with an effective mass. Concluding this discussion, for
temperatures $T$ below  $m_{\pi}$ and for densities $n\lsim 3n_0$ of our interest here, the temperature dependence can be included only in the nucleon and $\Delta$ isobar quasiparticle contributions and the pion kinetic energy terms, which  within the $\pi_{\rm free}$ model are described by the free dispersion law.

Our KVORcut03-based and MKVOR*-based models differ also by the density dependence of the symmetry energy coefficient.
On the right panel in fig. \ref{NuclEfMass} we show the baryon density dependence of the symmetry energy coefficient $\widetilde{\mathcal{E}}_{\rm sym}(n)$ derived in our models KVORcut03 and MKVOR* for ISM.
The density dependence of the symmetry energy in the KVORcut03-based models, where the cut-mechanism is implemented in the $\omega$ sector \cite{Kolomeitsev:2016ptu} for $n>n^* =3n_0$, is rather smooth (dashed line). In the MKVOR*-based models, where the cut-mechanism  is used
in the $\rho$ sector \cite{Kolomeitsev:2016ptu} responsible for the symmetry energy, the dependence of $\widetilde{\mathcal{E}}_{\rm sym}$ on $n$ becomes sharp for $n> (2.5-4)n_0$ (solid line). In \cite{Ma:2018xjw} the dramatic change in  the density dependence of the nuclear symmetry energy  for $n$ above some value $n_{1/2}$ varied in the interval $(2-4)n_0$ was associated with the change of the topology
mimicking the baryon-quark continuity. By the dash-dotted line in figure is shown the symmetry energy coefficient obtained in model \cite{Ma:2018xjw} for $n_{1/2}=4n_0$. The resulting density dependence is similar to that we obtain in the MKVOR-based models.

In the heated dense nuclear system formed in the collision of nuclei,
initial charge per baryon $n_Z$ is redistributed among all the involved electrically charged hadrons appeared for $T\neq 0$, \textit{cf.} \cite{Muller:1995ji,Li:1997px}, following minimum of the free energy expressed in $n,T$ variables. We use the following decomposition of the system charge:
\begin{gather*}
Y_Z =\frac{n^Q_B}{n} +\frac{n^Q_\pi}{n}\,,\quad  \frac{n^Q_B}{n}=Y_p +\frac{
	2 n_{\Delta^{++}}+n_{\Delta^{+}}-
	n_{\Delta^{-}}}{n}\,,\\ \frac{n^Q_\pi}{n}=\frac{n_{\pi^{+}}-n_{\pi^{-}}}{n}\,,
\label{ZA}
\end{gather*}
where  $Y_Z$ is the ratio of the  initially fixed number of protons  to the fixed total baryon number,  $Y_p =n_p/n$  is    the ratio of the number of protons to that of the baryons inside the thermal  system, ${n^Q_B}/{n}$ is the ratio of the total charge of the baryon subsystem to the total baryon density, that includes contribution of charged $\Delta$ isobars, and ${n^Q_\pi}/{n}$ is the ratio of the total charge of the pions to the total baryon density.

\begin{figure*}
	\centering
	\includegraphics[width=0.48\textwidth]{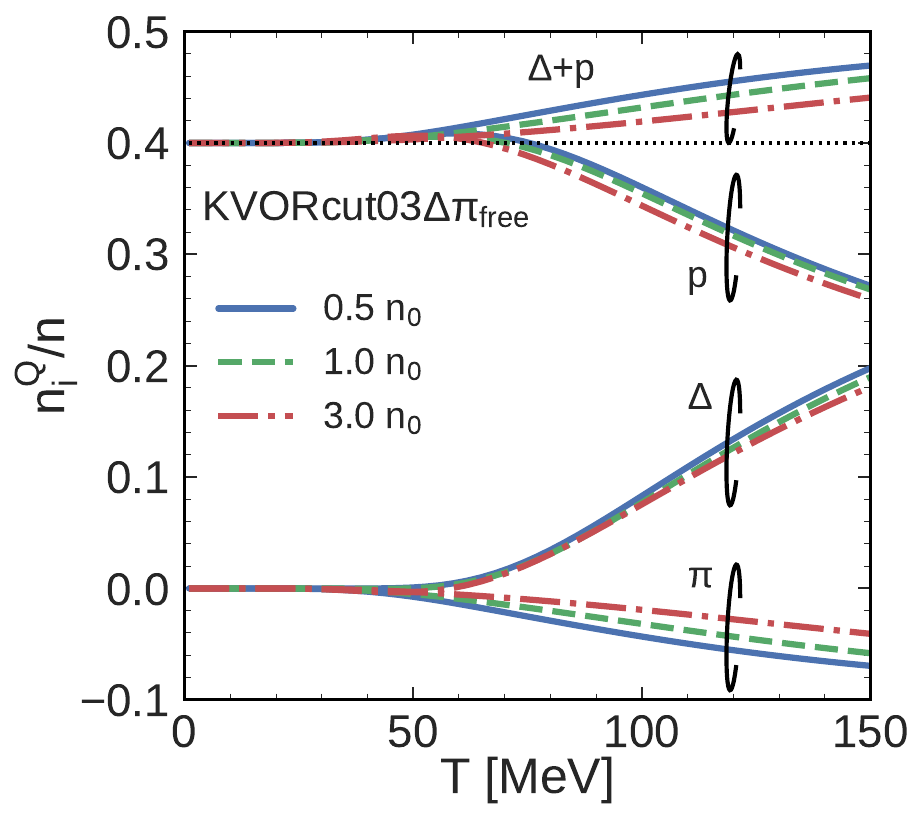}
	\includegraphics[width=0.48\textwidth]{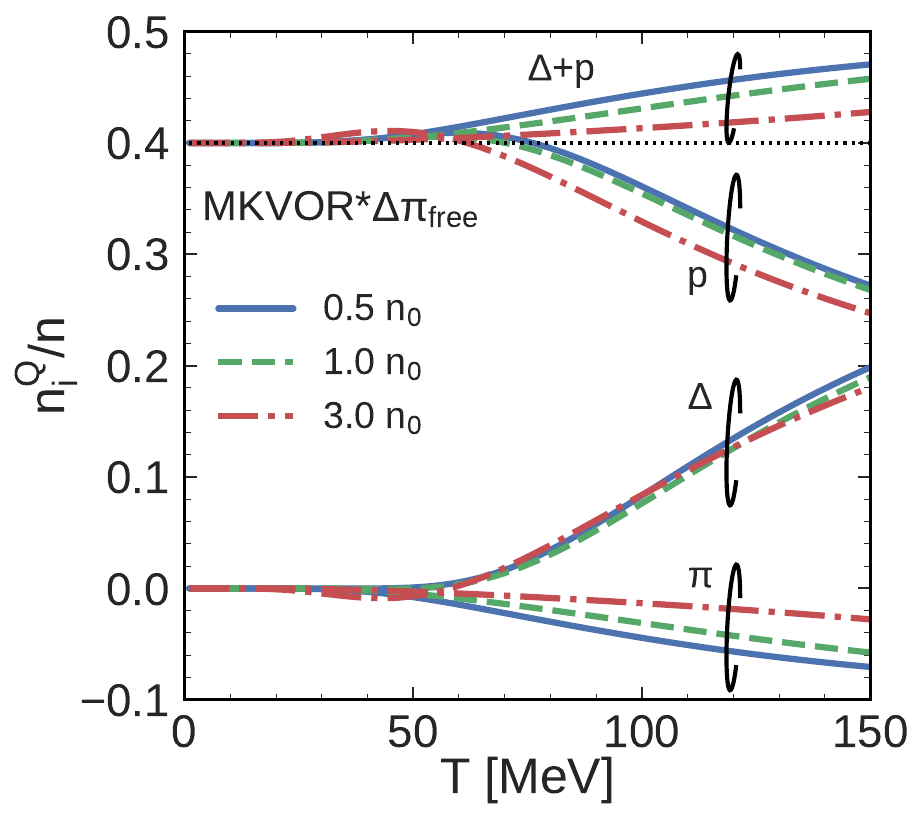}
	\caption{The ratio of the excess/deficiency of the positively charged hadrons of various species to the total baryon density at $Y_Z =0.4$ as a function of the temperature for the model KVORcut03$\Delta\pi_{\rm free}$ on left panel and for the model MKVOR*$\Delta\pi_{\rm free}$ on right panel. Solid curves are for $n=0.5 n_0$, dashed lines for $n= n_0$, and dash-dotted ones for $n=3 n_0$.
		See text for further details.    }
	\label{fig::charges}
\end{figure*}

In fig. \ref{fig::charges} we show the charge per baryon for various species for the model KVORcut03$\Delta\pi_{\rm free}$ on the left panel and for the model MKVOR*$\Delta\pi_{\rm free}$ on the right panel. We take $Y_Z =0.4$ as an example relevant to the matter formed in heavy-ion collisions. At  zero temperature  $Y_Z= Z/A = Y_p$, since the $\Delta$ isobars do not appear in both our models for $n\lsim 3\,n_0$ \cite{Kolomeitsev:2016ptu}, which we consider here, and there are no pions for $T=0$. With an increase of the temperature the abundance of $\Delta$ resonances and pions increases. For $Y_Z<0.5$ the charge chemical potential $\mu_Q$ is positive, and there appears an excess of $\pi^-$ respectively to $\pi^+$. The ratio $-n_Q^\pi/n \propto e^{-m_\pi /T}(e^{\mu_Q/T}-e^{-\mu_Q/T})/n$ (lines labeled $\pi$)  increases with the temperature $T$. The pion charge per unit of baryon density proves to be higher for smaller density.
Unlike in the neutron-star matter, the $\Delta^{++},\Delta^+, \Delta^0 , \Delta^-$ subsystem in heavy-ion collisions at $0.4 \lsim Y_Z \lsim 0.5$ remains  positively charged. Indeed, $n_Q^\Delta /n \propto e^{(-m^*_\Delta +m_N^* )/T}[2 e^{-2 \mu_Q/T} + e^{-\mu_Q/T} - e^{\mu_Q/T}]$.
Since the value of $\mu_Q$ remains rather small for $n \lsim 2\,n_0$, 
the $\Delta^{++},\Delta^+, \Delta^0$ and $\Delta^-$ abundances remain  close to each other.
Indeed, as we see in fig. \ref{fig::charges}  excess of the positively charged $\Delta$ resonances (lines labeled $\Delta$) increases with an increase of $T$  in agreement with above estimate. The value $Y_p$ (lines labeled by $p$) slightly increases with increase of $T$ for $T< \epsilon_{{\rm F},N}$ and then it sharply decreases for higher $T$. The reason of a slight increase of $Y_p$ for low $T$ is that an enhancement of the proton fraction should  compensate  a small  increase with $T$
of the negative pion charge  to fulfill the charge conservation condition. However, at $T \gsim \epsilon_{{\rm F},N}$ the $\Delta$ isobar concentration increases noticeably and the proton fraction decreases because of the baryon charge conservation.
The total charge of the baryon subsystem per baryon ${n^Q_B}/{n}$ is shown in fig. \ref{fig::charges} by the lines labeled ``$\Delta +p$".

For MKVOR*$\Delta\pi_{\rm free}$ model at $n \sim  3\,n_0$  for $30\, \mev \lsim T \lsim 50 \, \mev$ the $\Delta$ system is negatively charged. This happens because in the MKVOR* model the symmetry energy coefficient and, correspondingly, $\mu_Q$ are larger at such densities than in the KVORcut03 model, see fig. \ref{NuclEfMass}. Therefore in agreement with the estimate given above the density of $\Delta^-$ becomes greater than the sum of densities of  $\Delta^+$ and $\Delta^{++}$. The negative pion excess contributes less to the charge conservation at $n \sim  3\,n_0$ than for lower $n$ because the number of pions per baryon at a fixed temperature decreases with an increase of the baryon density.

 For $n\lsim 2n_0$ all the ratios obtained in both of our models are very close to each other.
Only for $n\sim 3\, n_0$ the contribution of  $\Delta +p$   to the charge excess computed within the KVORcut03$\Delta\pi_{\rm free}$ model proves to be a bit higher than that in MKVOR*$\Delta\pi_{\rm free}$ model.

Note that in MKVOR*$\Delta\pi_{\rm free}$ model a transition to the $\Delta$ resonance-enriched matter may occur in the dense medium. For $T=0$ and  for $U_\Delta (n_0) =-50$ MeV that we use in this work
the $\Delta$s appear by the crossover at $n\gsim n_{c,\Delta}\simeq 4.5 n_0$, \textit{cf.} fig. 11 in \cite{Kolomeitsev:2016ptu}.  
With an increase of the temperature in the MKVOR*$\Delta\pi_{\rm free}$ model $\Delta$s arise for a lower density and the phase transition becomes the transition of the first order. We found that for $U_\Delta(n_0) = -50\,\mev$ at $T = 55\,\mev$ the first-order phase transition occurs for
$n_{c,\Delta}\simeq 3.6\,n_0$. The critical density decreases very smoothly  with an increase of the temperature and we get $n_{c,\Delta}\simeq 3.3 \,n_0$ at $T \simeq 140\, \mev$, thus $n_{c,\Delta}$ remains  above $3n_0$, i.e, outside  the density range we consider in the given work.
However we should  point out that  some papers argue that $U_\Delta $ could be a  more attractive. For $U_\Delta (n_0) =-100$ MeV the first-order phase transition to the $\Delta$ resonance matter would occur  for $T = 0$ already at the density $n\gsim n_{c,\Delta}\simeq 2.5 n_0$, \textit{cf.} a discussion in \cite{Kolomeitsev:2016ptu}.
In this work we use $U_\Delta(n_0) = -50\,\mev$ and consider $n\lsim 3\,n_0$,  thereby the first-order phase transition to the $\Delta$ reach matter does not occur within  our MKVOR*$\Delta\pi_{\rm free}$ model.
In the KVORcut03$\Delta\pi_{\rm free}$ model the phase transition to the $\Delta$ resonance matter does not occur for all relevant values of $U_\Delta$ at densities and temperatures we are interested  in this work.

\section{Simplified  model for heavy-ion collisions at ${\cal{E}}_{\rm{lab}}<(1-2)$ $A$~GeV}\label{fireball}

\subsection{EoS and the system dynamics}
In a heavy-ion collision, nucleons can be subdivided on  “participants,” which intensively
interact with each other during the collision, and “spectators”, which remain
practically unperturbed. Baryons-participants form a nuclear fireball, which first is compressed (for $t<0$) and then (for $t>0$)  is expanded under action of the internal pressure.

The kinetic energy of the projectile nucleus per nucleon (per $A_{\rm p} =A$) in the laboratory system
${\cal E}_{\rm lab}$ is related to the kinetic energy per  nucleon in the
center-of-mass frame as follows
\begin{eqnarray}
{\cal E}_{\rm c.m.}=m_N \sqrt{1+\frac{2A_{\rm p} A_{\rm t}{\cal E}_{\rm lab}}
	{(A_{\rm p} +A_{\rm t})m_N}}-m_N \,,
\end{eqnarray}
$A_{\rm t}$ is the nucleon number of the target nucleus.
As in
\cite{Voskresensky:1989sn,Migdal:1990vm,Voskresensky:1993ud},  we  assume that at energies
less than a few $A$\,GeV in the laboratory frame the energy in the center-of-mass frame
of the nucleus-nucleus collision, ${\cal E}_{\rm c.m.}A_{\rm part}$, which corresponds
to the  nucleons-participants, is spent on the creation of an initially  quasi-equilibrium nuclear fireball resting in the center-of-mass frame at the end of the compression stage, for $t=0$. The energy per baryon, $E(n,T)/n$,
as a
function of the baryon density $n$ at fixed $T$ has the concave shape in  our KVORcut03$\Delta\pi$ and MKVOR*$\Delta\pi$ models, it decreases with increase of $n$, gets a minimum at $n=n_m$,  and then begins to increase,    see fig. \ref{fig::EScut03} below. Following \cite{Voskresensky:1989sn,Migdal:1990vm,Voskresensky:1993ud} we assume that the initial fireball state is
characterized  by the
temperature $T(0)=T_m$, and the baryon density $n(0)=n_m$ corresponding to the  minimum of
the energy per baryon, $E(n,T_m)/n$, as a function of the baryon density for $T=T_m$,
\begin{eqnarray}
E(n_m,T_m)/n_m ={\cal E}_{\rm c.m.}
+m_N
+{\cal E}_{\rm bind}\,.\label{relat}
\end{eqnarray} The quantity
${\cal E}_{\rm bind}$ is the binding energy per baryon in a cold nucleus of
the nucleon number $A_{\rm part}$, $-16\,{\rm MeV}\leq {\cal E}_{\rm bind}<0$. Below we will use values ${\cal E}_{\rm bind}$ which follow from
(\ref{edensity}) at ignorance of
surface and Coulomb effects, \textit{i.e.}, as would be for very heavy nuclei.
 Thus
we shall take
${\cal E}_{\rm bind} (Y_Z=0.5) \simeq -16\,{\rm MeV}$,  ${\cal E}_{\rm bind}(Y_Z=0.4) =-14.7$~MeV.

The specific entropy is a decreasing function of $n$,   see fig. \ref{fig::EScut03} below.
Thereby the state of the minimum of the energy on the
right branch of $ {\cal E}(n)$ (where ${\cal E}=E/n$ increases with $n$)
corresponds to the maximum of the entropy on the given isotherm  for the states belonging this branch.
Moreover, the state $T_m ,n_m$ also corresponds to the  maximum temperature among all available solutions of the eq. (\ref{relat}) for all   curves $E(T=\const,n)$  and respectively this state corresponds to the maximum of the stirring of the degrees of freedom possible at assumption of the full stopping of the matter in the center-of-mass frame for $t=0$.

\begin{figure*}
	\centering
	
	\includegraphics[height=8cm]{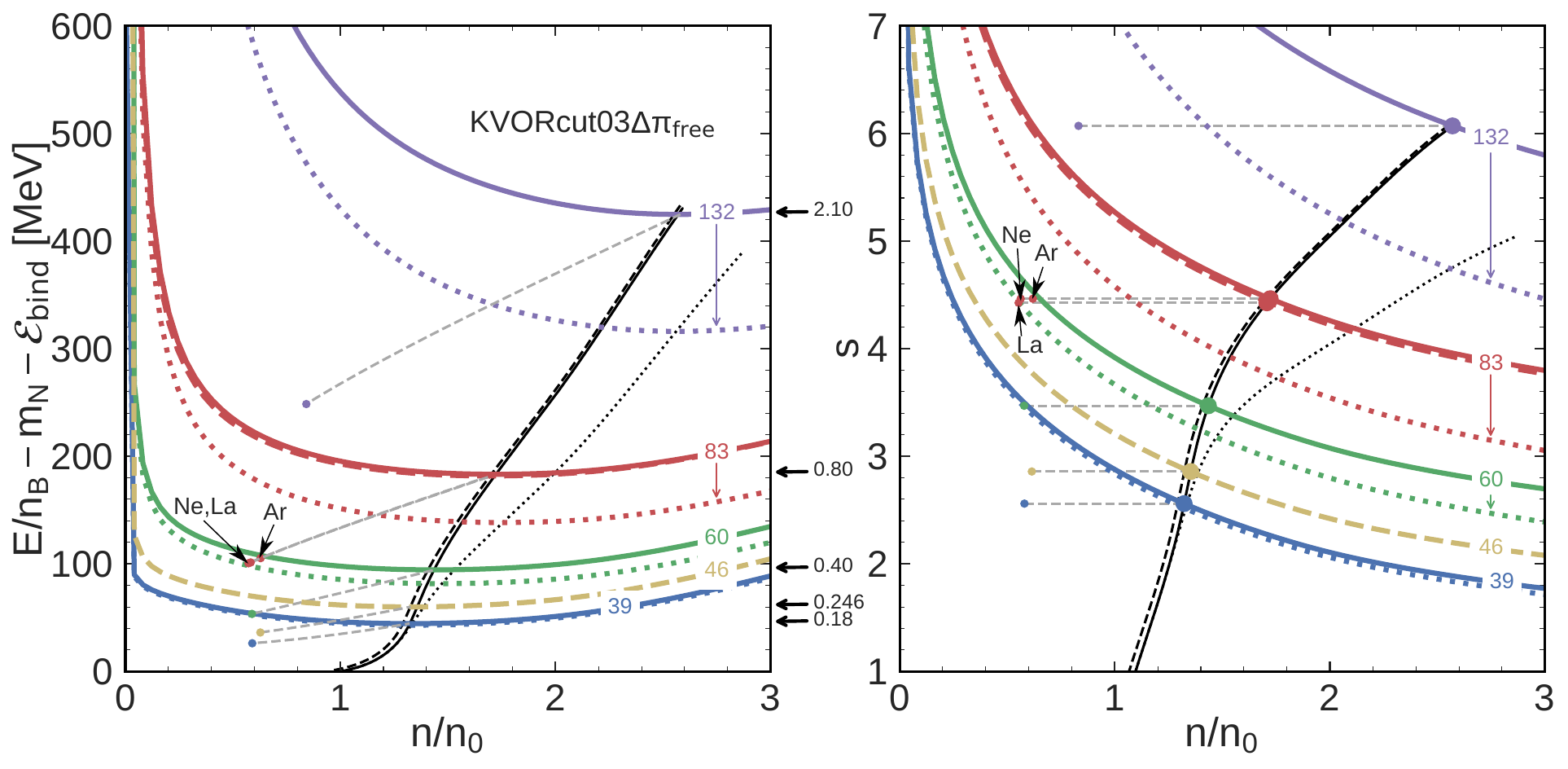}
	
	\caption{
		Energy per baryon (left panel)  and total entropy per baryon (right panel)
		in approximately ISM ($Y_Z =0.48$, as for Ar+Kcl collisions) for the KVORcut03$\Delta\pi_{\rm free}$ EoS as functions of the baryon density in units of
		$n_0=0.16$ fm$^{-3}$ for different temperatures indicated on the lines in MeV. The bold curves correspond to the set of  collision energies in laboratory system shown in $A$ GeV  by arrows at the right edge of the left panel indicating a position of the corresponding minimum of the $E(n, T)/n$.
		Solid bold curves are presented for the case $Y_Z = 0.48$ and two dashed bold curves are shown for the case $Y_Z = 0.4$, as for La+La collisions at $800A$ and  $246A$ MeV.  For comparison  by dotted bold curves we show the quantities at the same temperatures, as for the corresponding solid lines, but without the inclusion of $\Delta$~resonances.
		Thin lines indicate the initial fireball configurations constructed as described in the text.
		Thin dashed horizontal lines on the right panel denote  isoentropic
		trajectories and small dots are the break up points obtained by fitting of the $\pi^-$ production differential cross sections, \textit{cf.} in figs.  \ref{fig::fits}, \ref{fig::breakup} below.
	}
	\label{fig::EScut03}
\end{figure*}

\begin{figure*}
	\centering
	\includegraphics[height=8cm]{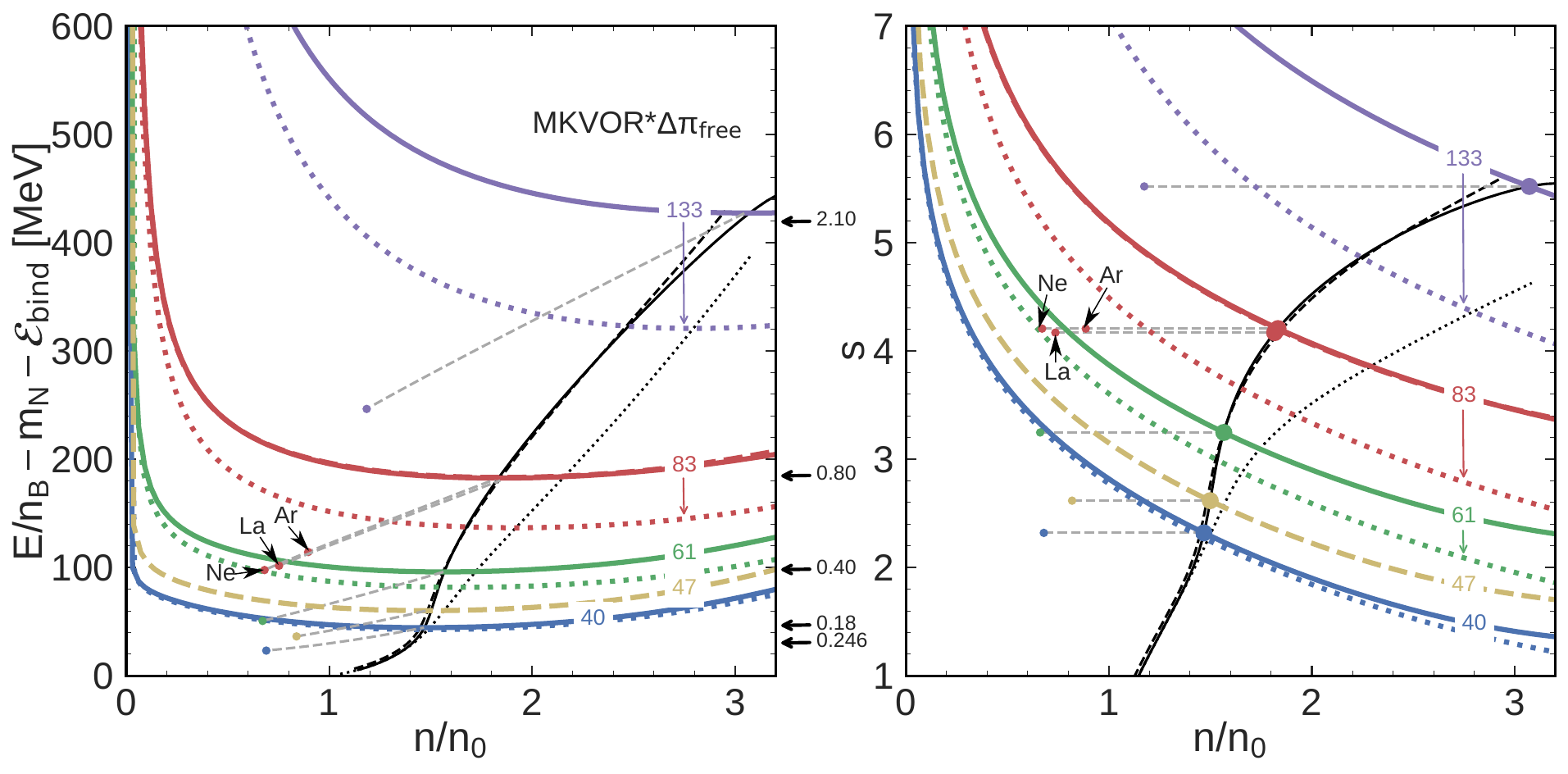}
	\caption{
		Same as fig. \ref{fig::EScut03} but for the MKVOR*$\Delta\pi_{\rm free}$ model.
	}
	\label{fig::ESmkv}
\end{figure*}

Note that for a weakly non-equilibrium system
moving with the velocity $\vec{v}(t,\vec{r})$
the local pressure
can be presented as \cite{Ivanov:2013uga}
$P_{\rm n.eq.}=P-\zeta \nabla \vec{v}$,
where
$\zeta$ is the bulk viscosity and $P$  is the
quasi-equilibrium pressure depending on the local temperature and density $T(t,\vec{r})$ and $n(t, \vec{r})$ following a given EoS. Since $\zeta$ is a positive-definite quantity,
the non-equilibrium correction to the pressure  is positive during the compression stage of the nuclear system and it is negative
during the  expansion stage, see also \cite{Voskresensky:2010qu}.
So, on the stage of expansion of the fireball in the heavy
ion collision the non-equilibrium pressure $P_{\rm n.eq.}$ is in reality smaller than the equilibrium one, $P$, being
used in the ideal hydrodynamics. This means that a realistic equilibrium EoS to be used in non-ideal hydrodynamical calculations should
be stiffer than the one used to fit experimental data within ideal hydrodynamical simulations. Besides, the entropy in the viscous process increases, whereas it stays constant within the ideal hydrodynamics.

The viscosity effects
prove to be rather small at energies
less than (1-2)$A$~GeV we study in this work.
This conclusion is supported by the analyses of heavy-ion collisions
performed in an expanding fireball model
\cite{Voskresensky:1989sn,Migdal:1990vm,Voskresensky:1993ud}, by
calculations used ideal hydrodynamics in a broad energy range
\cite{Ivanov:1991te,Mishustin:1991sp,Ivanov:2005yw,Arsene:2006vf,Buss:2011mx}, by
simulations done within transport codes \cite{Bass:1998ca,Arsene:2006vf} and by estimates of the
viscosity \cite{Khvorostukhin:2010aj}. In the ideal hydrodynamics dynamical trajectories of the system in heavy-ion collisions are
characterized
by constant initial values of the entropy per $A_{\rm part}$
(\textit{i.e.} by total entropy
density $S$ per net baryon density of baryons-participants, ${s}=S/n$). In our RMF approach with the contribution
of the ideal pion gas included we neglect the inelastic processes and thereby the
entropy is assumed to be conserved.
Thus we  have
\begin{gather}
s=S(n(t),T(n(t)))/n\simeq S(n_m, T_m)/n_m\,.\label{constentr}
\end{gather}
From this relation we obtain the dependence $T(n(t))$.
Note that actually all the results
of this model hold locally, and therefore are applicable to the case of either quasi-homogeneous fireball expansion, or inhomogeneous hydrodynamical expansion with $n=n(t, \vec{r})$
and $T=T(t, \vec{r})$ depending on the space point.
Just for the illustration purposes, as in
\cite{Voskresensky:1989sn,Migdal:1990vm,Voskresensky:1993ud}, we will further assume that the expansion is uniform.

We use the KVORcut03$\Delta\pi$ and MKVOR$^*$$\Delta\pi$ models. The
ideal pion gas  either is described  by the vacuum dispersion law in the $\pi_{\rm free}$ model or by the law (\ref{WT}) in the quasiparticle  model $\pi_{\rm WT}$ for IAM.
The energy density is given by
eq. (\ref{edensity}). As we have mentioned, from the relation (\ref{relat}) we unambiguously determine
quantities $n_m$ and $T_m$.
The values of the  energy per baryon and specific entropy ${s}$  are shown in fig. \ref{fig::EScut03} (left and right) for KVORcut03$\Delta\pi_{\rm free}$ model  and in fig. \ref{fig::ESmkv} (left and right) for MKVOR*$\Delta\pi_{\rm free}$ model as functions of the density at various temperatures. Being computed with the dispersion law (\ref{WT}) in $\pi_{\rm WT}$ model, thermodynamic quantities prove to be visually almost not distinguishable from those calculated in $\pi_{\rm free}$ model for $0.4\leq Y_Z \leq 0.5$. Therefore we do not show the curves for $\pi_{\rm WT}$ model in figs. \ref{fig::EScut03}, \ref{fig::ESmkv}. Values of the temperatures are indicated on lines in MeV. The horizontal arrows on the left panels denote the initial energy per baryon  in the laboratory system.
Thin lines   indicate  minima of the energies per baryon, which exist  in our models for all $T$ and $n$ corresponding to $\Elab \lsim 2.1A$~GeV.

With a knowledge of these quantities,
the dynamics of the expanding nuclear
fireball is determined by  the constant value of the entropy per baryon (thin horizontal dashed lines on right panel).  Solid bold curves are presented for $Y_Z = 0.48$ (as  for Ar+KCl collisions). Two dashed bold curves on each figure illustrate the case of 0.8$A$~GeV and 0.246$A$~GeV collisions  of La+La $(Y_Z =0.4)$. Comparing the bold dashed and solid curves for the collisions with $\Elab=0.8A$~GeV we see that the effect of the $Y_Z$ dependence is tiny for $0.4\leq Y_Z\leq 0.5$. This is because the symmetry energy of asymmetric matter is approximately  $\propto \beta^2$ and within the interval $0.4\leq Y_Z\leq 0.5$ the quantity $\beta^2$ changes from $0.04$ to 0 and the contribution remains negligible. For the  neutron-star matter $\beta^2 \simeq 0.9$ and thereby the symmetry energy gives significant contribution to the total energy.
Also in figs. \ref{fig::EScut03} and \ref{fig::ESmkv} by dotted bold lines we show the results for KVORcut03$\pi_{\rm free}$ and MKVOR*$\pi_{\rm free}$ models, \textit{i.e.} without the inclusion of $\Delta$ resonances. We see that the contribution of $\Delta$s becomes noticeable for all densities already at $T_m \gsim (40-50)$~MeV, which roughly corresponds to $\Elab \gsim (200-300)$~MeV. Compared to the case without $\Delta$s, with $\Delta$s included the values  $T_m$ and $n_m$ are lower for all $\Elab$, see the thin dotted line connecting minima of the energy. Contrary to that, the initial value of the entropy is larger for the models with $\Delta$s for all $\Elab$.

\begin{figure*}
	\centering
	\includegraphics[width=0.48\linewidth]{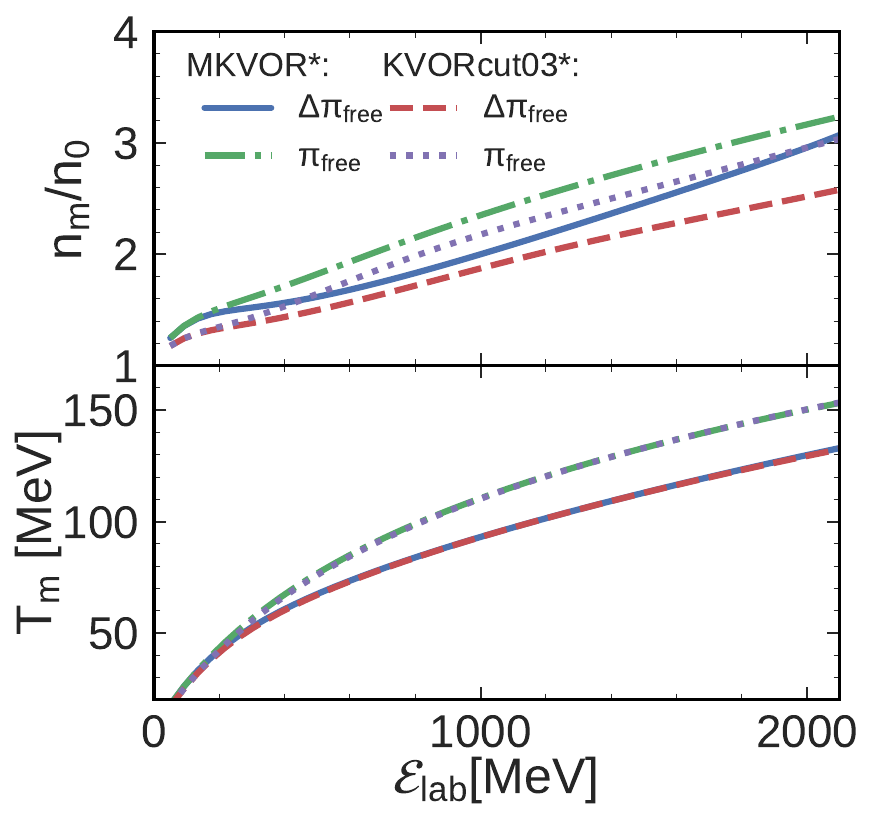}
	\includegraphics[width=0.46\linewidth]{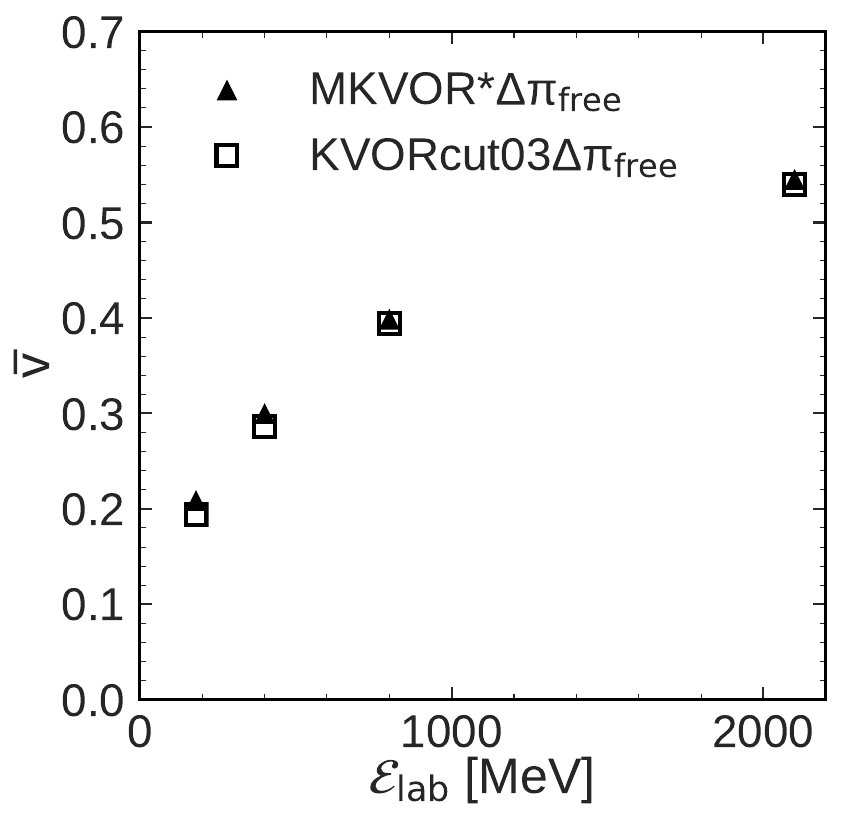}
	\caption{
		Left panel: The initial fireball density (upper left panel) and temperature (lower left panel) as functions of $\Elab$ for KVORcut03-based and MKVOR*-based models supplemented with ideal pion gas with vacuum dispersion law, with and without $\Delta$ included ($\Delta\pi_{\rm free}$ and $\pi_{\rm free}$, respectively). The curves $T_m(\Elab)$ visually coincide for both models within each set of included species, see the text for details. Right panel: Mean  flow velocity at the system breakup as a function of $\Elab$ for  the models KVORcut03$\Delta\pi_{\rm free}$ (squares)
		and MKVOR*$\Delta\pi_{\rm free}$ (triangles).
		Points correspond to fits of pion spectra in Ne+NaF collisions.
	}
	\label{fig::TmNm}
\end{figure*}

We assume  that in the  fireball, expanding with the velocity $v(t)$ (in reality with $v(\vec{r}, t)$),  the thermodynamical quasi-equilibrium  is
sustained up to a certain rather short breakup stage at which the nucleon and pion mean free paths become compatible with the fireball size, or more precisely, the typical expansion time
becomes comparable with the inverse collision frequency \cite{Mishustin:1983nv}).
After that the nucleon and pion momentum distributions
can be considered as frozen. We assume that the breakup stage is  characterized by the baryon
density $n_{\rm b.up}$ and temperature $T_{\rm b.up}$.
First fireball models estimated values of the freeze-out densities
in the interval $0.5\,n_0<n_{\rm b.up}\lsim n_0$ \cite{Gosset:1988na,DasGupta:1981xx,Nagamiya:1981sd,Barz:1982ed,Mishustin:1983nv}. Resonance gas model, \textit{cf.} \cite{Randrup:2006nr},   yields $0.3 n_0<n_{\rm b.up}\lsim n_0$ in the
whole interval of
available collision energies.
References \cite{Voskresensky:1989sn,Migdal:1990vm,Voskresensky:1993ud} argued that at
lowest SIS energies
$0.5 n_0<n_{\rm b.up}\lsim 0.8 n_0$ since for higher densities there appears a
significant contribution to the $NN$ scattering amplitude from  the $NN$  exchange
by soft pions with momenta $k\sim p_{{\rm F},N}$, region of larger $n$  is  usually called the
liquid phase of the pion condensate \cite{Voskresensky:1993ud}. Besides, the freeze-out densities
can be estimated from analysis of the HBT pion interferometry, \textit{cf.} fig. 2 in \cite{Mishra:2007xg}. In  figs. \ref{fig::EScut03} and \ref{fig::ESmkv}  the breakup moments are indicated by small dots. The choice of these points is explained further in the text.

The fireball expansion velocity is supposed to be zero at the initial moment and grows with time, since a part of the energy $E(n_m,T_m)/n_m -E(n(t), T(t))/n$
which is found using condition (\ref{constentr})
is transformed to the kinetic energy of particles.
Neglecting an energy loss due to the particle radiation in direct reactions
and surface radiation during the fireball expansion up to its breakup,
from an approximate conservation of the energy we may evaluate the velocity of the collective flow. They are slightly overestimated because of ignoring the mentioned  effects.
Resulting values of the mean  flow velocities at the freeze-out can be estimated as

\begin{gather}
\bar v = \frac{\sqrt{\Delta {\cal E} + 2 m \Delta{\cal E}}}{m + \Delta {\cal E}}, \quad \Delta {\cal E} = {\cal E}_m - {\cal E}_{\bup},
\label{eq::v_mean}
\end{gather}

where ${\cal E}_m,\, {\cal E}_\bup$ are respectively the maximum energy per participant nucleon reachable for a given $\Elab$ and the energy per participant nucleon at the breakup. Within the
KVORcut03$\Delta\pi_{\rm free}$ model we obtain $\bar{v}=0.19$, 0.29, 0.39, and 0.54 for collision energies $\Elab =0.18$, 0.4, 0.8, and 2.1$A$~GeV respectively and within the MKVOR*$\Delta\pi_{\rm free}$ model  we get values 0.21, 0.299, 0.40, 0.53, which differ only slightly from those obtained in the KVORcut03$\Delta\pi_{\rm free}$ model. Note that the values $\bar{v}$ obtained for $\gsim 1 \,A$~GeV are probably too high and would be smaller, if we included effects of the $p$-wave pion-nucleon interaction, {\em cf.} \cite{Voskresensky:1993ud}, see also results of a fit of the data for higher energies \cite{Adamczyk:2017iwn}.

On the left panel in fig. \ref{fig::TmNm} we show the initial baryon density and temperature, $n_m, T_m$, as functions of the collision energy in the laboratory system within MKVOR*-based models (solid and dash-dotted lines) and KVORcut03-based models (dashed and dotted lines).
We see that for all $\Elab$ the initial density and temperature for models without $\Delta$s (see dash-dotted and dotted lines) are larger than those with $\Delta$s (see solid and dashed lines). Also  for all $\Elab$ the values of $n_m$ in the MKVOR*$\Delta\pi_{\rm free}$ model are higher  than those in the KVORcut03$\Delta\pi_{\rm free}$ model. However, it is remarkable that the initial temperature dependence on $\Elab$ proves to be almost model independent within the same particle set.
Thus we see that in our approach to choosing the initial state the dependence of the initial thermodynamic state on the model for the EoS resides in the value of the maximum reachable baryon density, while the maximum fireball temperature depends only weakly on the employed model.
On the right panel in fig. \ref{fig::TmNm} we show the mean velocity of the fireball expansion at the breakup, $\bar{v}$,   as a function of the collision energy in the laboratory system evaluated within our models.
We see that values $\bar{v}$ evaluated in KVORcut03$\Delta\pi_{\rm free}$  and MKVOR*$\Delta\pi_{\rm free}$ models prove to be approximately the same.

Taking into account the non-zero particle velocities at the breakup leads to a modification of their spectra. Experimental slopes of the spectra, which are determined by effective temperatures at freeze out, for nucleons are a bit higher
than those for pions,
\textit{cf.} \cite{Nagamiya:1981sd}.
The mentioned difference is attributed to
the fact that differential cross sections of massive nucleons
are more affected by the presence of non-zero mean expansion velocity than differential cross sections of lighter pions, \textit{cf.} \cite{Siemens:1978pb}. Indeed, in a frame moving with the 3-velocity $\vec{v}$ the particle distribution is expressed through that in the rest frame by a replacement $p_0\to p_\nu u^\nu$, where $u^\nu$ is the 4-velocity of the frame. For non-relativistic particles and for $v\ll 1$ the transition to the moving reference frame is reduced to the replacement $(m+p^2/2m)/T\to (m+ (\vec{p}-m\vec{v})^2/2mT$ and nucleon distributions are more affected than pion ones since $m_N v\gg m_\pi v$.
In relativistic case in presence of the expansion velocity $\bar{v}$ the particle distributions are characterized by the effective temperatures $T_{{\rm ef},N}=T_{\rm b.up}
\sqrt{1-\bar{v}^2}$ and by shifted momenta. Below we focus on  pion distributions and determine the values $T_{\rm b.up}$ and $n_{\rm b.up}$ fitting the  pion distributions. For $p\gsim m_\pi$ we may neglect $m_\pi v \lsim 0.6 \, m_\pi $ compared to $p$.
Owing to this circumstance and taking into account that a slight decrease of $T_{{\rm ef},\pi}$ in comparison with $T_{\rm b.up}$ is partially compensated by the fact that $T_{\rm b.up}$ would be a bit higher, if we took into account an increase of the entropy in a realistic viscous expansion of the fireball, as in \cite{Voskresensky:1989sn,Migdal:1990vm,Voskresensky:1993ud}, we simplifying put $T_{{\rm ef},\pi}\simeq T_{\rm b.up}$.

The momentum-dependent pion free path length proves to be short for pions with momenta $k\gsim 1.5 \,m_\pi$  up to the breakup stage and thereby such  pions radiate from   the fireball breakup \cite{Voskresensky:1991uv}.  Oppositely, pions with momenta $k<(1-1.5)\,m_\pi$ have  larger mean-free path
\cite{Voskresensky:1991uv}
and  radiate from
an intermediate stage of the fireball expansion,
\textit{cf.} \cite{Voskresensky:1993mw,Voskresensky:1995wn}.
Moreover, a contribution to the pion yield comes from the decay of thermal $\Delta$-resonances at the breakup, or may be a bit later, if typical time of the reaction  $\tau_{\Delta\to N\pi}$ at $n\sim n_{\rm b.up.}$, $T\sim T_{\rm b.up.}$ is larger than $\tau_{\rm b.up.}$ for thermal pions.
Further we determine the values $n_{\rm{b.up}}({\cal{E}}_{\rm lab})$, $T_{\rm{b.up}}({\cal{E}}_{\rm lab})$   from
the best fit of the differential pion cross sections for the momenta $k\gsim 1.5 m_\pi$.

\subsection{Description of pion differential cross sections}

The   $\pi^-$ differential cross section in  inclusive processes reads \cite{Senatorov:1989cg,Voskresensky:1989sn}
\begin{gather}
\omega_k \frac{d\sigma}{d^3 k}=15.64 \,\frac{(2 \pi)^3}{V_{{\rm b.up}}}\frac{A^{5/3}}{n_{\rm b.up}}\,\frac{{\rm{mb}}\cdot {\rm{GeV}}}{{\rm{sr}}\cdot
	({\rm{GeV}}/c)^3} \Big\{\omega_k \frac{d^3N_{\pi^-}}{d^3k}\nonumber \\ + \omega_k \frac{d^3N_{\Delta}}{d^3k}\Big\}\,,
\label{eq::difcross}
\end{gather}
where $\omega_k =\sqrt{m_\pi^2 +{k}^2}$. The first term in the curly brackets is the contribution of thermal pions  at the breakup stage,  whereas the second term corresponds to the decay  $\Delta\to N+\pi$ occurring at the breakup stage with direct radiation of free pions. For simplicity we
consider collisions of nuclei with equal atomic weights $A=A_{\rm t}=A_{\rm p}$, $V_{{\rm b.up}}$ is the volume of the fireball at the breakup, $15.64 \,  A^{5/3}/n_{\rm{b.up}}$ is the geometric factor for inclusive processes. The numeric coefficient corresponds to all the quantities being measured in units of $m_\pi$, in particular $n_{\rm{b.up}}$ is measured in $m^3_\pi$, $n_0\simeq0.45\, m_\pi^3$.

We assume that for $t>t_0$ during the breakup stage, which lasts for $\tau_{\rm b.up}(n_{\rm b.up}, T_{\rm b.up})$, nucleons and pions  decouple and for $t>t_0 +\tau_{\rm b.up}$ pions, which before breakup stage were described by thermal distributions, can be considered as freely moving  particles. In the quasiparticle model the distribution of pions is described by \cite{Senatorov:1989cg,Voskresensky:1989sn}
\begin{eqnarray}
\frac{d^3N_{\pi^-}}{d^3k} =\frac{V_{{\rm b.up}}}{(2\pi)^3} \frac{ \Gamma_{{\pi}^{-}} (k)}{e^{[(\omega_{\pi^-} (k) - \mu_Q)/T]} - 1}\,.\label{eq::difcrosspion}
\end{eqnarray}
The contribution from $\mu_Q \equiv \mu_n -\mu_p$
appears for $Y_Z \neq 0.5$ and distinguishes the $\pi^-$ from other pion species. The quantity $\om (k)$ coincides with $\om_k$, if one uses the vacuum dispersion law, and $\om (k)$ is given by eq. (\ref{WT}) in the $\pi_{\rm WT}$ model.
 In the model of the sudden breakup (if the typical time for the pion sub-system breakup   is $\tau_{\rm b.up}\ll 1 / |\om_k -\om (k)|$) the value of $\Gamma_k$ is given by \cite{Senatorov:1989cg}
\begin{gather*}
\Gamma_{{\pi}^{-}} (k)= \frac{2\om_k}{\left(2\om (k) -\dfrac{\partial \mbox{Re}\Pi_{{\pi}^{-}}}{\partial \om}|_{\om =\om_{{\pi}^{-}} (k)}\right)}\,.
\end{gather*}
Using the spectrum (\ref{WT})  $\Gamma_k \simeq 1+O((n_n-n_p)^2/n_0^2)$
we can put $\Gamma_k \simeq 1$ at the breakup in both $\pi_{\rm WT}$ and $\pi_{\rm free}$ models.
In the limit of a slow breakup, $\tau_{\rm b.up}\gg 1 / |\om_k -\om (k)|$, one should use eq. (\ref{eq::difcrosspion}) with $\om (k)$ replaced by $\om_k$. In present work where we disregard effects of the $p$-wave pion-baryon interaction the limit $\tau_{\rm b.up}\ll 1 / |\om_k -\om (k)|$ looks as more realistic  since effects of the $s$-wave interaction remain weak for $0.4\leq Y_Z\leq 0.5$ and thereby $1 / |\om_k -\om (k)| \gg 1/m_\pi$, whereas a typical value of $\tau_{\rm b.up}$ is estimated as few $1/m_\pi$. However note that for $\om (k)$ close to $\om_k$ at breakup conditions the difference between pion distributions calculated in the limit of the prompt breakup  and the slow breakup is not as significant.

\begin{figure*}[t]
	\centering
	\includegraphics[width=.48\textwidth,clip=true]{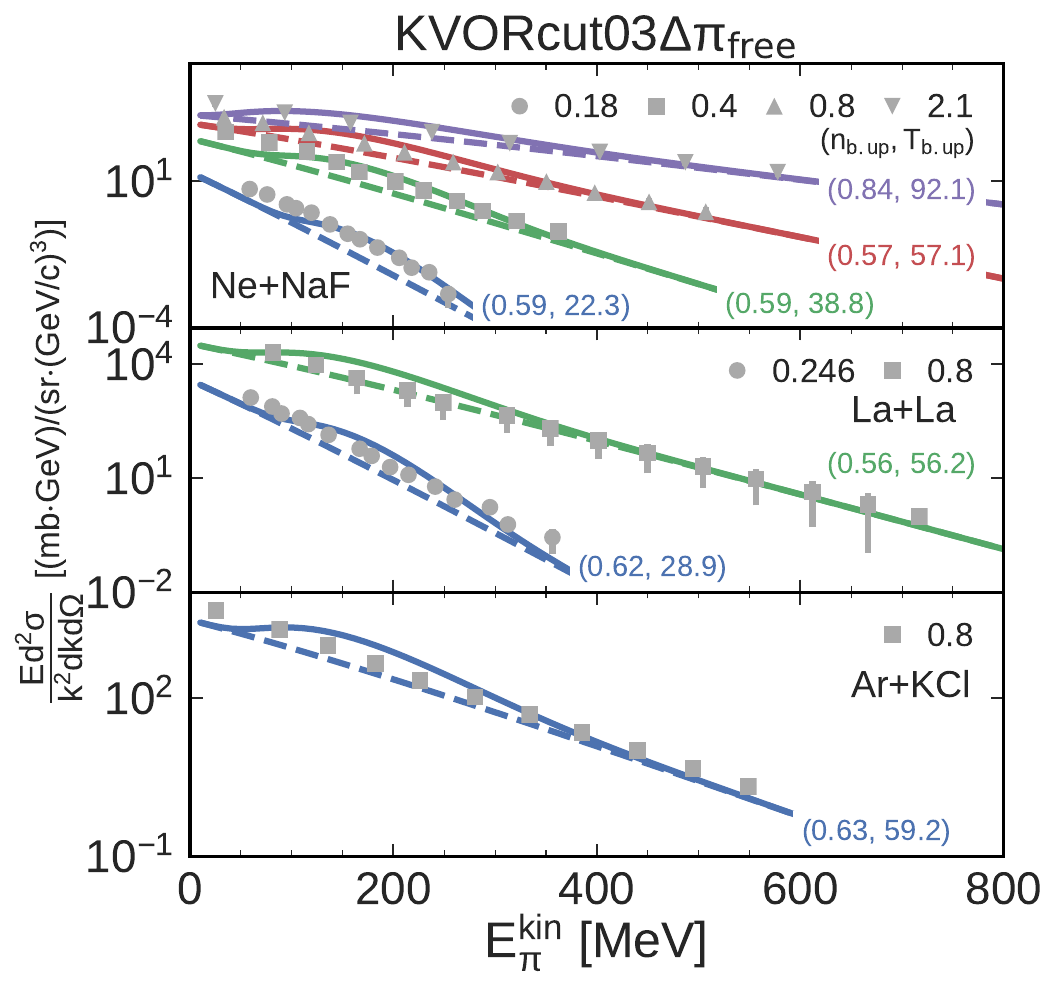}
	\includegraphics[width=.48\textwidth,clip=true]{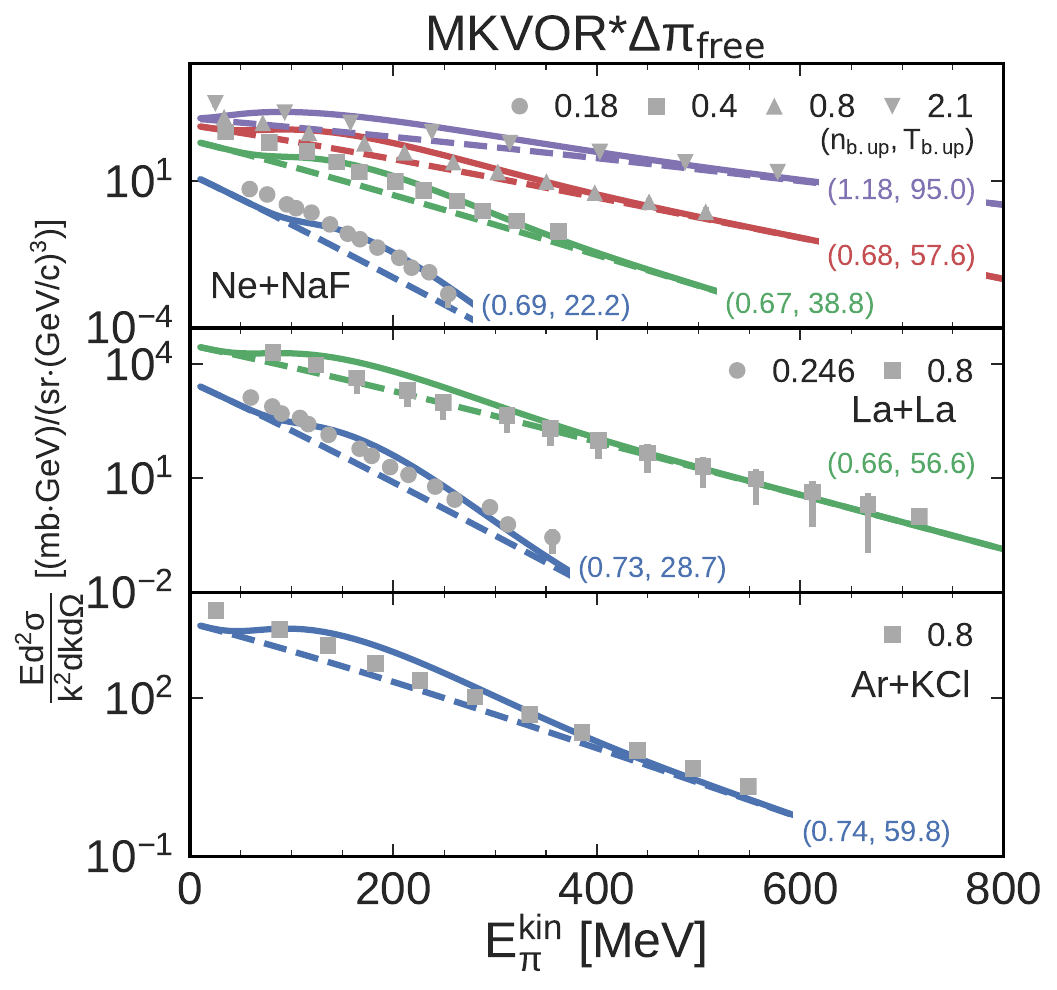}
	\caption{Invariant differential cross sections of the $\pi^-$ production for the models KVORcut03$\Delta\pi_{\rm free}$ (left panel) and MKVOR*$\Delta\pi_{\rm free}$    (right panel).  The data are shown  for  Ne+NaF collisions at ${\cal E}_{\rm lab}=$~(180, 400, 800, 2100)$A$~MeV (top panels), for La+La reactions (middle panels) at ${\cal E}_{\rm lab}=246 A$~MeV and $800 A$~MeV, and for Ar+KCl collisions (lower panels) at ${\cal E}_{\rm lab}=800A$~MeV. Data are taken
		from \cite{Nagamiya:1981sd,Miller:1987zz,Hayashi:1988en} for $\theta_{\rm c.m}=\pi/2$. $\Elab/A$ in the legends are indicated in GeV. Values  $(n_{\rm b.up}/n_0\, , T_{\rm b.up}/\mev)$ are shown near the curves. Solid lines demonstrate  calculations done within $\Delta\pi_{\rm free}$ models, dashed lines are  presented for  models with the $\Delta$ contribution switched off.
	}
	\label{fig::fits}
\end{figure*}

For the contribution of $\Delta$  quasiparticles described with effective mass to the $\pi^-$ momentum distribution one gets \cite{Ivanov:2005yw}
\begin{gather}
\omega_k \frac{d^3N_{\Delta}}{d^3k} = \frac{V_{{\rm b.up}}}{(2\pi)^3} \sum_{r}  \sum_{n=1}^{\infty} \frac{(-1)^{n-1} b_r g_r m_r^\s2}{k \, \Lambda^{1/2}} \nonumber \\ \times\Big[\frac{n ET + T^2}{n^2}  e^{-[n(E - \mu_r^*)/T]}\Big]^{E^-}_{E^+},
\label{eq::difcrossDelta}\\
\nonumber
\lambda(x, y, z) = (x - y - z)^2 - 4 y z,\, \Lambda \equiv \lambda(m_r^\s2, m_\pi^2, m_N^\s2), \\\nonumber
E^{\pm}(k) = \frac{1}{2 m_\pi^2}\Big[ (m_r^\s2 + m_\pi^2 - m_N^\s2) E_\pi(k) \pm \Lambda^{1/2} k \Big].
\end{gather}
Here $r$ runs through ($\Delta^-,\,\Delta^0$). The branching ratios $b_r = (1,\,1/3)$ arise since  $|N \pi> = \frac{1 }{\sqrt{3}} |p \pi^-\rangle + \frac{2 }{\sqrt{3}} | n \pi^0\rangle$ and therefore the rate of probabilities of reactions
$\Delta^0 \to n + \pi^0$, $\Delta^0 \to p + \pi^-$ is $P(n \pi^0) / P(p \pi^-) = 2$. The factor $g_r = 4$ is the resonance spin degeneracy.

The in-medium effective masses should be used in (\ref{eq::difcrossDelta}), if the breakup time for pions to become freely moving particles  and the time step at $n=n_{\rm b.up}$ and  $T=T_{\rm b.up}$  for $\Delta$s to decay to pions and nucleons are shorter than the time typical for the nucleons to become freely moving particles. Otherwise during the breakup stage the  in-medium $\Delta$s first transit to their vacuum branch and only after that decay into the freely moving nucleons and pions. In the latter case $m_r^*$ and $m_N^*$
in (\ref{eq::difcrossDelta}) should be replaced to $m_r$ and $m_N$.

Pion differential cross sections and yields in heavy-ion collisions at Bevalac and lowest GSI SIS energies were measured in
\cite{Nagamiya:1981sd,Miller:1987zz,Hayashi:1988en,Pelte:1997rg,
	Sandoval:1980bm,Harris:1987md,Reisdorf:2006ie,Hong:2005at}
with Ne, Ar, Ru, Zr,
La, Au and Ru beams.
In fig. \ref{fig::fits} we show invariant differential cross sections of $\pi^-$ production in reactions Ne+NaF, Ar+KCl ($Y_Z \simeq 0.48$) and for  La$+$La ($Y_Z \simeq 0.4$), computed within KVORcut03-(left panel) and MKVOR*-based (right panel) models with $\Delta\pi_{\rm free}$ particle set. Solid lines show results computed following eqs. (\ref{eq::difcross}-\ref{eq::difcrossDelta}). To be specific we use eq. (\ref{eq::difcrossDelta}) for $\Delta$s described with effective masses. The values  $n_\bup, T_\bup$ lie on the $s = \const$ lines indicated on the right panels in figs. \ref{fig::EScut03} and \ref{fig::ESmkv} and correspond to the best fit of the pion differential cross sections performed for $k>1.5 m_\pi$. Dashed lines show the result obtained for the same $n_\bup, T_\bup$ provided the contribution of $\Delta$ resonances is turned off. We see that $\Delta$ decays give a contribution centered around $E_\pi^{\rm kin} = (100-200)$~MeV, which is significant even for low temperatures due to the in-medium mean fields acting on the baryons.
As expected, the cross sections for low pion kinetic energies $E_\pi^{\rm kin}\lsim (1-1.5)m_\pi$ are underestimated because the direct emission of low-momentum pions and the $p$-wave polarization effects are not taken into account. The differential cross sections computed for La $+$ La reactions for KVORcut03$\Delta\pi_{\rm WT}$   and MKVOR*$\Delta\pi_{\rm WT}$ models   (\textit{i.e.} using the dispersion law (\ref{WT}) and that  $\mu_{{\pi}^{-}} =\mu_n -\mu_p$)  are not shown since the difference with the distributions computed using the corresponding $\pi_{\rm free}$ models is not seen visually. More specifically, for ${\cal{E}}_{\rm lab}=800A$~MeV the difference changes within $(2-7)\%$, for lower collision energies it increases, and for ${\cal{E}}_{\rm lab}=246A$~MeV the difference reaches $16\%$, although on the logarithmic plot such a difference is almost not seen.
So we conclude that our models with taking into account the $\Delta$ decays allow to describe the pion differential cross sections rather appropriately, even within $\pi_{\rm free}$ models, except for soft pions at low collision energies. Here we once more point out that we did not include a contribution \cite{Voskresensky:1993mw,Voskresensky:1995wn} of  soft  pions produced in direct reactions before the breakup and the  $p$-wave polarization effects \cite{Voskresensky:1989sn,Migdal:1990vm,Voskresensky:1993ud}, which inclusion should  improve the description.

\subsection{Description of ratios of pion yields to those of positively charged baryons}

\begin{figure*}[t]
	\centering
	\includegraphics[width=.45\textwidth]{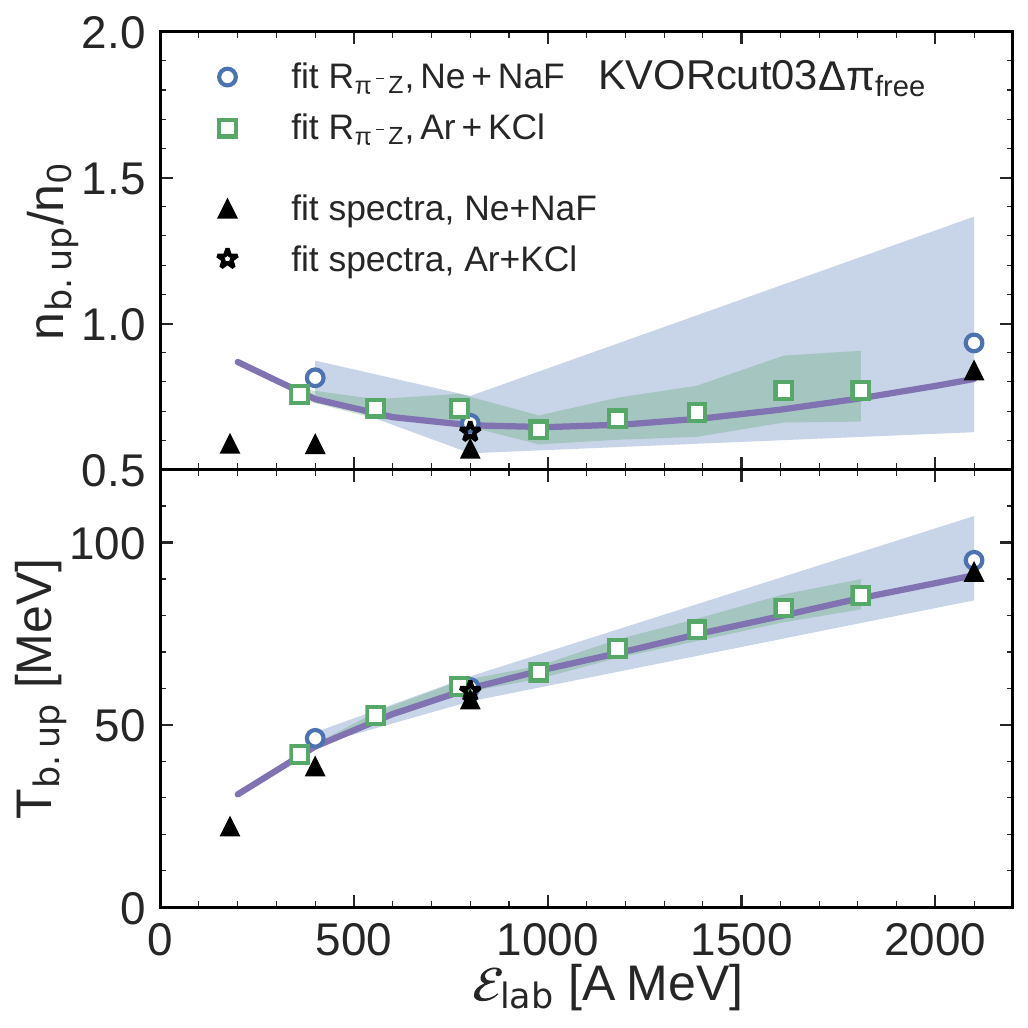}
	\includegraphics[width=.45\textwidth]{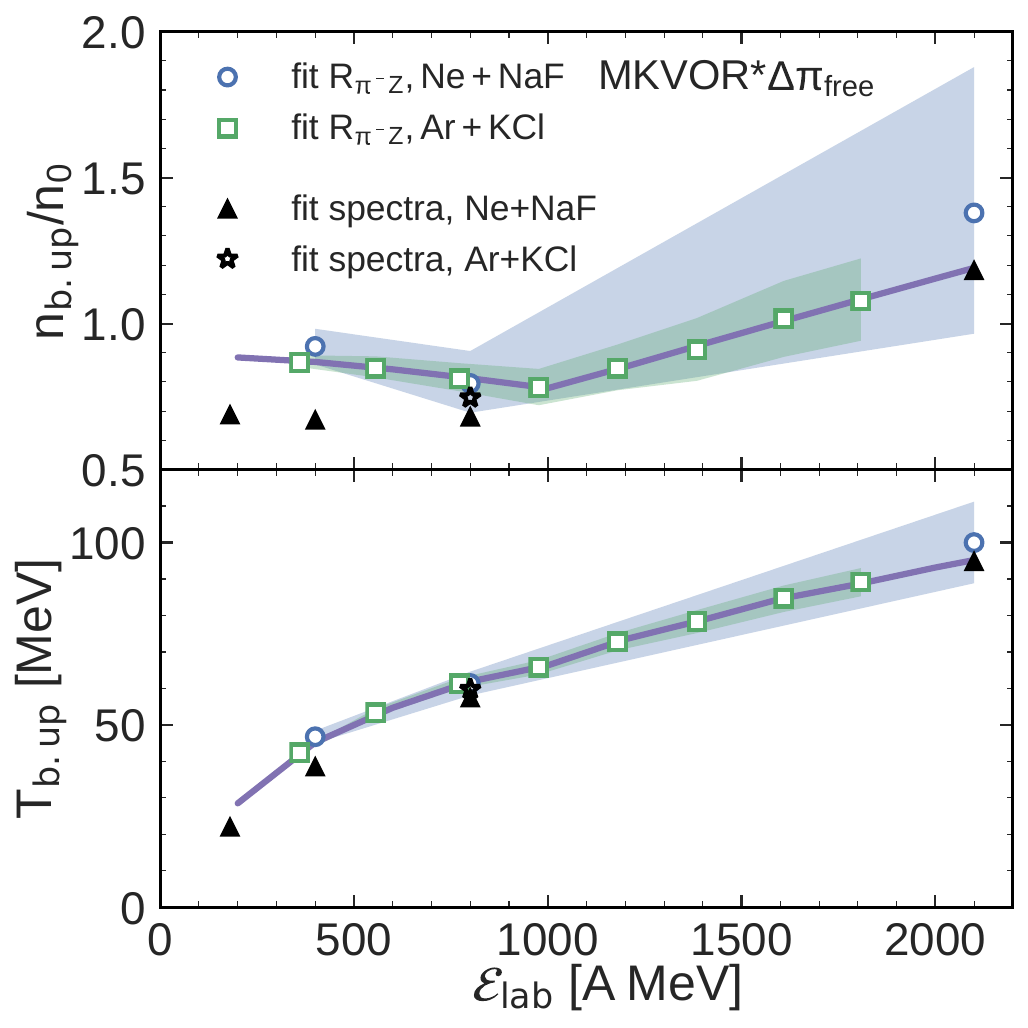}
	\caption{Breakup density and  temperature as functions of $\Elab$ computed within models KVORcut03$\Delta\pi_{\rm free}$ (left panel) and MKVOR$^*\Delta \pi_{\rm free}$ (right panel). Circles and squares denote the best fits to the $R_{\pi^-Z}$ data, and the shaded areas are the uncertainties allowed by the fits within the data error bars. Other symbols label the values $(n_\bup, T_\bup)$ found from the fit of the pion spectra shown in fig. \ref{fig::fits}. Lines show the parameterization given by eq. \ref{interp}, which lies within the uncertainties.}
	\label{fig::breakup}
\end{figure*}
The mean number of protons per baryon measured after an event is given by
$$\frac{n_{\rm p-like}}{n}=Y_p +\frac{1}{3n} (n_{\Delta^0} +2n_{\Delta^+} + 3 n_{\Delta^{++}}),$$
which includes the contribution from the $\Delta$ decays with corresponding branching factors. Here we assume that all $\Delta$s at the breakup stage are transformed to pions and nucleons.
The ratio of $\pi^-$-like to  $p$ -like particle concentrations at the breakup is as follows
\begin{gather}
R_{\pi^- Z} = \frac{1}{n_{\rm p-like}} \Big[n_{\pi^-} + \sum_{r} b_r n_r \Big]\,,
\label{eq::RpiZ}
\end{gather}
where the first term in the brackets shows the contribution of thermal pions computed with the help of eq. (\ref{eq::difcrosspion}), which at breakup stage are transformed to freely moving pions, and
the second term  given by (\ref{eq::difcrossDelta}) takes into account  the decay of the thermal $\Delta$s into the freely moving pions and the nucleons at the breakup stage. As in eq. \eqref{eq::difcross}, $r$ runs through $(\Delta^-, \Delta^0)$ and the respective values $b_r$ are $(1, 1/3)$.
In our model, where we disregard the surface  effects, the $R_{\pi^- Z}$ ratio does not depend on the fireball volume. In a case of the ISM we get  $R_{\pi^-Z} = 2 (n_{\pi^-} + \frac 1 3 n_\Delta) / n$ and in the Boltzmann limit we have approximately:
\begin{gather}
\frac{2}{3}\frac{N_\Delta}{Z} \simeq \frac{2e^{-\Delta m /T_{\rm{b.up}}}}{3Y_p
}\left(\frac{m^*_\Delta (n_{\rm{b.up}},
	T_{\rm{b.up}})}{m^*_N(n_{\rm{b.up}},
	T_{\rm{b.up}})}\right)^{3/2}\,,
\label{deltapi1}
\end{gather}
where $\Delta m = m^*_\Delta (n_{\rm{b.up}},T_{\rm{b.up}}) - m^*_N (n_{\rm{b.up}},T_{\rm{b.up}})$ when we use eq. (\ref{eq::difcrossDelta}) with in-medium baryon masses,  and $\Delta m = m_\Delta - m_N$, if we explore  (\ref{eq::difcrossDelta}) with vacuum baryon masses.

\begin{figure*}
	\centering
	\includegraphics[width=.45\textwidth]{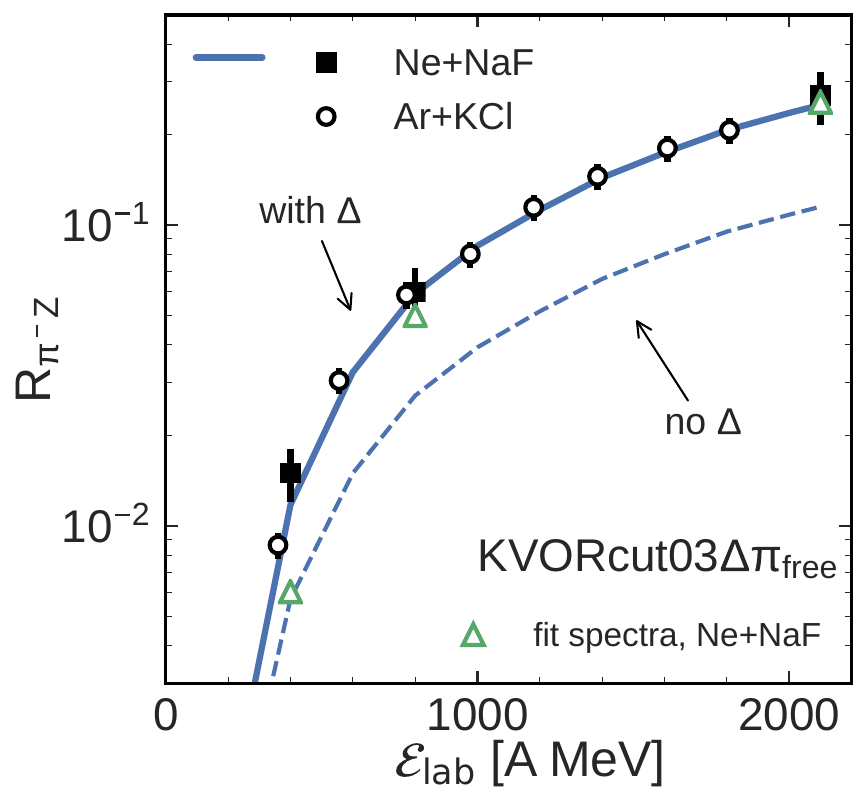}
	\includegraphics[width=.45\textwidth]{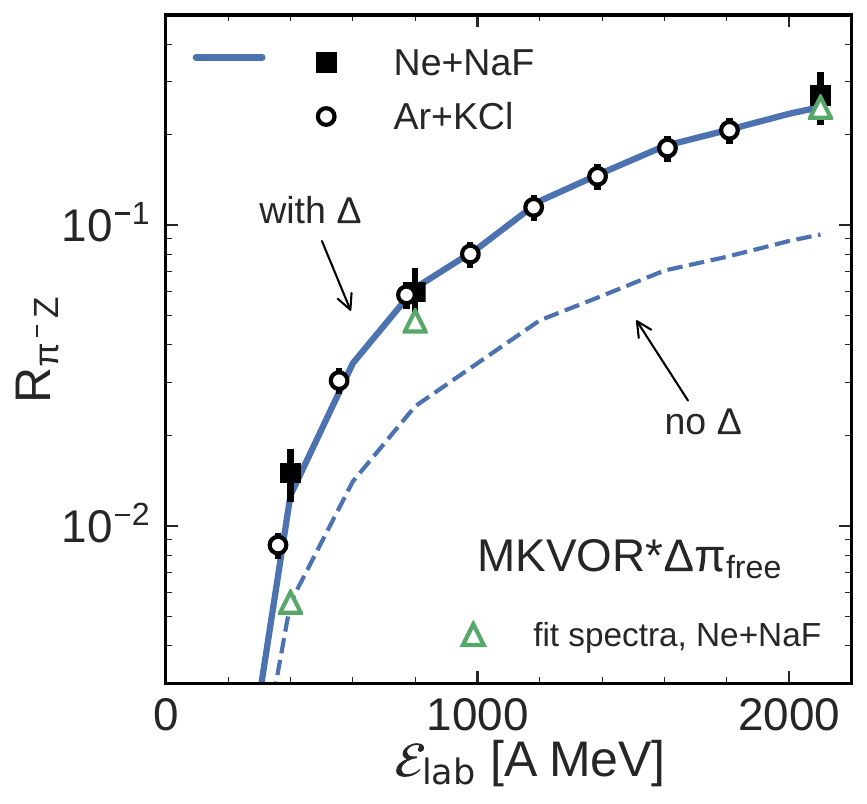}
	\caption{Left panel: Ratios of $\pi^-$ yields to those of protons coming to the detector, $R_{\pi^- Z}$,
		as a function of the collision energy in the
		laboratory system calculated according to eq. (\ref{eq::RpiZ}) within the models KVORcut03$\Delta\pi_{\rm free}$ (left panel) and MKVOR*$\Delta\pi_{\rm free}$ (right panel). Solid lines are for the calculation with $\Delta$ included and dashed lines denote the results with only the first term in \ref{eq::RpiZ} taken into account. The breakup parameters follow the solid lines in fig. \ref{fig::breakup}. Data points are taken from refs. \cite{Sandoval:1980bm,Nagamiya:1981sd}. Triangles show the breakup parameters obtained from the fits to the pion spectra in Ne+NaF collisions shown in fig. \ref{fig::fits}. }
	\label{fig::RpiZ}
\end{figure*}

In fig. \ref{fig::breakup} we show the results for the breakup density $n_\bup$ (upper panels) and $T_\bup$ (lower panels) within our models. The symbols denote the best fits using \eqref{eq::RpiZ} to the available data on $R_{\pi^- Z}$ in Ne+NaF and Ar+KCl collisions, and the shaded areas demonstrate the associated uncertainties within the data error bars. The triangles and a star show the corresponding results for the differential cross-section given by eq. \eqref{eq::difcross}. We see that at $\Elab \gsim 800$~MeV for each model both fits for Ne+NaF and Ar+KCl generally agree with each other within the error bars. For $\Elab \lsim 800$~MeV the breakup densities and corresponding temperatures required by $R_{\pi^- Z}$ ratio are larger than those following from the fit of the differential cross sections demonstrated in fig. \ref{fig::fits}.
To parameterize a possible
breakup density dependence on $\Elab$  for the energies  $200\, \mev \lsim \Elab \lsim 2100 \, \mev$  of our interest here we use the interpolation formula
\begin{gather}
\frac{n_\bup(\Elab)}{n_0} \simeq n_1 + \frac{a}{\rm MeV^{1/2}} \sqrt{\Elab} + \frac{b}{\mev} \, \Elab  \nonumber \\+ \frac{c}{\mev} \, (\Elab - {\cal E}_1) \theta(\Elab - {\cal E}_1), \label{interp} \\\nonumber
\mbox{KVORcut03$\Delta\pi_{\rm free}$: }\nonumber\\\, n_1=1.39, \,\, a =-4.80\cdot10^{-2}, \,\, b = 7.69\cdot10^{-3}, \nonumber \\ c = {\cal E}_1 = 0, \nonumber \\
\mbox{MKVOR*$\Delta\pi_{\rm free}$:}\nonumber\\ n_1= 0.838, \,\, a = 7.38\cdot10^{-3}, \,\, b = -3.02 \cdot 10^{-4}, \nonumber \\ \nonumber
c = 2.59 \cdot 10^{-4}, \,\, {\cal E}_1 = 1.01 \,{\rm GeV},
\end{gather}
where $\theta(x)$ is the step function. Lines given by \eqref{interp} pass through both the regions for Ne+NaF and Ar+KCl, and are shown for each model by solid lines in fig. \ref{fig::breakup}.

In fig. \ref{fig::RpiZ} we show the fitted ratios of $\pi^-$-like  yields to those of $p$-like yields, $R_{\pi^- Z}$, for particles
emitted at the breakup of the nuclear fireball within the KVORcut03$\Delta\pi_{\rm free}$ model (left panel) and MKVOR*$\Delta\pi_{\rm free}$ model (right panel). Solid curves are calculated according to eq. (\ref{eq::RpiZ}) assuming $Y_Z = 0.48$ with the
$\Delta$ contribution  calculated using the prompt breakup assumption, so all the
$\Delta$s with effective masses $m^*_\Delta (n_{\rm b.up}, T_{\rm b.up})$ are supposed to be transformed at the fireball breakup to free pions and nucleons with masses $m^*_N (n_{\rm b.up}, T_{\rm b.up})$, which after switching off the mean fields go  on the  mass shell. The breakup parameters follow the solid curves in fig. \ref{fig::breakup} given by eq. \eqref{interp}, and we see that the $R_{\pi^- Z}$ data are indeed well described for such breakup parameters.
Thin dashed lines show the $R_{\pi^- Z}$ ratios calculated taking into account only the thermal pion contribution to $R_{\pi^- Z}$ (first term in (\ref{deltapi1})) for the same breakup parameters.  We see that $\Delta$s contribute essentially in $R_{\pi^- Z}$ ratios in the whole interval of energies  $200\, \mev \lsim \Elab \lsim 2100 \, \mev$ demonstrated in figures. As in fig. \ref{fig::breakup}, the triangles show the results with the breakup parameters following the best fits of the spectra.
Data points are extracted from refs. \cite{Nagamiya:1981sd,Sandoval:1980bm,Harris:1987md}.
These data show that the pion yields are approximately linear functions of the participant number and that $R_{\pi^- Z}$ are approximately the same for different
colliding nuclei with the same $Y_Z$.
We see that the best fit values $(n_\bup, T_\bup)$ extracted from the $\pi^-$ spectra noticeably underestimate the $R_{\pi^-Z}$ ratios compared with the experimental data.

Figures \ref{fig::breakup} and \ref{fig::RpiZ} demonstrate that there is substantial difference in the breakup parameters inferred from $R_{\pi^-Z}$ ratios and the $\pi^-$ spectra. There are two reasons of this discrepancy within the current implementation of the thermodynamical model. It is known \cite{Voskresensky:1989sn,Migdal:1990vm,Voskresensky:1993ud} that, if the in-medium modifications of the pion spectra due to the $p$-wave pion-baryon interaction are taken into account, the pion yields would increase, especially at low temperatures (for low $\Elab$). Another reason is a lack of soft pions with long mean-free paths in our calculations, being emitted in direct reactions before the fireball breakup. Their inclusion would not change the results of the fit of the spectra, because the fit is done for pions with kinetic energies $\gsim 1\,m_\pi$. However, the direct pions will contribute substantially to $R_{\pi^- Z}$, which are largely underestimated, if the breakup parameters following from the $\pi^-$ spectra are used. As we expect, if the pion distributions and yields were calculated with taking into account mentioned effects,  the values $(n_\bup, T_\bup)$ obtained in those fits would coincide.

\subsection{Ratios of $\pi^{-}/\pi^{+}$ yields}

\begin{figure*}
	\centering
	\includegraphics[width=0.45\textwidth]{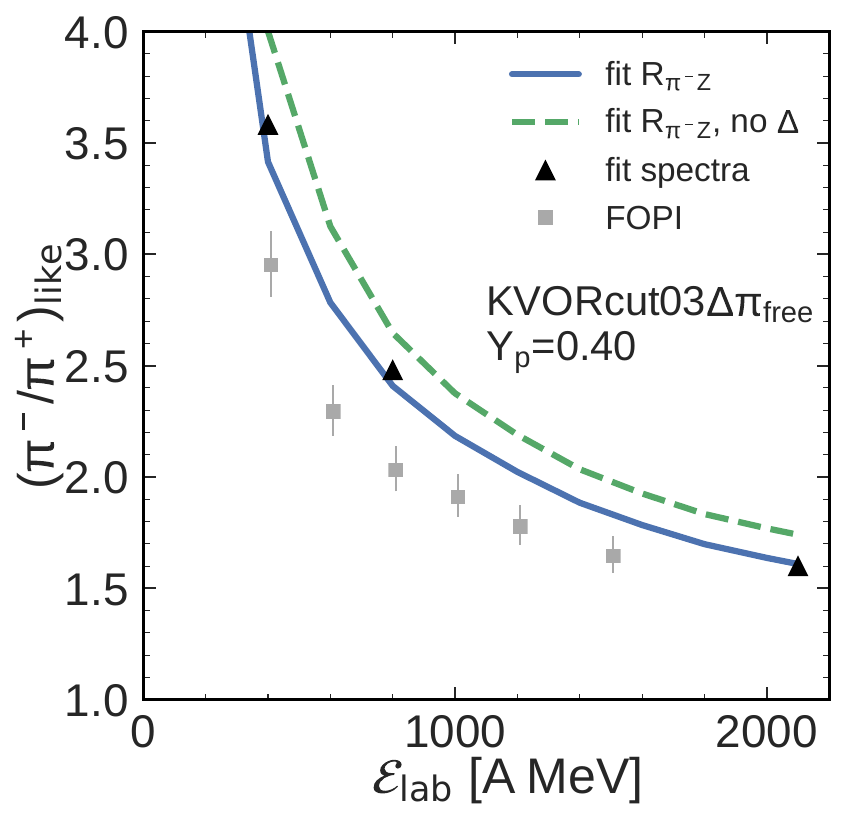}
	\includegraphics[width=0.45\textwidth]{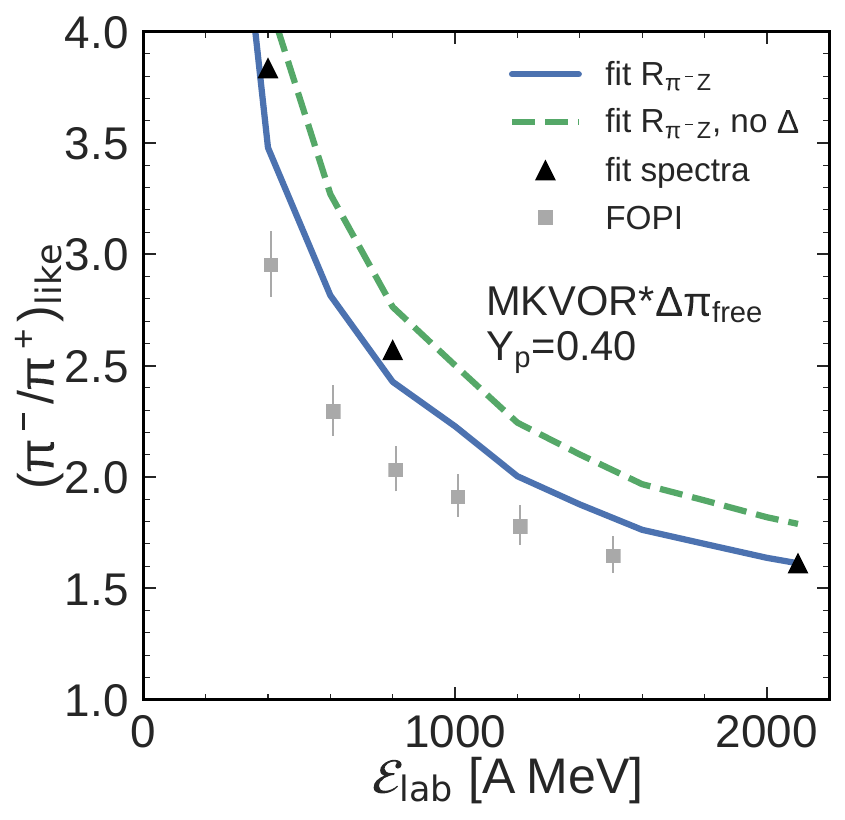}
	\caption{Ratios of $(\pi^{-}/\pi^{+})_{\rm like}$ yields as functions of the collision energy in the laboratory system computed in the KVORcut03$\Delta\pi_{\rm free}$ model (left panel) and in MKVOR*$\Delta\pi_{\rm free}$ model (right panel). Data (squares) are for Au$+$Au collisions from \cite{Reisdorf:2006ie}. Solid lines show result of our calculation performed  with the help of eqs. (\ref{eq::difcrosspion})-(\ref{eq::difcrossDelta}) for $Y_Z =0.4$. Dashed lines demonstrate calculations with  switched off the $\Delta$ contribution. Triangles show  results obtained using parameters taken from fit of the differential cross sections shown in fig. \ref{fig::fits}.
	}
	\label{fig:piratios}
\end{figure*}
Experimental data demonstrate that ratio of the  $(\pi^{-}/\pi^{+})$ yields
is larger than unity and increases with a decrease of the collision energy. In fig. \ref{fig:piratios}  experimental ratios are shown by the squares. Previously $(\pi^{-}/\pi^{+})$ ratios have been computed within different models, \textit{cf.} \cite{Xu:2013aza,Hong:2013yva} and refs. therein. A role of effects of $\Delta$ and nucleon optical potentials, $s$-wave and $p$-wave interactions has been discussed.

Since the $\Delta$ decays to nucleons and pions contribute to the ratio of the $(\pi^{-}/\pi^{+})$ we should compare the ratios with the data using the following relation
\begin{equation}\left(\frac{\pi^-}{\pi^+} \right)_{\rm like}  = \frac{n_{\pi^-} + n_{\Delta^-} + \frac{1}{3} n_{\Delta^0}}{n_{\pi^+} + n_{\Delta^{++}} + \frac 1 3 n_{\Delta^+}}\,.
\label{pimpip}
\end{equation}
As before, we assume that the  breakup is prompt  and the $\Delta$ densities are calculated with the in-medium effective masses and take into account the finite $\mu_Q>0$ for $Y_Z\neq 0.5$.

The solid curves in fig. \ref{fig:piratios} show results of our calculations
performed using eqs. \eqref{pimpip} for $Y_Z =0.4$. We use the same $n_\bup(\Elab)$ given by \eqref{interp} which we used to describe $R_{\pi^-Z}$ ratios. Triangles denote the results for $(n_\bup, T_\bup)$ obtained by fitting the differential cross sections shown in fig. \ref{fig::fits}. We employ the $\pi_{\rm free}$ model. If we used the WT model with spectra following eq. (\ref{WT}), the $s$-wave pion-nucleon interaction would affect the $\left(\frac{\pi^-}{\pi^+} \right)_{\rm like}$ ratios only slightly, since $e^{-(n_n-n_p)/2f_{\pi}^2 T}\simeq 1$ in the collision energy interval we show in fig. \ref{fig:piratios}.  The dashed lines demonstrate the results obtained within the models with the $\Delta$ contribution artificially suppressed. We see that without $\Delta$s the $(\pi^{-}/\pi^{+})$  ratios prove to be significantly overestimated. Results of calculations in both of our models KVORcut03$\Delta\pi_{\rm free}$  and  MKVOR*$\Delta\pi_{\rm free}$  reproduce the experimentally observed increase of the $(\pi^-/\pi^+)_{\rm like}$ ratios with a decrease of $\Elab$.  However, the results of our calculations are roughly by $30 \%$ higher than the experimental data. This can be again attributed to a lack of directly emitted soft pions and to the lack of $p$-wave pion-baryon interaction effects in our calculations.

\section{First-order liquid-gas phase transition in KVORcut03 and MKVOR*- based  models}\label{liquidgas}
The first-order nuclear LG phase transition
in the matter prepared in low energy
heavy-ion collisions may occur for densities $n <n_0$ and  temperatures $T<T_{cr}^{\rm LG} <(15-23)$ MeV \cite{Ropke:1982vzx}, see \cite{Dutra:2018amp} and refs. therein. The values of the critical density and temperature depend on the model of the EoS. The
$\Delta$ isobars do not contribute  at such low densities and temperatures,
and contribution of all thermal excitations of mesons (pions, $\sigma$, $\omega$,
$\rho$ mesons and heavier ones)
is negligibly small.
The temperature dependence remains essential only in nucleon terms. Although we continue to use KVORcut03$\Delta\pi_{\rm free}$ and MKVOR*$\Delta\pi_{\rm free}$ models the results for densities $n<n_0$ and for temperatures $T\lsim  (20-30)$ MeV are the same as within KVORcut03 and  MKVOR models.

A general physical picture of the dynamics of the nuclear first-order LG phase transition has been formulated long ago, \textit{cf.} \cite{Schulz:1983pz}, where the EoS of ISM was considered in
the original Walecka RMF model.
The non-ideal hydrodynamics of  the nuclear
LG phase  transition was then studied in
\cite{Skokov:2008zp,Skokov:2009yu,Skokov:2010dd,Voskresensky:2010gf}  on an example of the
van der Waals EoS. The case of the ISM has been studied first.
Depending on the collision energy, the system may enter the region
of the supercooled gas, overheated liquid or the region of the spinodal instability.
Thus depending on the experimental conditions, the system may undergo the
hydrodynamical instability, the first-order phase transition from the supercooled gas
to the liquid, or the transition from the  overheated liquid to the gas.
For ISM one deals with one conserved baryon charge.
In case of IAM one deals with two
conserved charges, the baryon charge and the isospin, that adds
specific features  to the problem, \textit{cf.} \cite{Muller:1995ji,Ducoin:2005aa,Alam:2017krb}.

\subsection{Liquid-gas  phase transition in ISM}
\subsubsection{RMF consideration}
\begin{figure}
	\centering
	\includegraphics[height=6.5cm,clip=true]{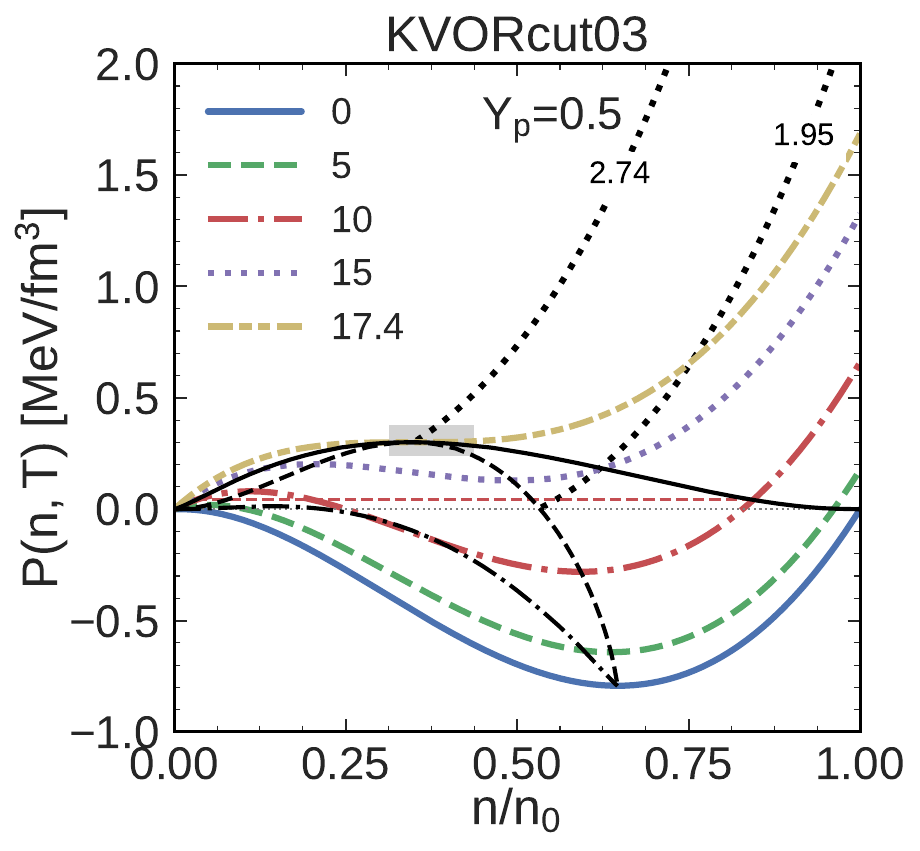}
	\includegraphics[height=6.5cm,clip=true]{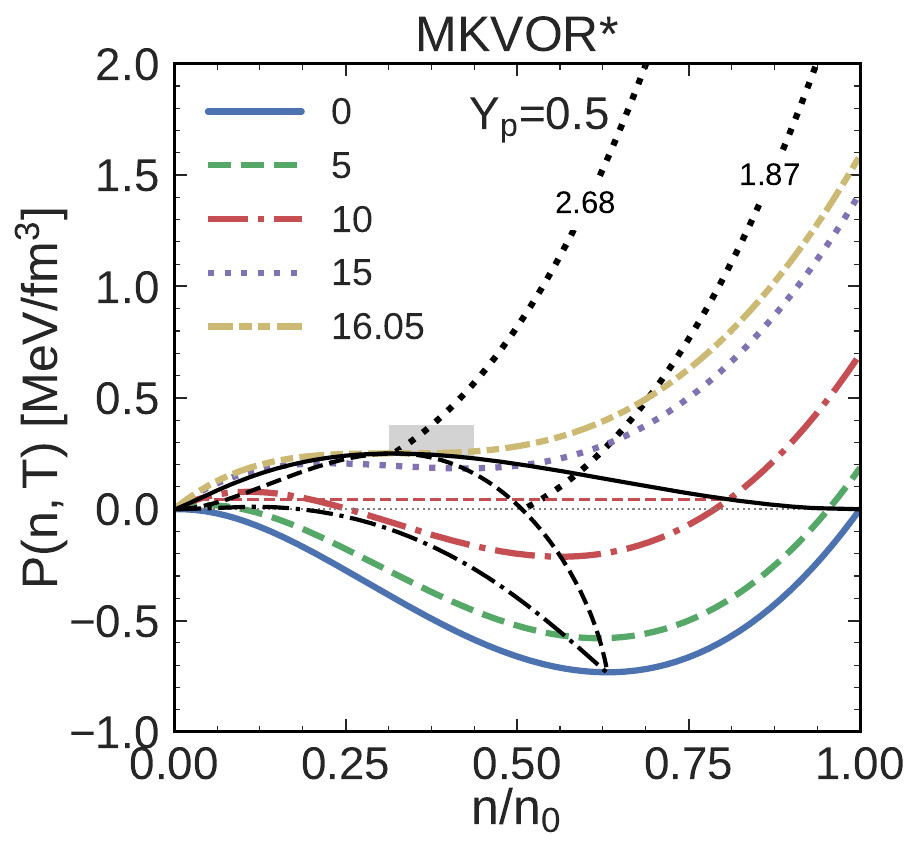}
	\caption{  The pressure-density isotherms for ISM. Upper panel: KVORcut03 EoS, lower panel: MKVOR* model. The isothermal spinodal (ITS) region is bounded
		by a  dashed line. The horizontal dashed line shown for $T=10$ MeV is an example of the Maxwell construction line. Two isoentropic expansion trajectories
		(two dotted lines) are labeled by the values of the entropy per baryon ${s}=S/n$. Numbers in the legend indicate values of the temperature
		for the corresponding isotherms in MeV. The adiabatic spinodal (AS) region  is bounded
		by a  dash-dotted line. Shaded rectangle denotes the critical parameters extracted from analysis of the experiments \cite{Elliott:2013pna}.
	}
	\label{Pn}
\end{figure}

First let us consider the LG phase transition in ISM within our models of EoS.
The pressure isotherms as functions of the density in the region of
the first-order LG phase transition are demonstrated in fig. \ref{Pn} for the  models KVORcut03 (upper panel) and for MKVOR* (lower panel).
The Gibbs conditions determining equilibrium of two phases (I-liquid  and II-gas)  are as follows:
\begin{gather*}
P(n^{\rm I},T)=P(n^{\rm II},T)\,, \quad \mu_B (n^{\rm I},T) =\mu_B (n^{\rm II},T)\,,
\end{gather*}
where $\mu_B =\mu_n =\mu_p$, $n=n_n + n_p$, $Y_p =1/2$ and thereby $n=2n_n$.
The phase separation boundary in fig. \ref{Pn}
is denoted by a solid line. This line connects the endpoints of the isothermal Maxwell constructions.
The isothermal spinodal (ITS) line, which delimits the hydrodynamically
unstable region, is denoted by a dashed line.  On the Maxwell construction (shown for an example by a horizontal dashed line for $T=10$~MeV) the total baryon density is composed from the densities of the liquid and gas phases as
\begin{gather}
n=\chi n^{\rm I} +(1-\chi) n^{\rm II}\,,
\end{gather}
where the liquid phase fraction $\chi$ changes along the Maxwell
construction horizontal line from $0$ at  $n =n^{\rm II}$ up to 1 for
$n =n^{\rm I}$.
The location of the critical point, which satisfies the conditions $(\partial P/ \partial n)_{T}=0$,
$(\partial^2 P/ \partial n^2)_{T}=0$, is described by the critical values of
$T=T_{cr}$, $n=n_{cr}$, $s={s}_{cr}$. For KVORcut03 model $T_{cr} = 17.4\,\mev$, $n_{cr}= 0.054\,{\rm fm}^{-3}$, ${s}_{cr}=2.74$, $P_{\rm cr} = 0.30 \, \mev/{\rm fm}^3$,
and  for MKVOR* model  $T_{cr} = 16.05 \, \mev$, $n_{cr}=0.051 \, {\rm fm}^{-3}$, ${s}_{cr}=2.68$, $P_{\rm cr} = 0.25 \, \mev/{\rm fm}^3$. These values lie within the $T_{\rm cr} = (16.4 \pm 2.3)\,\mev$ band given by microscopic calculations based on chiral nucleon-nucleon potentials \cite{Carbone:2018kji}. The shaded region denotes the experimental bound taken from \cite{Elliott:2013pna}:
$n_{\rm cr} = (0.06 \pm 0.02) \, {\rm fm^{-3}}$, and $P_{\rm cr} = (0.3 \pm 0.1) \, \mev/{\rm fm}^3$,   $T_{\rm cr} = (17.9 \pm 0.4)\,\mev$.  Both the KVORcut03 and MKVOR* models pass the  constraints for critical density and pressure and KVORcut03 model passes marginally the constraint for critical temperature from \cite{Elliott:2013pna}.  The  critical temperature calculated in the MKVOR* model is lower than the experimental bound extracted in  \cite{Elliott:2013pna}.  Note that the  traditional RMF models do not fulfill the existing experimental constraints, \textit{cf.} fig. 2  in \cite{Dutra:2018amp}.  Two  dotted lines demonstrate trajectories of the system characterized by the values of the constant specific entropy and are labeled by the values of $s$. The boundary of the hydrodynamically
unstable region, as it occurs at constant value of the specific entropy -- adiabatic  spinodal (AS), is shown by a dash-dotted line.

\begin{figure}
	\centering
	\includegraphics[height=6.5cm,clip=true]{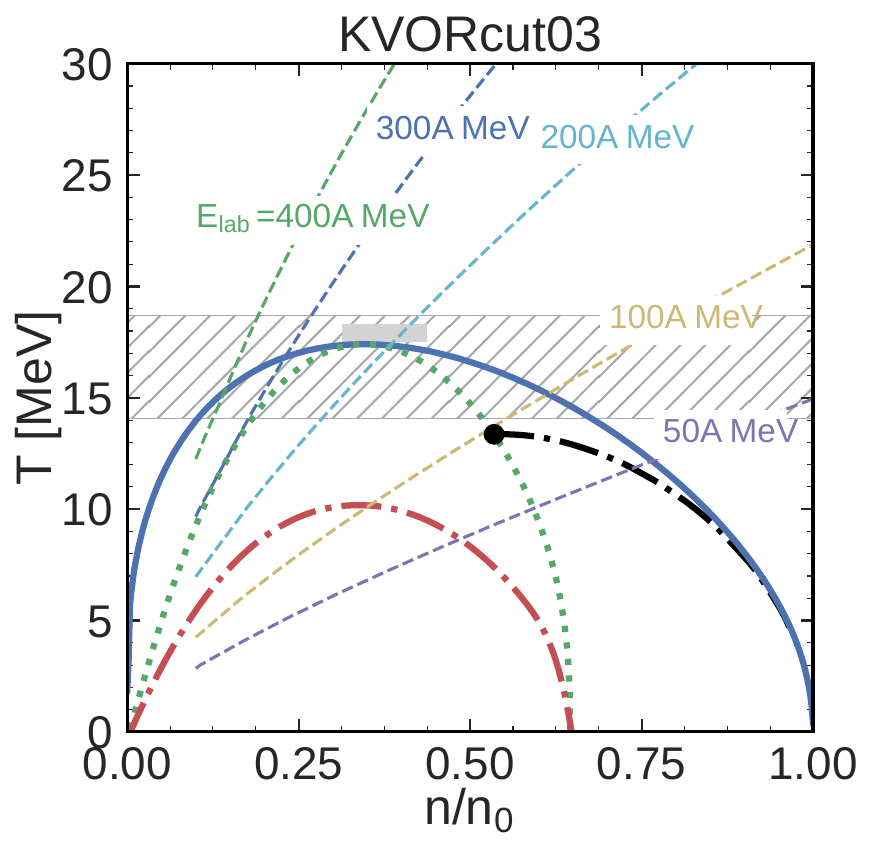}
	\includegraphics[height=6.5cm,clip=true]{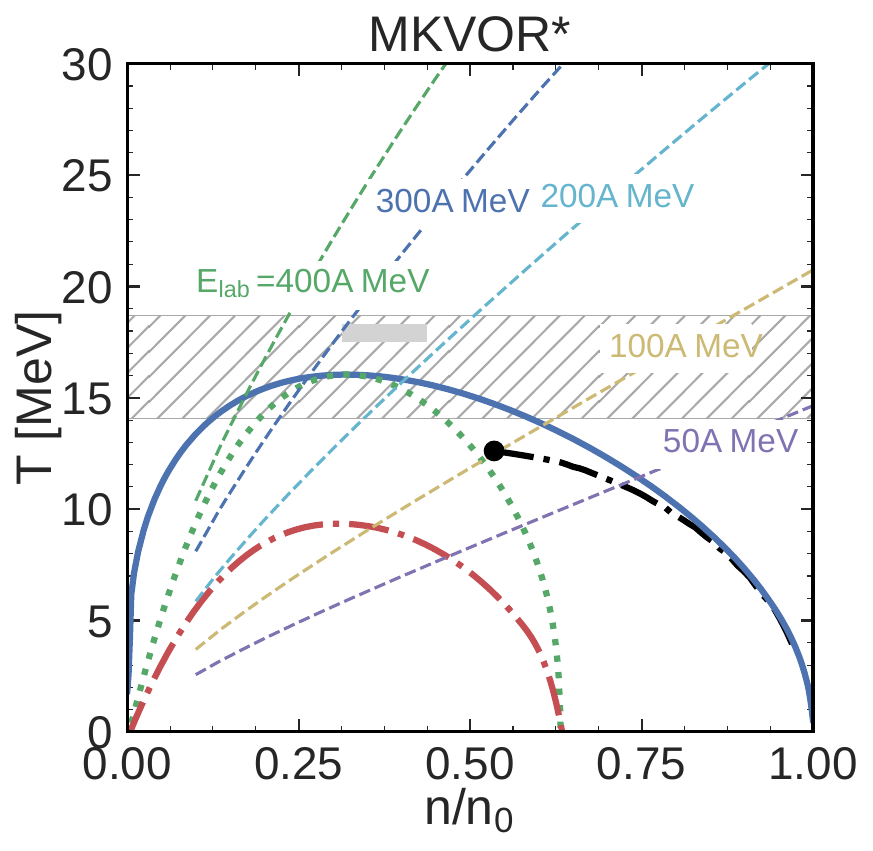}
	\caption{ Phase diagram on $T-n$ plane for ISM. Upper panel: KVORcut03 EoS, lower panel: MKVOR* model. Solid and dotted  curves
		demonstrate boundaries of the Maxwell construction and the ITS  regions,
		respectively. The dash-dotted curves indicate  boundaries of  the AS  region. Dashed lines labeled by values of the collision energy in the laboratory frame show adiabatic trajectories of the system evolution computed
		in the RMF approximation. Dash-dotted  lines ended with full dots show values of the temperature  $T_{\rm comp}(n_{\rm comp})$, at
		which the pressure  reaches zero
		at the maximum  $n$  among three roots on each isotherm. After the system has reached this line, under a subsequent slow decrease of the temperature it  may come to  the stable state with $n=n_0$ by radiating particles.
		Shaded rectangle denotes the results of an experimental   analysis of reactions with formation of a compound nucleus  \cite{Elliott:2013pna} and the hatched band denotes result of calculations using  chiral forces \cite{Carbone:2018kji}.
	}
	\label{Tn}
\end{figure}

In fig. \ref{Tn} we show the phase diagram on $T(n)$ plane for ISM within the KVORcut03 (upper panel) and MKVOR* (lower panel) models.
By solid and dotted  curves we show boundaries of the Maxwell construction and the ITS  regions, respectively.
The dash-dotted curves indicate  boundaries of  the AS  region. Dashed lines labeled by values of the laboratory collision energy  show adiabatic trajectories of the system evolution computed within the approach  described in the previous section. Dash-dotted  lines ending with full dots show values of the temperature  $T_{\rm comp}(n_{\rm comp})$, at
which the pressure  reaches zero
at the maximum  $n$  among three roots on each isotherm.
For $T<T_{\rm comp}$ a compound nucleus can be formed, which
by radiation of nucleons may reach the ground state. Shaded rectangle denotes the constraint from the experimental  analysis of reactions with formation of a compound thermal nucleus, which emits neutrons, protons and heavier charged
fragments \cite{Elliott:2013pna}, and the hatched band denotes the result of  calculations performed with chiral forces \cite{Carbone:2018kji}.

After the ${s} \simeq \const$ trajectory of the expanding system crosses the border of the Maxwell construction,
the system comes into a metastable region. One may expect that the typical  fireball
expansion time is  less than the typical time for which in the system,
being in the supercooled gas (overcooled liquid) state, an overcritical liquid drop (gaseous bubble) appears.
At such assumptions we
can still apply the thermodynamical description
until the system trajectory did not reach  the ITS line. Beyond the ITS line the system enters the hydrodynamically unstable region.
The $s\simeq \const$ system trajectory crosses the ITS line already for the collision energies $ E_{\rm lab}\lsim 250-300$ MeV within our models, however the system breakup occurs at higher temperatures than $T_{cr}$, as we have seen from figs. \ref{fig::EScut03}, \ref{fig::ESmkv}.  Thereby, within our models the system breakup  occurs in the region of the spinodal instability only for $E_{\rm lab}\lsim 150$ MeV. For $E_{\rm lab}\lsim 100$ MeV the system trajectory passes the region of $T-n$, where the pressure reaches zero and a compound nucleus can be formed.
Experimentally such a region can be attained not only in heavy-ion collisions but also in collisions of light nuclei and particles with a heavier nucleus, \textit{cf.} \cite{Elliott:2013pna}.

The isothermal and adiabatic sound velocities,
$u_{T}$ (at ${T} = \const$) and $u_{{s}}$ (at ${s} =\const$), satisfy the
thermodynamic relation, \textit{cf.} \cite{Skokov:2010dd},
\begin{equation}
u_{{s}}^2  =u_T^2  + \frac{T}{nm_N c_V}   \left[ \left(\frac{\partial P}{\partial
	T}\right)_n \right]^2 . \label{UC}
\end{equation}
Here
$u_{{s}}^2 =(\partial P/\partial E)_{{s}}$ and
$u_T^2=(\partial P/\partial E)_T$,
$c_V$ is the specific heat  at  fixed volume $V$.   The  quantity $u_{{s}}$
characterizes
propagation of sound waves in ideal hydrodynamics. In non-ideal
hydrodynamics with finite values of the shear and bulk viscosities,
$\eta$ and $\zeta$, and thermal conductivity, $\kappa$,
the  propagation of sound waves is defined by the interplay between  $u_T$
and $u_{{s}}$, \textit{cf.} \cite{Skokov:2010dd}.
Condition $u_T =0$  defines  the ITS line  on the  $T-n$ plane and
$u_{{s}}=0$ determines the AS line.
The maximum temperature points on these lines are the critical
temperature $T_{cr}$ (for the $P(n)$ isotherms $T=\const$), and the adiabatic maximum
temperature $T_{P,max}$ (for $P(n)$ at ${s}=\const$).

If the evolution is studied within the ideal hydrodynamics (for ${s}=\const$),  then the instability occurs at the crossing of the AS line,
where $u_{{s}} =0$. Note that $u_T^2 < 0$ at the point $u_{{s}} =0$, whereas $u_T^2$ reaches zero first, when the system trajectory crosses the ITS line.  Note that  \cite{BS,Panagiotou:1984rb} and a number of  subsequent
works exploiting ideal hydrodynamics considered the crossing of the AS line by the  trajectories
${s}$=const  as the starting point for
the clustering.
In contrast,  authors ~\cite{Schulz:1983pz} and later
~\cite{FRS,FRS1,PR,LL,Skokov:2009yu,Skokov:2010dd} argued
that clustering in realistic systems, described by non-ideal hydrodynamics,  may appear already at higher temperatures after the system trajectory crosses the ITS line.

\subsubsection{Effects of fluctuations in ISM}
Throughout this work we use the RMF framework. However, an inclusion of
fluctuations \cite{Skokov:2009yu,Skokov:2010dd} results in the  divergence of the specific heat
at $T=T_{cr}$. Thereby it can be
concluded from eq. (\ref{UC}), that with taking fluctuations into consideration
both temperatures $T_{cr}$
and $T_{P,max}$   should coincide, \textit{cf.} \cite{FRS,FRS1}.

The  heat capacity $C_V$ is related to the variance of the energy as
$$T^2 C_V=\overline{E^2}-\bar{E}^2$$ within the grand canonical description.
Papers ~\cite{Stephanov:1999zu} suggested that, if at some incident energy the trajectory passes in the vicinity of the critical point, the system may linger longer in this region due to strong thermodynamical fluctuations resulting in the divergence of the susceptibilities. For example, one could expect that in heavy-ion reactions the vicinity of the critical point could manifest via abnormal fluctuations of the energy. However ref. \cite{Skokov:2009yu} argued  that fluctuation effects in the vicinity of
the critical point  can hardly be pronounced in actual heavy-ion collisions, since the typical time for the development of critical fluctuations diverges in the critical point. Thereby, all relevant processes prove to be frozen near the critical point, whereas the system passes this
region in a rather short time.  Concluding this discussion, the time which the system spends in a vicinity of the critical point in the course of the heavy ion collisions might be insufficient for the formation
of critical  fluctuations.

Now we  argue that the ITS line can be manifested in abnormal fluctuations
of the conserved charges, such as the baryon number.
In equilibrium one-component systems having the mean density $n=
\overline{\hat n}({\bf r})
$,
the static structure factor describing fluctuations of the conserved particle number
can be written in terms of the density derivative of the chemical potential at fixed
temperature \cite{LL1980,Roepke:2017bad},
\begin{eqnarray}
S({\bf q}\to 0)=(
\overline{\hat
	n^2}({\bf r})
-\overline{\hat n}^2)V=T\left\{\left(\frac{\partial \mu_B}{\partial n}\right)_T\right\}^{-1}\,.
\label{EqStstr0}
\end{eqnarray}
Thereby the normalized variance of the nucleon number $N$ for the ISM
(in absence of other baryons except nucleons) is given by,
\begin{align}
{w}=
\frac{\overline{(\Delta N)^2}}{N}= \frac{T}{n}\left(\frac{\partial n}
{\partial \mu_B}\right)_{T}\,.
\label{var-in}
\end{align}

Since $\left(\frac{\partial P}{\partial n}\right)_{T}=
n\left(\frac{\partial \mu_b}{\partial n}\right)_{T} =0$  on the whole ITS line $T(n)$,
the static structure factor and the variance of the baryon number diverge on this line.
Thus fluctuations of the baryon number  grow when the system trajectory comes close to
the ITS line, which can be seen in corresponding event-by-event observables.
\begin{figure}
	\centering
	\includegraphics[height=6.5cm,clip=true]{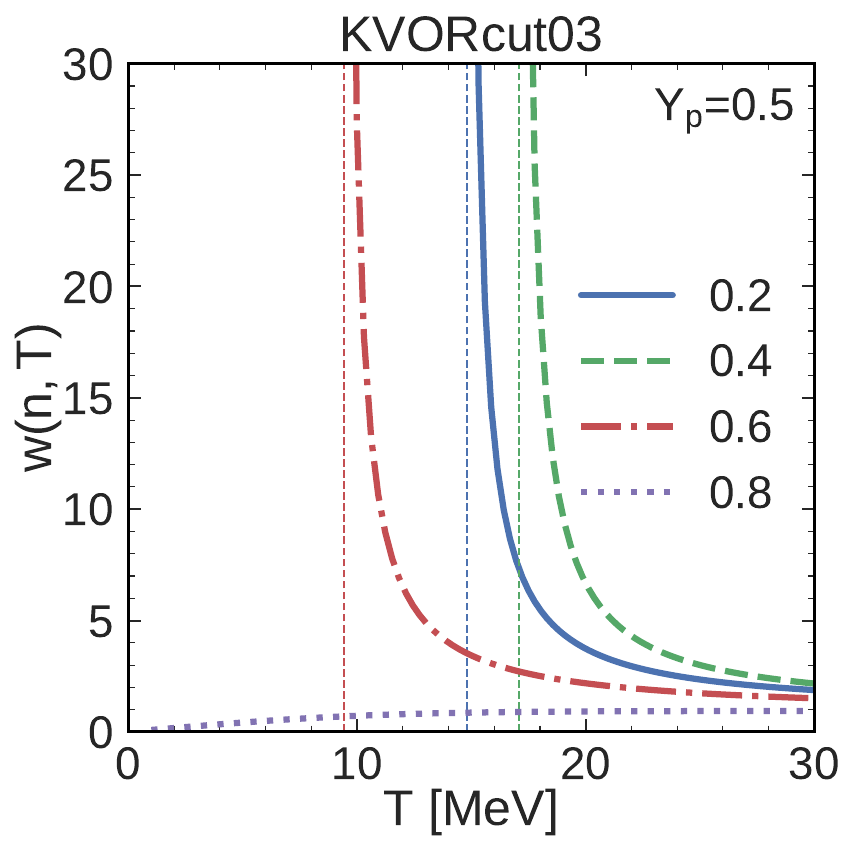}
	\includegraphics[height=6.5cm,clip=true]{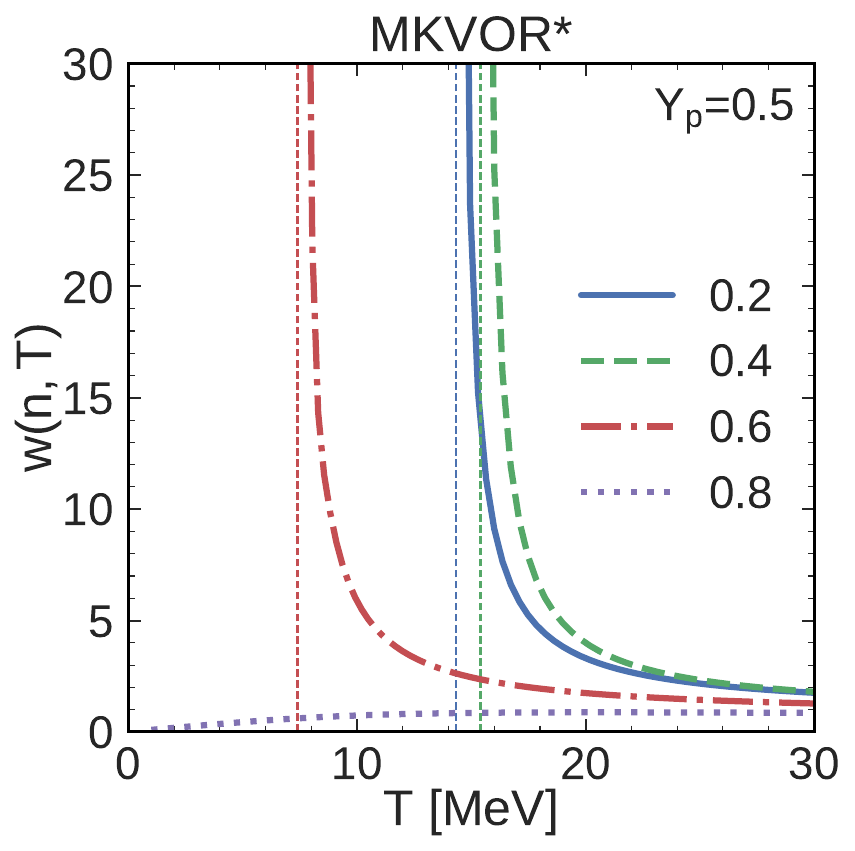}
	\caption{ Normalized variance $w$ as a function of temperature for the ISM in models KVORcut03 (on upper panel) and MKVOR* (on lower panel)
		for several values of $n$ (shown in legend in units of $n_0$) in the region of the LG first-order phase transition. }
	\label{Varience}
\end{figure}

In fig. \ref{Varience}  for  KVORcut03 model (upper panel) and MKVOR* model (lower panel)  we show the dependence of $w(T)$ in ISM
that we computed numerically
for several values of $n$ in the region of the LG first-order phase transition.
As we see, the normalized variance diverges on the ITS line, {\em cf.}  fig. \ref{Tn}.  The value  $n=0.8 n_0$ is outside the ITS region, thereby $w$ for $n=0.8n_0$ shown by the dotted line remains a smooth function of $T$.

\subsection{Liquid-gas  phase transition in IAM}

Recently two reactions $^{124}$Xe $+^{112}$Sn and $^{136}$Xe $+^{112}$Sn have been experimentally studied
at 32$A$~MeV and 45$A$~MeV  bombarding energies to produce quasi-fusion hot
nuclei, which undergo multifragmentation \cite{Borderie:2018fsi}. Using charge correlations the fossil signature of spinodal instabilities, \textit{i.e.} the abnormal presence of equal-sized fragments was  established at a confidence level of around 6-7 sigma  for both reactions at 32$A$~MeV incident energy.

In low-energy heavy-ion collisions, the baryon and electric charges obey two independent conservation laws, implying that the proton fraction $Y_p\simeq Y_Z$ is conserved. Therefore for the description of the LG first-order phase transition in the  IAM formed in heavy-ion collisions  we introduce two chemical
potentials of the baryon charge $\mu_B =\mu_n$ and the electric charge $\mu_Q =\mu_n -\mu_p$.   First, following the analysis performed  in the literature in different models, consider an occurrence of the first-order LG phase transition at assumption that there is no  surface tension on the spatial border of phases, \textit{cf.} \cite{Ducoin:2005aa,Alam:2017krb,Glendenning:2001pe,Poberezhnyuk:2018mwt}.

\begin{figure}
	\centering
	\includegraphics[height=.7\linewidth]{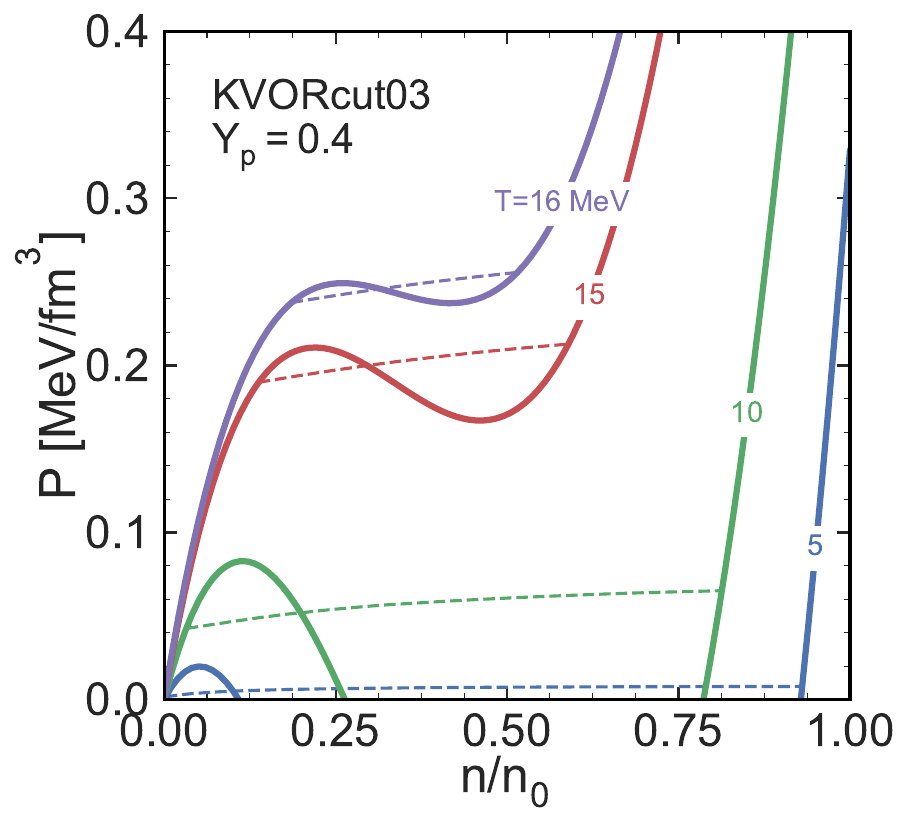}
	\includegraphics[height=.7\linewidth]{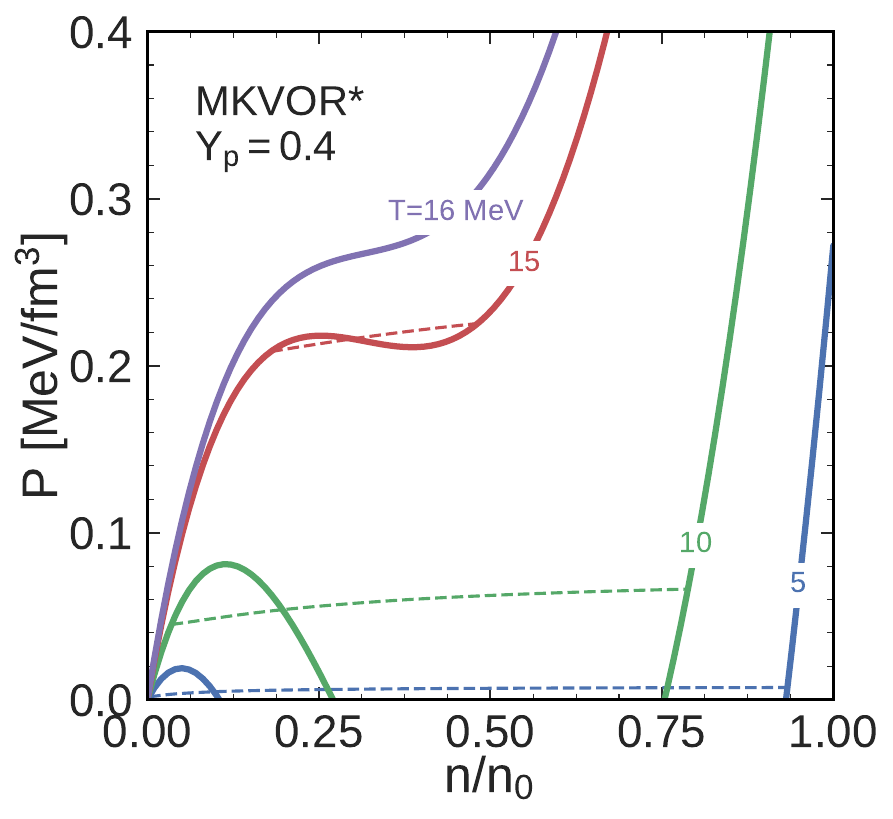}
	\caption{ The pressure-density isotherms for IAM at $Y_p =0.4$.
		Numbers near the curves indicate values of the temperature
		for the corresponding isotherms. Equilibrium states within mixed phase are connected by dashed lines.
		Upper panel: KVORcut03 EoS, lower panel: MKVOR* EoS.
	}
	\label{PnYp}
\end{figure}

As in fig. \ref{Pn}, in fig. \ref{PnYp}  we show the $P(n)$ isotherms, but now for $Y_p =0.4$ within KVORcut03 EoS (upper panel) and MKVOR* EoS (lower panel). $P(n)$ isotherms are shown by solid lines. Numbers near the curves indicate values of the temperature in MeV. Contrary to the $Y_Z=0.5$ case, the equilibrium  states within mixed phase are  not described by horizontal Maxwell lines, but  by the curves shown in fig. \ref{PnYp} by dashed lines. This is a general feature of the description of the  first-order phase transitions in multi-component systems \cite{Glendenning:2001pe}.  Transitions in systems with more than one conserved charge are offen  called Gibbs phase transitions or  non-congruent phase transitions \cite{Iosilevskiy:2010qr,Hempel:2013tfa}.

\begin{figure}
	\centering
	\includegraphics[height=.8\linewidth]{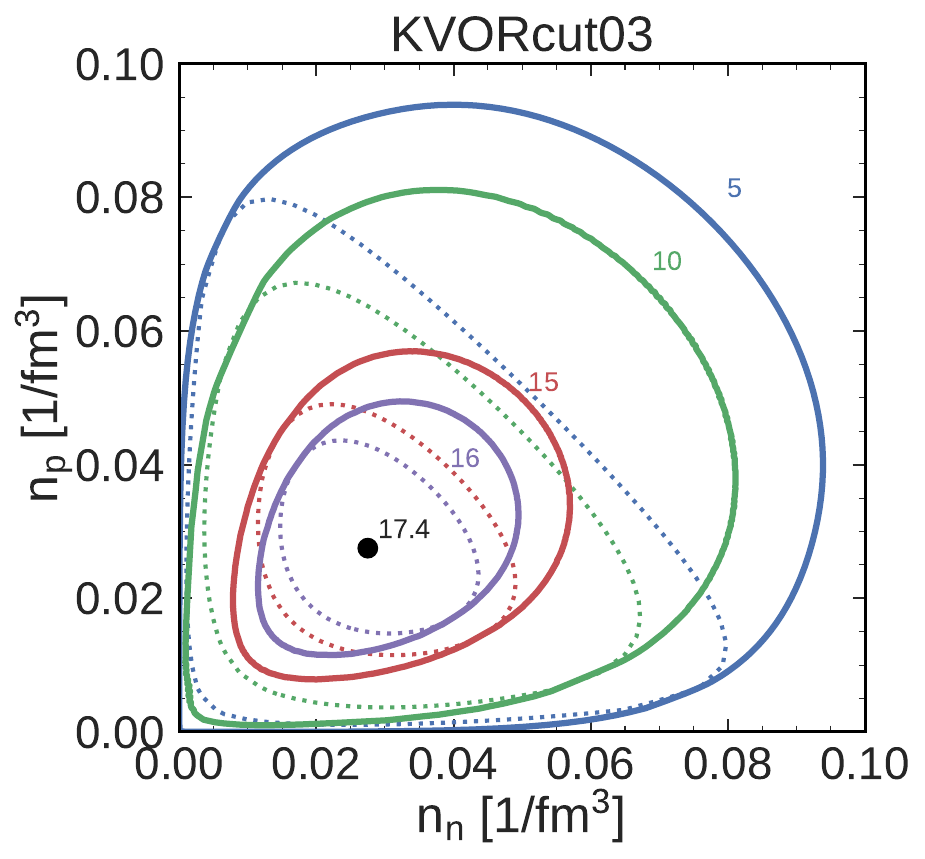}
	\includegraphics[height=.8\linewidth]{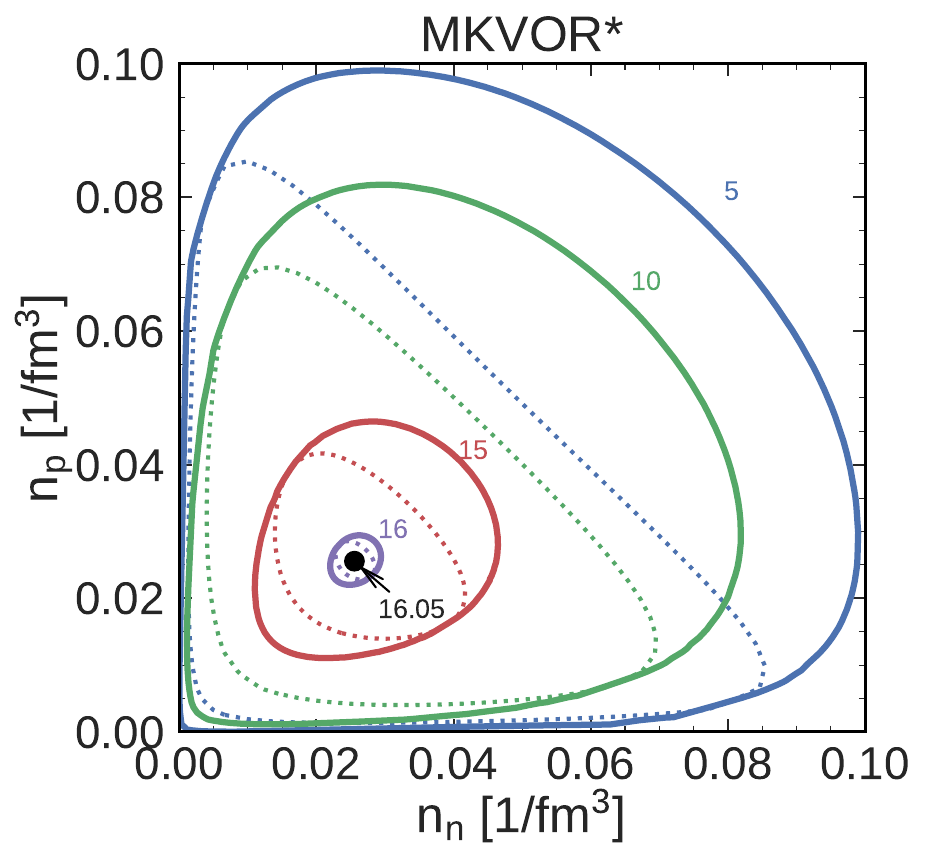}
	\caption{The coexistence
		border (solid lines) and the isothermal spinodal region border (dashed lines) in the proton-neutron density plane,  for KVORcut03 (upper panel) and MKVOR* (lower panel) models. Bold dots show critical points. Numbers are values of $T$ in MeV.}
	\label{npnn}
\end{figure}

In fig. \ref{npnn} by solid lines we show the phase coexistence
borders in the proton-neutron density plane for KVORcut03 (upper panel) and MKVOR*  (lower panel) models. Bold dots show critical points. Numbers near the curves are the corresponding values of $T$ in MeV. Dashed lines denote borders of the isothermal spinodal instability regions, defined as the line where the matrix
\begin{gather}
{\cal C}_{ij} = \Big(\frac{\pt \mu_i}{\pt n_j}\Big)_T, \quad i,j = \{n, p\}
\label{eq::C}
\end{gather}
becomes singular. This corresponds to the lowest eigenvalue becoming negative, which is the known condition for the presence of the instability \cite{Chomaz:2003dz}. We see that the allowed region of the LG phase transition on the proton-neutron density plane is a bit broader in MKVOR* model than in KVORcut03 model, whereas the critical temperature is a bit higher in the latter model.

On  the pressure $P(n)$ isotherm for $T<T_{cr}$ the mixed phase
begins at the point $G_{\rm eq}$
and ends at the point $L_{\rm eq}$. The point $G_{ \rm eq}$
corresponds to the gaseous phase being in equilibrium with an infinitesimal fraction of the liquid phase $L_{0}$, which has a higher density at the same pressure and both chemical potentials:
$$P_{\rm G_{\rm eq}}(\mu_B ,\mu_Q,T)=P_{\rm L0}(\mu_B ,\mu_Q,T).$$
The point $L_{ \rm eq}$
corresponds to the liquid  phase, being at equilibrium with the gaseous  phase $G_0$ of infinitely small fraction and a lower density at the same pressure and chemical potentials:
$$P_{\rm L_{\rm eq}}(\mu_B ,\mu_Q,T)=P_{\rm G0}(\mu_B ,\mu_Q,T).$$  At the points ${\rm G_{\rm eq}}$ and ${\rm L_{\rm eq}}$ the values of the pressure, as well as of  both  chemical potentials, are different, \textit{cf.} similar discussion in \cite{Ducoin:2005aa}. The baryon and electric charge densities in the mixed phase are connected to those in each phase as
\begin{eqnarray}
n=(1-\chi)n^{\rm G} (\mu_B , \mu_Q,T)+\chi n^{\rm L} (\mu_B , \mu_Q,T),
\label{barch}
\end{eqnarray}
\begin{eqnarray}
n_Q=(1-\chi)n_Q^{\rm G} (\mu_B , \mu_Q,T)+\chi n_Q^{\rm L} (\mu_B , \mu_Q,T),
\label{cheq}
\end{eqnarray}
where $n^{\rm G}$, $n_Q^{\rm G}$ and $n^{\rm L}$, $n_Q^{\rm L}$
are the baryon and charge densities of the gaseous and the liquid fractions,
respectively, the liquid phase fraction is $0<\chi <1$, $Y_p \simeq n_Q/n$. For a fixed proton fraction $Y_p = \const$ the system in the phase coexistence region follows the critical line, that allows us to determine the fraction of the liquid phase in the coexistence region as
\begin{gather}
\chi = \frac{Y_p n^{\rm G} - n_Q^{\rm G}}{n_Q^{\rm L} - n_Q^{\rm G} - Y_p (n^{\rm L} - n^{\rm G})}.
\end{gather}

\begin{figure*}
	\centering
	\includegraphics[width=0.37\textwidth]{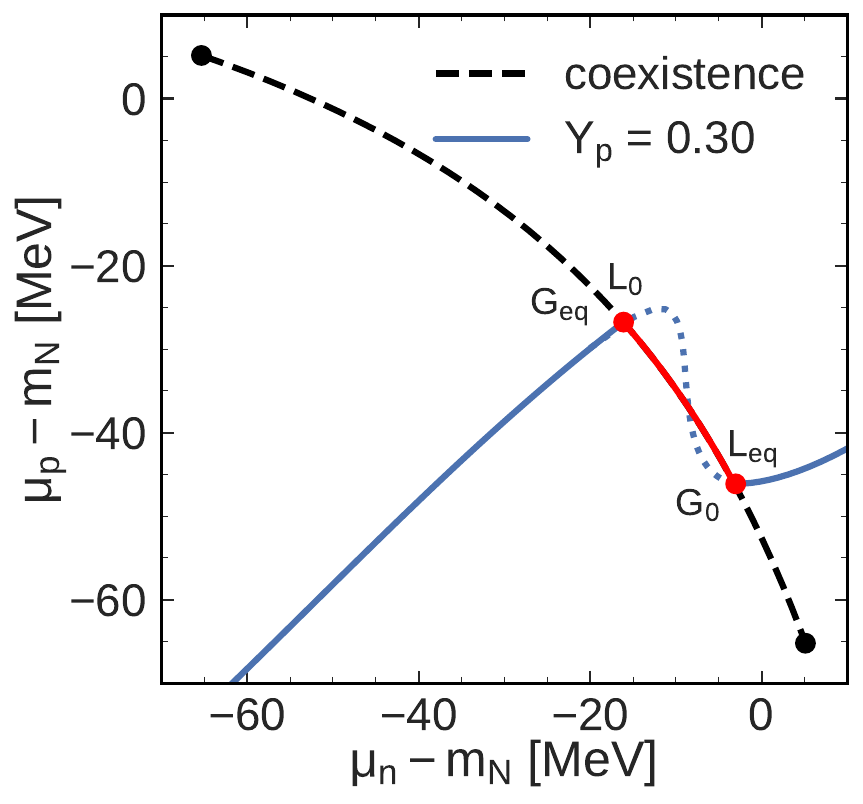}
	\includegraphics[width=0.4\textwidth]{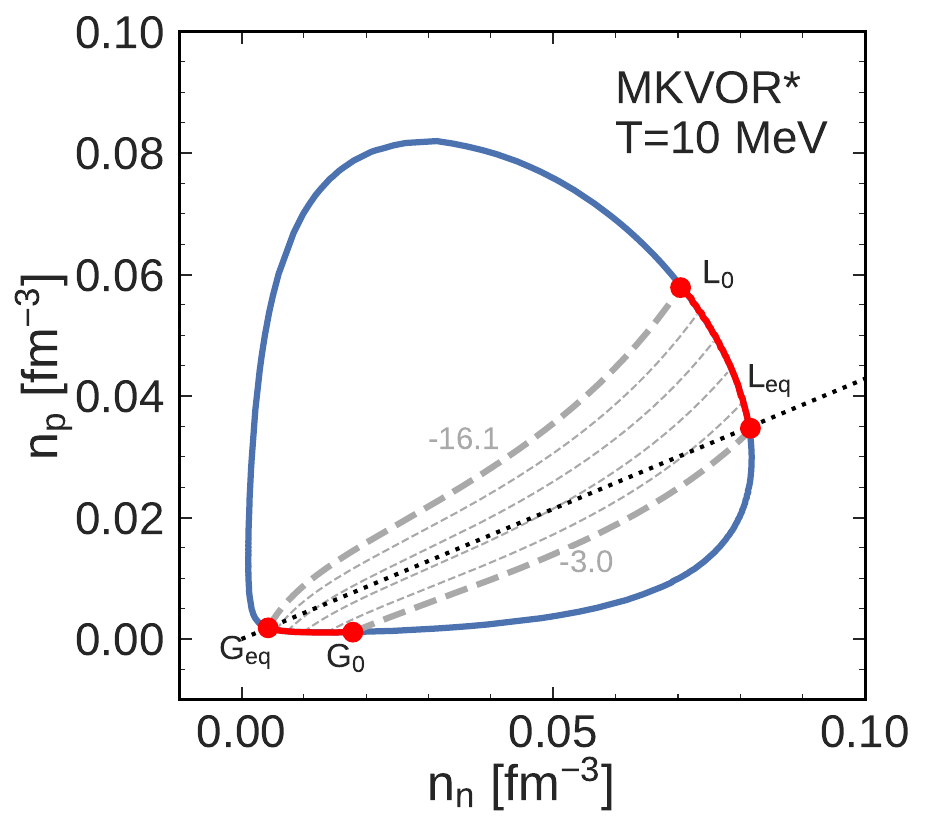}
	\caption{Example of the LG phase transition  for the MKVOR* model at $T = 10$~MeV. Left panel: $\mu_p - \mu_n$ plane. Full solid  line with the dotted segment is $\mu_p - \mu_n$  isotherm for $T=10$ MeV calculated at fixed $Y_p=0.3$. Solid line connects equilibrium states $G_{\rm eq},L_0$ and $L_{\rm eq},G_0$ within mixed phase at $Y_Z =0.3$. Dashed curve shows the border of the coexistence region. Right panel: $n_p - n_n$ plane, $T = 10$~MeV. Solid line denotes the border of the coexistence region, and within this region the dashed lines are shown for fixed $\mu_n$. The thicker dashed lines correspond to the minimum and maximum values of $\mu_n - m_N$ indicated in MeV.  Paths $G_{\rm eq}-G_0$ and $L_{\rm eq}-L_0$ refer to a transformation at $Y_Z =0.3$. Dotted line connects equilibrium states within mixed phase at $Y_Z =0.3$.}
	\label{fig::mumu_nn}
\end{figure*}

In fig. \ref{fig::mumu_nn} we demonstrate the description of the LG phase transition in IAM in terms of $\mu_p, \mu_n$ (left) and $n_p, n_n$ for comparison (right). As an example, we present the results of  calculations performed within the MKVOR* model at $T= 10$~MeV and $Y_p = 0.3$ for  easier comparison  with the result of calculation  \cite{Ducoin:2005aa} presented there in fig. 8 for the same values of $T$ and $Y_p$.
Inside the coexistence region, the system with a
given proton fraction $Y_p=0.3$ is decomposed into two phases,
located at the intersections of the coexistence
curve with the corresponding isotherm (left panel) and with constant $\mu_n -m_N$ curve (right panel). The constant $Y_p$ transformation does not exhibit a transition from liquid to gas at a single value of $\mu_n$. The intensive parameters change smoothly as the system is driven along the coexistence line with an increasing density. We can see on the right panel in fig. \ref{fig::mumu_nn} that the liquid phase of a neutron-rich matter is closer to ISM than the gaseous phase, so our models exhibit the known isospin-distillation phenomenon \cite{Muller:1995ji, Colonna:2002ti, Ducoin:2005aa}.  When $Y_p$
is kept constant, the system is forced to follow the first-order phase-transition line
\cite{Ducoin:2005aa}. When
the system reaches the coexistence border (point $G_{\rm eq}$), a liquid phase of infinitesimal fraction appears in point $L_0$
at the same
values of $\mu_n$, $\mu_p$ and $T$. The gaseous phase goes along coexistence line from the point $G_{\rm eq}$ to point $G_0$, as it is indicated on right panel in fig. \ref{fig::mumu_nn},
while the dense phase goes on
the other side of the coexistence border from $L_{\rm eq}$ to $L_0$. When the system reaches the state $L_{\rm eq}$
the gas is entirely transformed
into a liquid, the  phase transition from gas to liquid is over and the $Y_Z=\const$ transformation corresponds
to a homogeneous system again.

\subsubsection{Effects of fluctuations in IAM}

For the IAM  the static structure factor can be presented
as \cite{Burrows:1998cg,Roepke:2017bad}
\begin{equation}
S({\bf q}\to 0)=S_{nn}({\bf q}\to 0)+S_{pp}({\bf q}\to 0)+2S_{np}({\bf q}\to 0)\,,
\end{equation}
provided that $S_{np}=S_{pn}$. Partial normalized variances are
\begin{equation}
w_{\tau,\tau'}=\frac{S_{\tau,\tau'}({\bf q}\to 0)}{n}=\frac{T}{n}\left(\frac{\partial n_\tau}{\partial \mu_{\tau'}}\right)_{T} = \frac{T}{n} C_{\tau \tau'}^{-1},
\label{Stf}
\end{equation}
where $n_{\tau}=\frac{\partial P[\mu_\tau , \mu_{\tau'},T]}{\partial \mu_\tau}$ and the matrix $C_{\tau\tau'}$ is defined by \eqref{eq::C}. These expressions for $w_{\tau \tau'}$ become manifestly equivalent if all $\partial n_\tau / \partial \mu_{\tau'}$ are evaluated with all other $\mu_i, \, i \neq\tau'$ being fixed, and in turn $\partial \mu_\tau/\pt n_{\tau'}$ are evaluated with all other $n_i, i\neq\tau'$ being held constant.
Similarly we may introduce normalized variances of the baryon number
\begin{equation}
w_B=w_{pp}+w_{nn}+2w_{np}=\frac{\overline{(n_n+n_p)^2} -n^2}{n}\,,\nonumber
\end{equation}
and the charge $$\quad w_Q = w_{pp}= \frac{\overline{n_p^2} -n_p^2}{n}\,.$$

\begin{figure}
	\centering
	\includegraphics[height=6.8cm,clip=true]{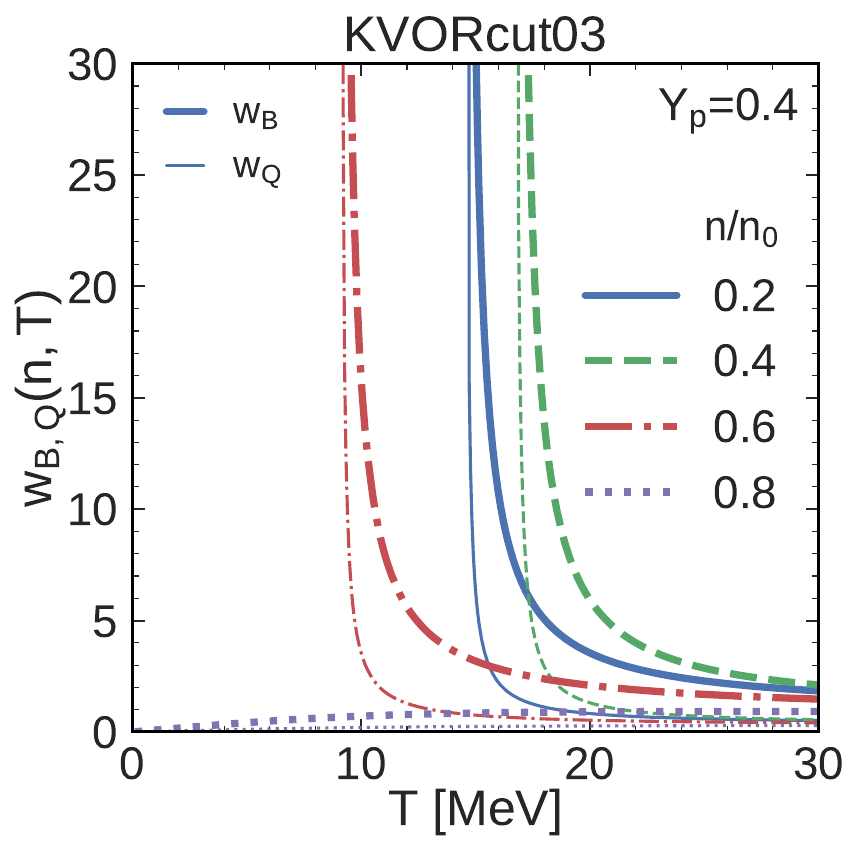}
	\includegraphics[height=6.8cm,clip=true]{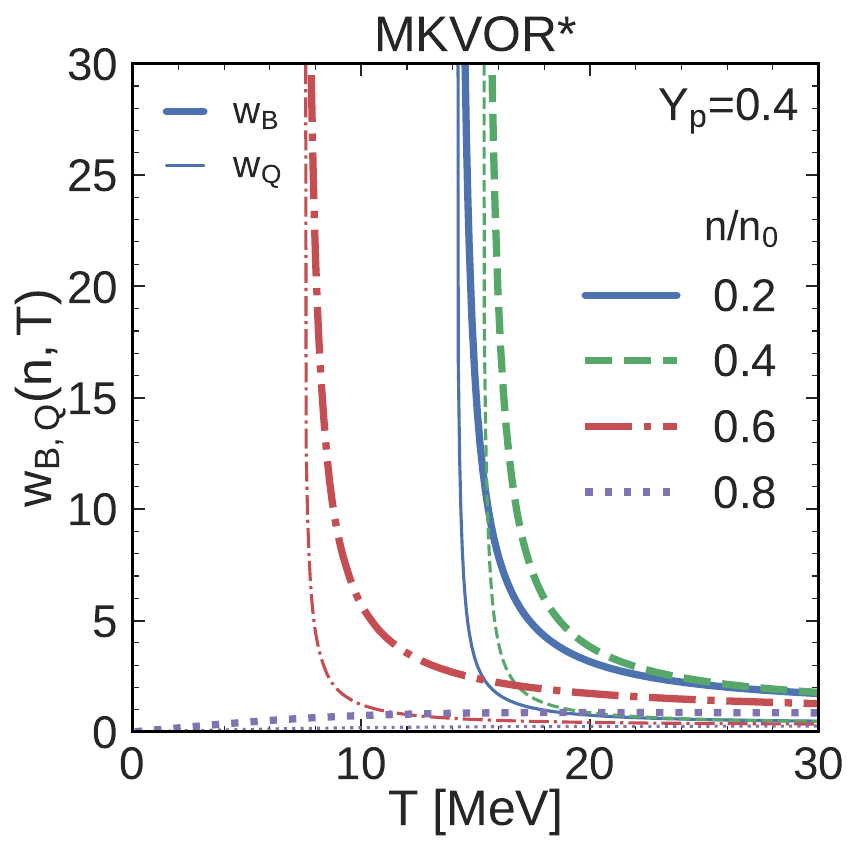}
	\caption{The dependence of $w_{B, Q}(T)$ for $Y_p =0.4$ in KVORcut03 (upper panel) and MKVOR* (lower panel) models for several values $n$ in the region of the LG first-order phase transition. Bold lines stand for $w_B$ and thin lines denote $w_Q$.}
	\label{VarienceYp}
\end{figure}

In fig. \ref{VarienceYp} we show our results for $w_{B}$ (bold lines) and $w_Q$ (thin lines) as functions of the temperature for several values of the baryon density $n$ within the PT region. We see that both $w_B$ and $w_Q$  are divergent at the border of the isothermal spinodal instability region, because the matrix $C_{ij}$ becomes singular there.

\subsection{Features of pasta phase in systems formed in heavy-ion
	collisions}
The conservation of the electric charge should be considered not locally, but globally \cite{Glendenning:2001pe}.
In the previous subsection following the standard description of the LG phase transition in IAM in finite-size systems in absence of the charge neutrality, \textit{cf.} \cite{Ducoin:2005aa},  effects of a non-zero surface
tension $\sigma$ on the LG phase boundary were disregarded. Such a
description assumes that $\sigma =0$. In reality $\sigma (n\sim
n_0,T=0)\sim 1$~MeV$/$fm$^2$, \textit{cf.} \cite{Baym:1971ax}. In presence of
$\sigma \neq 0$ the uniform BEM can be structured
\cite{Voskresensky:2001jq}.
We will show  that
taking into account the surface tension between liquid and gas
structures significantly modifies the physical picture of the LG
transition phenomenon in isospin-asymmetric systems of finite size
(here, in absence of the charge neutrality of the system).

Typical radius of the  fireball is $$R_{\rm f}(t)\simeq \frac{1}{m_\pi}
\left(\frac{n_0}{n(t)}\right)^{1/3}
A_{\rm part.}^{1/3}$$  for a spherical expansion, and thereby $R_{\rm
	f}(t)\lsim  (15-30)$ fm
for central collisions of the heaviest atomic nuclei, and for $n\sim
(0.1 -0.5) n_0$.
We will for simplicity assume that before  the
first-order LG phase transition occurs the fireball is  homogeneously
charged with the proton charged density $n_Q $.
In order for the pasta phase to be formed it should be at least
$R_{\rm WZ} <R_{\rm f}$, where $R_{\rm WZ}$ is the radius of the
Wigner-Seitz cell
for  the given $d$ geometry, $d=3$ for droplets/bubbles, $d=2$ for rods
and $d=1$   for slabs.
The  droplet fills the sphere  $r_3 =\sqrt{x^2+y^2+z^2}<R$, the rod
fills space of the cylinder
$r_2=\sqrt{x^2+y^2}<R$,
where $R$ is now the transversal radius of the rod, and the slab fills
the layer $r_1=|x|<R$. We assume $R<R_{\rm WZ}<R_{\rm f}$.
Moreover, we will imply that the fireball expansion is  so
slow that there is enough time to prepare the pasta structures,  which
we further on  consider
in the static approximation. Thus, simplifying  consideration we assume
that all mentioned  conditions are fulfilled although in actual
heavy-ion collisions it might be not the case.

After the pasta phase has appeared,
the charge density is redistributed as $$n_{Q} = n_{Q} +\delta n_{Q}\,,$$
$n_Q =nY_p$, $\delta n_{Q}=\delta n_{Q}^{\rm I}$ for $r_d <R$, and
$\delta n_{Q}=\delta n_{Q}^{\rm II}$ for $R<r_d <R_{\rm WZ}$, where we
for specificity assumed that the interior of a structure is in the liquid phase
and the exterior is in the gaseous phase. Thereby we start with consideration of the
denser phase as the minor phase (for  $0<\chi\lsim 0.5$).

From eq. (\ref{cheq}) we find the relation
\begin{equation}
\chi \delta n_{Q}^{\rm I}+ (1-\chi) \delta n_{Q}^{\rm II}=0\,,\quad \chi
=(R/R_{\rm WZ})^d\,,
\label{elneutr}
\end{equation}
that is similar to the global charge neutrality condition in the pasta
phase realized in BEM of neutron stars.

Then we present the electric potential well as $V=V_0 +\delta V$, with
$V_0$ and $\delta V$ obeying
the Poisson equations
\begin{equation}
\Delta V_0=4\pi e^2 n_{Q}\,,\quad \Delta \delta V=4\pi e^2 \delta n_{Q}\,,
\label{elwell}
\end{equation}
$e>0$ is the proton charge. Simplifying, we assume  the quantities
$\delta n_Q^{\rm I, II}$ to be spatially uniform in each region. With the solution of  eq.
(\ref{elwell}) at the condition (\ref{elneutr})
we are able to find the electric field
contribution to
the free-energy density (for fixed $T$)
due to appearance of the Coulomb pasta
\cite{Pethick:1983eft,Heiselberg:1992dx}, \textit{cf.} also
\cite{Norsen:2000wb,Voskresensky:2002hu},
\begin{equation}
\delta F_{\rm C}^d =2\pi e^2 (\delta n_Q^{\rm I}-\delta n_Q^{\rm II})^2
R^2 \chi \Phi_d (\chi)\,,
\label{Coulpasta}
\end{equation}
where $$\Phi_d (\chi) =\frac{2(d-2)^{-1}(1-d\chi^{1-2/d}/2)+\chi}{d+2}\,,$$
and thereby $\Phi_3 (\chi) =\frac{2-3\chi^{1/3} +\chi}{5}$,
$\Phi_2(\chi) =\frac{\ln (1/\chi)+\chi - 1}{4}$, and
$\Phi_1(\chi) =\frac{(1-\chi)^2}{3\chi}$.

The surface free-energy density contribution is
\begin{equation}
\delta F_{\rm S}^d =\frac{\chi \sigma d}{R}\,.
\label{surf}
\end{equation}
For $T\neq 0$ the surface tension $\sigma$ is given by
\cite{Ravenhall:1984ss,Schulz:1983pz}, $\sigma =\sigma_0
\left(\frac{T_c^2 -T^2}{T_c^2 +T^2}\right)^{5/4}$, $\sigma_0 \sim
1$~MeV$/$fm$^2$.

The total (Coulomb plus surface) contribution to the free-energy density
due to Coulomb pasta effects $\delta F_{\rm C.pasta}$  is
given by $$\delta F_{\rm C.pasta}^d =\delta F_{\rm C}^d +\delta F_{\rm
	S}^d\,.$$
Now we are able to minimize
$\delta F_{\rm C.pasta}^d (R)$ in $R$. Thus using the condition (\ref{elneutr})
we find the optimal size of a structure with a
given geometry parameter $d$
and $\delta F_{\rm C.pasta}^d (R_m^d)$,
\begin{gather}
\delta F_{\rm C.pasta}^d (R_m^d)=\frac{3[2\pi e^2 (\delta n_Q^{\rm I})^2
	]^{1/3} (\sigma d)^{2/3}
	\chi \Phi_d^{1/3}}{2^{2/3} (1-\chi)^{2/3}}\,,\label{Rm} \\
R_m^d =\left[\frac{\sigma d (1-\chi)^2}{4\pi e^2 (\delta n_Q^{\rm I})^2
	\Phi_d}\right]^{1/3}\,.
\nonumber
\end{gather}
A simple perturbative way to take into account finite size effects is as
follows
\cite{Glendenning:2001pe,Voskresensky:2001jq,Voskresensky:2002hu}.
First for the given mean values $n$,  $n_Q =n Y_p$  at fixed
$T<T_{cr}$   following eqs. (\ref{barch}), (\ref{cheq}) we find values
$n^{\rm G}$, $n^{\rm G}_Q =n_Q+\delta n^{\rm II}_Q$,  $n^{\rm L}$,
$n^{\rm L}_Q =n_Q+\delta n^{\rm I}_Q$, $\chi$, without including finite
size contribution. Then,
implying that the correction to the free-energy density owing to  the
finite-size term $\delta F_{\rm C.pasta}^d (R_m^d)$
is small we add this  term not modifying $\chi$.

Comparing $\delta F_{\rm C.pasta}^d (R_m^d)$ for $d=1,2,3$ at a given
$\chi$ and
$\delta n_Q^{\rm I}$ we determine the energetically favorable geometry
of the structures. The following structures are energetically favorable:
\begin{gather*}
\begin{tabular}{lc}
droplets for &$0\,\,\, < \chi < 0.22$,\\
rods for &$0.22 < \chi < 0.35$,\\
slabs for &$0.35 < \chi < 0.5$.
\end{tabular}
\end{gather*}
The result for $\chi >0.5$ is obtained by the replacement $\chi \to
1-\chi$ and describes a bubble phase. Note
\cite{Voskresensky:2001jq,Voskresensky:2002hu} that the Coulomb limit is
actually valid only for $R_m^d \ll \lambda_{\rm D}^{\rm I,II}$, where
$1/\lambda_{\rm D,I}^2 =4\pi e^2 \frac{\partial n_Q^{\rm I}}{\partial
	\mu_Q}$, $1/\lambda_{\rm D,II}^2 =4\pi e^2 \frac{\partial n_Q^{\rm
		II}}{\partial \mu_Q}$ are taken for $V=0$. Otherwise a spatial
inhomogeneity  of the charge distribution in both liquid and gas phases
should be taken into account.
Complete numerical calculations can be performed similar to
those \cite{Maruyama:2005vb,Maruyama:2005tb} done  for BEM in neutron stars.

For $Y_p =1/2$ the condition (\ref{cheq})  coincides with (\ref{barch}).
In this case
$Y_p^{\rm I} =Y_p^{\rm II} =1/2$  and eq. (\ref{elneutr}) is fulfilled
for $\delta n_Q^{\rm I}=\delta n_Q^{\rm I}
=0$. Solutions with $\delta n_Q^{\rm I,II}\neq 0$ are energetically not
profitable due to the symmetry energy. For the given mean density
$n^{\rm II}<n<n^{\rm I}$ one finds the fraction
$\chi$, now $\chi =(R/R_{\rm f})^d$, $R_{\rm WZ}=R_{\rm f}$. Thus one
defines $R(\chi (n))$. Therefore
for $Y_p =1/2$ there is no Coulomb pasta. The situation is rather similar to
that occurring in the description of BEM in neutron stars  by  the
Maxwell construction. Then there is only one boundary
(following the Maxwell construction) that
separates the phase I from the phase II.
The minor phase
(I for $\chi <1/2$) occupies
the drop, the rod or the slab provided $\delta F_{\rm C pasta}^d (R(\chi
(n)))$
is the smallest among the drop, rod or slab geometries, respectively.

For $Y_p \neq 1/2$  for $R<R_{\rm WZ}<R_{\rm f}$ the pasta phase is
energetically preferable compared to the case of the only one boundary
(for $R_{\rm WZ}=R_{\rm f})$ since $\delta F_{\rm C pasta}^d(R_m^d)$ is
less than that for $R\neq R_m$
(at least provided $R\ll R_{\rm f}$). It would be interesting to look for
possible consequences of the formation of the pasta structures in actual
heavy-ion reactions.

\section{Conclusion}\label{Conclus}
In this paper we constructed the equation of state (EoS) of the nuclear matter
within the relativistic mean-field (RMF) framework with hadron masses and coupling constants
dependent on the mean scalar field. We considered a range of thermodynamic parameters relevant for description of heavy-ion collisions at the laboratory energies per baryon $ \Elab \lsim (1-2)A$~GeV, namely temperatures $T$ below  $m_{\pi}$,  densities $n\lsim 3n_0$, and the isospin asymmetry in the range  $0.4\leq Y_Z \leq 0.5$.
We included the $\Delta$ isobars as the most important baryon resonances for this range of temperatures and densities with their effective masses. In the meson sector we  included   the lightest triplet of pions as an ideal gas of particles with either the vacuum dispersion law or (for $Y_Z\neq 0.5$) with  the $s$-wave pion-nucleon interaction taken into account.

We used the fact, \textit{cf.} ~\cite{Maslov:cut}, that within an RMF model the EoS becomes stiffer for $n >n^* >n_0$, if a growth of the scalar field as a function of the
density is quenched and the nucleon effective  mass becomes weakly dependent on the density for $n >n^*$. In refs. ~\cite{Maslov:cut,Maslov:2015msa,Maslov:2015wba,Kolomeitsev:2016ptu} it was demonstrated how this cut-mechanism can be realized in the $\sigma$, $\om$, or $\rho$  sectors. In  KVORcut-based models the cut-mechanism is realized in $\om$ sector, in MKVOR-based models the cut-mechanism is realized in $\rho$ sector. Thereby the density dependence of the symmetry energy proves to be essentially different in KVORcut-based and MKVOR-based models, see fig. \ref{NuclEfMass}. These models allowed to pass multiple constraints from properties of cold nuclear matter, neutron star observations,  even with an inclusion of hyperons and $\Delta$ isobars,  and heavy-ion collision flow analysis. In the given paper we applied the KVORcut03-based and MKVOR*-based models \cite{Kolomeitsev:2016ptu} for the description of the matter formed in heavy-ion collisions. The generalization to the finite temperature case was done in the standard way by introducing the temperature dependence in the fermion distribution functions and adding the pion thermal excitation term. The MKVOR* extension of the MKVOR model \cite{Kolomeitsev:2016ptu}  prevents vanishing of the effective nucleon mass at high density.

We found a redistribution of the charge initially concentrated in the proton subsystem of colliding nuclei between components in isospin-asymmetric systems. It was found that with an increase of the temperature the light pion subsystem becomes more isospin asymmetric, while the baryon subsystem becomes more  isospin symmetric, see fig. \ref{fig::charges}.

Thermodynamical characteristics on the $T-n$ plane were considered for $0.4\leq Y_Z\leq 0.5$ and the  energy--density   and entropy--density  isotherms were constructed, shown in the figs. \ref{fig::EScut03}, \ref{fig::ESmkv}. We further applied our results to the description of heavy-ion collisions for the collision energies $E\lsim 2A$~GeV. To understand the physical picture of the phenomena clearer, we assumed validity of the expanding fireball model although our main results remain valid locally and can be used in hydrodynamical calculations.
As in \cite{Voskresensky:1989sn,Migdal:1990vm,Voskresensky:1993ud},  we  assumed that at energies less than a few $A$~GeV the energy in the center-of-mass system
of the nucleus-nucleus collision, ${\cal E}_{\rm c.m.}A_{\rm part}$, which corresponds
to the nucleons-participants, is spent on the creation of a quasi-equilibrium nuclear fireball at the end of the compression stage (labeled $t=0$),
characterized  by the
temperature $T(0)=T_m$ and the baryon density $n(0)=n_m$ corresponding to the  minimum of
the energy per baryon, $E(n,T_m)/n$,
as a function of the baryon density for $T=T_m$.
Then exploiting an assumption of approximately isoentropic expansion of the system we found the temperature as a function of the baryon density in an expanding fireball and performed a best fit to describe the $\pi^-$ production differential spectra, see fig. \ref{fig::fits}, the ratios of $\pi^-$ to proton multiplicities $R_{\pi^- Z}$, see fig. \ref{fig::RpiZ}, and $\pi^-$ to $\pi^+$ ratios, see fig. \ref{fig:piratios}.  The effects  of taking into account the $\Delta$ isobars and  the $s$-wave pion-nucleon interaction (for $Y_Z\neq 0.5$) on pion differential cross sections, pion to proton and $\pi^-/\pi^+$ ratios were studied. At the assumption of a prompt breakup the contribution to the $\pi^-$ yields from the $\Delta$ decays was evaluated using the in-medium effective masses of $\Delta$s and nucleons. Compared to the works mentioned above,  
we extended our consideration to the case of the charge ratio $Y_Z \neq 0.5$, for instance taking $Y_Z \simeq 0.4$ in case of Au+Au and La+La collisions.
This isospin asymmetry, despite being small compared to that allowed  in  neutron stars or supernovae, plays a role in description of the heavy-ion collisions.
For instance, taking into account  the difference in the  chemical potentials of  neutrons and protons for $Y_Z\neq 0.5$ leads to a noticeable increase of the negative pion yields at the conditions of the fireball breakup.

The maximum and breakup temperatures of the fireball proved to be almost  model independent quantities, see figs. \ref{fig::TmNm}, \ref{fig::breakup},  whereas the maximum reachable density $n_m(\Elab)$ and the breakup density $n_\bup(\Elab)$ proved to be higher in the MKVOR*-based model than in the KVORcut03-based model.
The resulting values of the breakup temperatures and densities
deduced from the pion spectra for the collision energy $\Elab \gsim 800A \, \mev$ proved to be
lower than the ones required to describe the $R_{\pi^- Z}$. The reason of this is twofold. First,  as was shown in \cite{Voskresensky:1993mw}, a large yield of low-momentum pions originates from the reactions occurring before the fireball breakup, because of larger mean-free paths for soft pions. Taking into account the direct reactions could essentially affect the $R_{\pi^-Z}$ rates making values $n_\bup$ and $T_\bup$ deduced from $R_{\pi^-Z}$ lower. Second, we ignored effects of the $p$-wave pion-baryon interaction.
Inclusion of $p$-wave effects leads to a substantial change of the temperature dependence of the pion distribution function \cite{Voskresensky:1993ud}, which should be especially noticeable at low $T$ corresponding to low $\Elab$.
Our evaluations show 30$\%$ larger $\pi^-/\pi^+$ ratios compared with the data for Au$+$Au collisions, that can be again attributed to ignoring the direct pion emission and $p$-wave effects in our calculations.

We also investigated  various characteristics
of the liquid-gas first-order phase transition in isospin symmetric and asymmetric systems within the same KVORcut03-based and MKVOR*-based models. Such a transition occurs for $T\lsim 20$ MeV and for $n<n_0$. At these conditions $\Delta$ isobars and pions do not contribute. Thermodynamical characteristics were presented on  $P-n$, $T-n$, $n_n -n_p$, $\mu_n - \mu_p$ planes, where  $P$ is the pressure, $n_n$,  $n_p$ and $\mu_n$ and $\mu_p$  are neutron and proton densities and   chemical potentials.

The following values characterize the critical point in isospin-symmetric matter  for KVORcut03 (MKVOR*) models:
$n_{\rm cr} = 0.34 (0.32) \, n_0$, $T_{\rm cr} = 17.4 \, (16.05) \,\mev$, and $P_{\rm cr} = 0.30  \,(0.25)\,$MeV$/{\rm fm}^3$. These values lie within the $T_{\rm cr} = (16.4 \pm 2.3)\,\mev$ band given by microscopic calculations with use of the chiral nucleon-nucleon potentials \cite{Carbone:2018kji}. Besides, the KVORcut03 model passes the  constraints on the critical density and pressure (and marginally on critical temperature)  extracted in experimental analysis of reactions going through the compound nuclear decay and  multi-fragmentation \cite{Elliott:2013pna}:
$n_{\rm cr} = (0.06 \pm 0.02) \, {\rm fm^{-3}}$, and $P_{\rm cr} = (0.3 \pm 0.1) \, \mev/{\rm fm}^3$,   $T_{\rm cr} = (17.9 \pm 0.4)\,\mev$. The MKVOR* model passes the constraints for the critical density and pressure, whereas the predicted critical temperature is lower than the  bound extracted in  \cite{Elliott:2013pna}. We demonstrated the system trajectories for various heavy-ion collision energies on $P-n$ and $T-n$ planes demonstrating at which conditions the system passes  instability regions, see figs. \ref{Pn}, \ref{Tn}.

We studied specifics of the liquid-gas phase transition in isospin-asymmetric matter within our models of the EoS. First, disregarding possible effects of the surface tension, following \cite{Ducoin:2005aa} we solved the Gibbs conditions and constructed the equilibrium pressure in the mixed phase, see fig. \ref{PnYp}. Our results are demonstrated in the $n_n - n_p$ and $\mu_n - \mu_p$ planes, see fig. \ref{npnn}, \ref{fig::mumu_nn}.

We evaluated the scaled variances of the baryon and electric charges within the phase transition region and demonstrated that they diverge at the onset of spinodal instabilities, see figs.\ref{Varience}, \ref{VarienceYp}.

In addition, taking into account the non-zero surface tension we formulated the novel possibility of the formation of a structured pasta phase in the  isospin-asymmetric finite nuclear systems in the region of the liquid-gas first-order phase transition.

Concluding, in the given work we demonstrated efficiency of the   KVORcut03-based and MKVOR*-based models of EoS, which have passed the check  for the  description of cold isospin asymmetric nuclear matter \cite{Maslov:2015msa,Maslov:2015wba,Kolomeitsev:2016ptu,Kolomeitsev:2017gli}, now  for the description of heavy-ion collisions at the collision energies below few $A$~GeV.
In the subsequent work we expect  to probe our EoSs in actual hydrodynamical calculations. Besides, we will  take into account effects of the $p$-wave pion-baryon interactions. Moreover, we will generalize our consideration to  higher temperatures, densities and collision energies.

	\section*{Acknowledgements} We would like to thank D. Blaschke, Yu. B. Ivanov, E. E. Kolomeitsev and A. S.
Khvorostukhin  for valuable discussions and M. Borisov and P. Lukyanov for the interest to this work.   This work (Sect. 1-3) has been supported by the
Russian Science Foundation under grant No. 17-12-01427. Work of K.A.M. on Sect. 4 was
supported by the Foundation for the Advancement of Theoretical Physics and Mathematics ``BASIS''. The work of  D.N.V. on Sect. 4 was  supported by   the Ministry of
 Science and High Education of the Russian Federation within the state assignment,
project No 3.6062.2017/6.7.

\section*{Appendix}\label{Appen}
In the paper body we have discussed only effects of the $s$-wave pion-nucleon interaction which are  weak for $0.4\leq Y_Z\leq 0.5$, which we study in this work.
 The important role of the $p$-wave pion-baryon and kaon-baryon interactions has been
 intensively studied in the literature, \textit{cf.}
\cite{Baym:1975tm,Migdal:1978az,Migdal:1990vm,Voskresensky:2016oee,Voskresensky:2018ozf,Kolomeitsev:1995xz,Kolomeitsev:2002pg} and refs. therein.

With taking into account the nucleon-nucleon hole,  the $\Delta$ isobar-nucleon hole,
and   a residual interaction, the Dyson equation for the retarded pion Green function describing
propagation of particles with pion quantum numbers is as follows
\begin{gather}
\includegraphics[width=\linewidth]{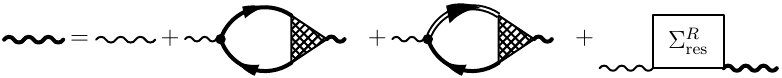}\,.
\nonumber
\end{gather}
Hatched vertices take into account $NN$ and $\Delta N$ correlations.

\begin{figure}
	\centering
	\includegraphics[height=5.5cm,clip=true]{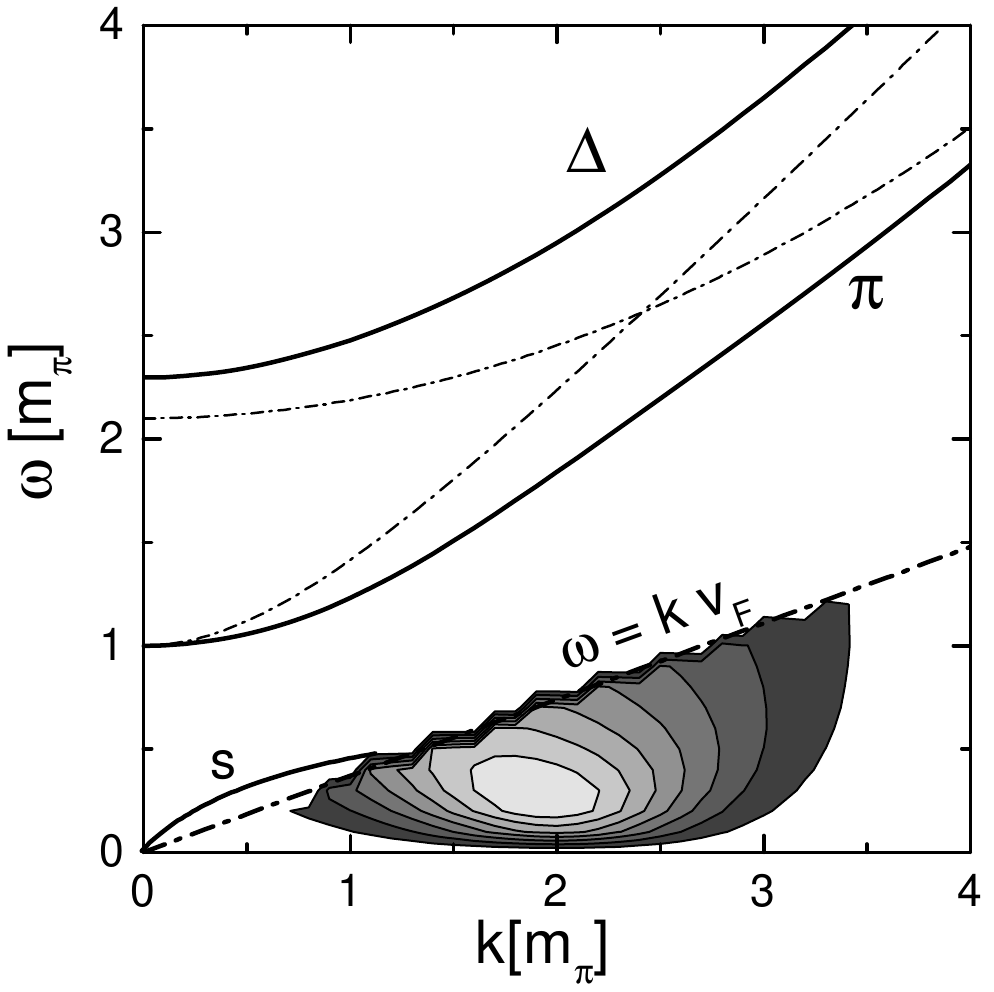}
	\caption{Pion spectrum in ISM for $n=n_0$, $T=0$, \textit{cf.} \cite{Kolomeitsev:2000ie}.}
	\label{fig::piSpectrum}
\end{figure}

 On the pion energy-momentum, $\omega -k$, plane there
  exist regions, where pions with a good accuracy can be treated as  quasiparticles, and regions,
 where the pion spectral function differs significantly
 from the $\delta$ function \cite{Voskresensky:1989sn,Migdal:1990vm,Voskresensky:1993ud}. There are three quasiparticle branches in the pion spectrum in ISM:
 the pion branch $\omega =\omega_\pi (k),
 \omega_\pi \to m_\pi$ for  $k\to 0$ and for $n\ll n_0$, the $\Delta$ branch $\omega =
 \omega_\Delta (k)$,
 $\omega_\Delta (k)\to m_\Delta - m_N$ for  $k\to 0$ and for $n\ll n_0$, and the spin-sound branch
 $\omega =
 \omega_s (k)$, $\omega_s (k)\simeq vk$ for $k\ll p_{{\rm F},N}$. A region, where the spectral function has a large width ($\omega <kv_{{\rm F},N}$
  and $k\sim p_{{\rm F},N}$), describes virtual pions.
 Thereby the spectrum of particles with pion quantum numbers differs significantly from the dispersion
law of the free pions, $\omega_k =\sqrt{m^2_{\pi}+k^2}$. Thus to consider pions in the baryon matter as obeying the vacuum dispersion law is an oversimplification. With taking into account the polarization in the ISM  for $n=n_0$, $T=0$, the pion spectrum is shown in fig. \ref{fig::piSpectrum},\textit{ cf.} \cite{Migdal:1990vm,Voskresensky:2016oee,Voskresensky:2018ozf,Kolomeitsev:2000ie}. 

The medium effects on the pion emission from the fireball depend on a relation between the breakup time-scale $\tau_{\rm b.up}$ and the characteristic time of the quasiparticle absorption. The pion quasiparticle has enough time to transit to the vacuum spectrum branch during the system breakup, if
$$\tau_{\rm b.up} \gsim
{|\omega_i (k; n_{\rm{b.up}}, T_{\rm{b.up}})-\omega_k|}^{-1},$$
where $\omega_i (k; n_{\rm{b.up}}, T_{\rm b.up})$ is the pion energy on the $i$-th quasiparticle branch.   In the opposite limit
 $$\tau_{\rm b.up} < 
{|\omega_i (k; n_{\rm{b.up}}, T_{\rm{b.up}})-\omega_k|}^{-1},$$ the breakup is prompt for pions from those branches and pion polarization should be included as before as at the fireball breakup stage, \textit{cf.} \cite{Senatorov:1989cg}. We point out that  before the fireball breakup (for $n>n_{\rm{b.up}} $)  the pion polarization should be taken into account in any case. Despite that, in the given work we focus on other features of our model and include only $s$-wave pion-nucleon effects. Generalization will be considered in the future work.

\end{document}